\documentclass[pre,eqsecnum,preprint,showkeys,preprintnumbers,amssymb,amsfonts,amsmath]{revtex4}

\input epsf

\usepackage{graphicx}
\usepackage{bm}

\newcommand{\beq}{\begin{equation}}
\newcommand{\eeq}{\end{equation}}
\newcommand{\beqs}{\begin{eqnarray}}
\newcommand{\eeqs}{\end{eqnarray}}

\newtheorem{theo}{Theorem}[section]
\newtheorem{cor}{Corollary}[section]
\newtheorem{lemma}{Lemma}[section]
\newtheorem{defi}{Definition}[section]

\newtheorem{propo}{Proposition}[section]

\newcommand{\EndProof}{\hfill \rule{1ex}{1ex}}

\begin{document}

\title{Hamiltonian paths on the Sierpinski gasket}

\author{Shu-Chiuan Chang$^{a,b}$} 
\email{scchang@mail.ncku.edu.tw} 

\author{Lung-Chi Chen$^{c,d}$} 
\email{lcchen@math.fju.edu.tw}

\affiliation{(a) \ Department of Physics \\
National Cheng Kung University \\
Tainan 70101, Taiwan} 

\bigskip

\affiliation{(b) \ Physics Division \\
National Center for Theoretical Science \\
National Taiwan University \\
Taipei 10617, Taiwan} 

\bigskip

\affiliation{(c) \ Department of Mathematics \\
Fu Jen Catholic University \\
Taipei 24205, Taiwan}

\bigskip

\affiliation{(d) \ Taida Institute for Mathematical Sciences \\
Taipei 10617, Taiwan}


\begin{abstract}

We derive exactly the number of Hamiltonian paths $H(n)$ on the two dimensional Sierpinski gasket $SG(n)$ at stage $n$, whose asymptotic behavior is given by $\frac{\sqrt{3}(2\sqrt{3})^{3^{n-1}}}{3} \times (\frac{5^2 \times 7^2 \times 17^2}{2^{12} \times 3^5 \times 13})(16)^n$. We also obtain the number of Hamiltonian paths with one end at a certain outmost vertex of $SG(n)$, with asymptotic behavior $\frac {\sqrt{3}(2\sqrt{3})^{3^{n-1}}}{3} \times (\frac {7 \times 17}{2^4 \times 3^3})4^n$. The distribution of Hamiltonian paths on $SG(n)$ with one end at a certain outmost vertex and the other end at an arbitrary vertex of $SG(n)$ is investigated. We rigorously prove that the exponent for the mean $\ell$ displacement between the two end vertices of such Hamiltonian paths on $SG(n)$ is $\ell \log 2 / \log 3$ for $\ell>0$.


\end{abstract}

\maketitle

\section{Introduction}
\label{introduction}

The origin of the name, Hamiltonian cycle or Hamiltonian circuit, is invented in 1857 by Sir William Rowan Hamilton when he asked for the construction of a cycle containing all the vertices of a dodecahedron. In general, a Hamiltonian cycle, or a closed Hamiltonian walk, is a cycle in a graph which visits each vertex exactly once and also returns to the starting vertex. A Hamiltonian path, or an open Hamiltonian walk, is a path in a graph which visits each vertex of the graph exactly once \cite{fh, boll}. Determining whether such paths and cycles exist in graphs is difficult in general \cite{garey}. It is one of the oldest problems in graph theory, and is closely related to the traveling salesman problem. A Hamiltonian path corresponds to a complete self-avoiding walk, that can be considered to represent one of the possible configurations of a close-packed, unbranched polymer in a solution \cite{cloizeaux, vanderzande}. It has been also used in the study of protein folding \cite{lua}.  The enumeration of Hamiltonian cycles or paths is a fundamental problem in physics \cite{kasteleyn, duplantier} and computer science \cite{rubin, rahman}. 

It is of interest to consider Hamiltonian paths and cycles on self-similar fractal lattices which have scaling invariance rather than translational invariance. Fractals are geometric structures of non-integer Hausdorff dimension realized by repeated construction of an elementary shape on progressively larger length scales \cite{mandelbrot, Falconer}. For a lattice $\Lambda$ with $v(\Lambda)$ vertices, the number of Hamiltonian paths, $H_\Lambda$, grows as $\omega_\Lambda^{v(\Lambda)}$ for large $v(\Lambda)$, where the connectivity constant $\omega_\Lambda$ is defined as
\beq
\ln \omega_\Lambda = \lim_{v(\Lambda) \to \infty} \frac{\ln H_\Lambda}{v(\Lambda)} \ .
\eeq
It is known that $\omega_\Lambda$ is the same for Hamiltonian cycles and Hamiltonian paths \cite{schmalz, orland}.
Some recent studies on the enumeration of closed and open Hamiltonian walks on various fractal lattices, including $n$-simplex, Sierpinski gasket and its generalization, were carried out in Refs. \cite{bradley89, stajic, hadzic}, 
where the connectivity constants and scaling forms of $H_\Lambda$ were determined. However, except the 3-simplex lattice, the exact expressions of $H_\Lambda$ on the fractal lattices were never obtained. The purpose of this paper is to derive the number of Hamiltonian paths on the two-dimensional Sierpinski gasket exactly in Sections \ref{sectionH0} and \ref{sectionH} that complements the previous asymptotic studies. Furthermore, we should consider the Hamiltonian paths with one end at a certain outmost vertex as defined in Section \ref{distribution}, and investigate the probability distribution of the location of the other end in Sections \ref{fnxm}-\ref{fnxg}. Especially, we show the upper and lower bounds for the mean $\ell$ displacement for $\ell>0$ between these two end vertices and obtain its exponent in Section \ref{fnxg}.

\section{Preliminaries}
\label{preliminary}

We first recall some relevant definitions for graphs and the Sierpinski gasket in this section. A connected graph (without loops) $G=(V,E)$ is defined by its vertex (site) and edge (bond) sets $V$ and $E$ \cite{fh,bbook}.  Let $v(G)=|V|$ be the number of vertices and $e(G)=|E|$ the number of edges in $G$.  The degree or coordination number $k_i$ of a vertex $v_i \in V$ is the number of edges attached to it.  A $k$-regular graph is a graph with the property that each of its vertices has the same degree $k$. 

\begin{figure}[htbp]
\unitlength 0.9mm \hspace*{3mm}
\begin{picture}(108,40)
\put(0,0){\line(1,0){6}}
\put(0,0){\line(3,5){3}}
\put(6,0){\line(-3,5){3}}
\put(3,-4){\makebox(0,0){$SG(0)$}}
\put(12,0){\line(1,0){12}}
\put(12,0){\line(3,5){6}}
\put(24,0){\line(-3,5){6}}
\put(15,5){\line(1,0){6}}
\put(18,0){\line(3,5){3}}
\put(18,0){\line(-3,5){3}}
\put(18,-4){\makebox(0,0){$SG(1)$}}
\put(30,0){\line(1,0){24}}
\put(30,0){\line(3,5){12}}
\put(54,0){\line(-3,5){12}}
\put(36,10){\line(1,0){12}}
\put(42,0){\line(3,5){6}}
\put(42,0){\line(-3,5){6}}
\multiput(33,5)(12,0){2}{\line(1,0){6}}
\multiput(36,0)(12,0){2}{\line(3,5){3}}
\multiput(36,0)(12,0){2}{\line(-3,5){3}}
\put(39,15){\line(1,0){6}}
\put(42,10){\line(3,5){3}}
\put(42,10){\line(-3,5){3}}
\put(42,-4){\makebox(0,0){$SG(2)$}}
\put(60,0){\line(1,0){48}}
\put(72,20){\line(1,0){24}}
\put(60,0){\line(3,5){24}}
\put(84,0){\line(3,5){12}}
\put(84,0){\line(-3,5){12}}
\put(108,0){\line(-3,5){24}}
\put(66,10){\line(1,0){12}}
\put(90,10){\line(1,0){12}}
\put(78,30){\line(1,0){12}}
\put(72,0){\line(3,5){6}}
\put(96,0){\line(3,5){6}}
\put(84,20){\line(3,5){6}}
\put(72,0){\line(-3,5){6}}
\put(96,0){\line(-3,5){6}}
\put(84,20){\line(-3,5){6}}
\multiput(63,5)(12,0){4}{\line(1,0){6}}
\multiput(66,0)(12,0){4}{\line(3,5){3}}
\multiput(66,0)(12,0){4}{\line(-3,5){3}}
\multiput(69,15)(24,0){2}{\line(1,0){6}}
\multiput(72,10)(24,0){2}{\line(3,5){3}}
\multiput(72,10)(24,0){2}{\line(-3,5){3}}
\multiput(75,25)(12,0){2}{\line(1,0){6}}
\multiput(78,20)(12,0){2}{\line(3,5){3}}
\multiput(78,20)(12,0){2}{\line(-3,5){3}}
\put(81,35){\line(1,0){6}}
\put(84,30){\line(3,5){3}}
\put(84,30){\line(-3,5){3}}
\put(84,-4){\makebox(0,0){$SG(3)$}}
\end{picture}

\vspace*{5mm}
\caption{\footnotesize{The first four stages $n=0,1,2,3$ of the two-dimensional Sierpinski gasket $SG(n)$.}} 
\label{sgfig}
\end{figure}
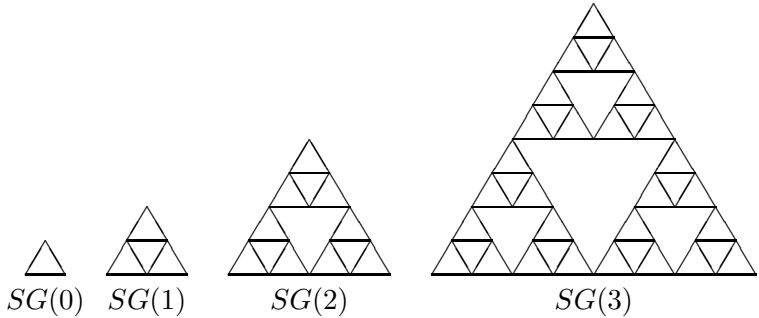

The construction of the two-dimensional Sierpinski gasket $SG(n)$ at stage $n$ is shown in Fig. \ref{sgfig}. At stage $n=0$, it is an equilateral triangle; while stage $(n+1)$ is obtained by the juxtaposition of three $n$-stage structures. The numbers of edges and vertices for $SG(n)$ are given by 
\beq
e(SG(n)) = 3^{n+1} \ , \qquad
v(SG(n)) = \frac32 (3^n+1) \ .
\eeq
Except the three outmost vertices which have degree two, all other vertices of $SG(n)$ have degree four, such that $SG(n)$ is 4-regular in the large $n$ limit. 

\bigskip

Use the notation $o$ (origin), $a_n$ and $b_n$ for the leftmost, rightmost and topmost vertices of $SG(n)$, respectively, and the notation for general vertices will be given in Section \ref{distribution}. The total number of Hamiltonian paths on $SG(n)$ will be denoted as $H(n)$, while its subset, the number of Hamiltonian paths with one end at $o$, is denoted as $H_0(n)$. The number of Hamiltonian cycles on $SG(n)$ will be denoted as $HC(n)$. Let us define the quantities to be used as follows.

\bigskip

\begin{defi} \label{defif} Consider the two-dimensional Sierpinski gasket $SG(n)$ at stage $n$. (i) Define $f_1(n)$ as the number of Hamiltonian paths whose end vertices are located at two certain outmost vertices, e.g. $o$ and $a_n$. (ii) Define $f_2(n)$ as the number of Hamiltonian paths whose one end vertex is located at a certain outmost vertex, e.g., $o$, while the other end vertex is not at the other two outmost vertices. (iii) Define $f_3(n)$ as the number of Hamiltonian paths whose both end vertices are not located at any of the outmost vertices. 
\end{defi}

It follows that
\beq
H_0(n) = 2f_1(n) + f_2(n) \ ,
\label{H0}
\eeq
\beq
H(n) = 3f_1(n) + 3f_2(n) + f_3(n)
\label{H}
\eeq
for the numbers of Hamiltonian paths and
\beq
HC(n) = f_1(n-1)^3
\label{S}
\eeq
for the number of Hamiltonian cycles. To calculate $f_1(n)$, $f_2(n)$ and $f_3(n)$, we need the following definitions.

\begin{defi} \label{defig} Consider the two-dimensional Sierpinski gasket without one certain outmost vertices, e.g. $SG(n) \setminus \{b_n\}$. (i) Define $g_1(n)$ as the number of Hamiltonian paths whose end vertices are located at the remaining two outmost vertices, e.g. $o$ and $a_n$. (ii) Define $g_2(n)$ as the number of Hamiltonian paths whose one end vertex is located at one certain remaining outmost vertices, e.g., $o$, while the other end vertex is not at the other outmost vertex. 
\end{defi}

\begin{defi} \label{defign} Divide the vertices of the two-dimensional Sierpinski gasket into two subsets $V_1(n)$ and $V_2(n)$. One of them contains two outmost vertices while the other contains only one outmost vertex, e.g. $o, a_n \in V_1(n)$, $b_n \in V_2(n)$. (i) Define $g_3(n)$ as the number of two Hamiltonian paths separately visiting all the vertices of $V_1(n)$ and $V_2(n)$, such that one of them has end vertices located at the two outmost vertices in $V_1(n)$, e.g. $o$ and $a_n$, while the other one has one end vertex located at the other outmost vertex in $V_2(n)$, e.g. $b_n$. (ii) Define $g_4(n)$ as the number of two Hamiltonian paths separately visiting all the vertices $V_1(n)$ and $V_2(n)$, such that one of them has only one end vertex located at a certain outmost vertex in $V_1(n)$, e.g. $o$, while the other one has one end vertex located at the outmost vertex in $V_2(n)$, e.g. $b_n$.
\end{defi}

\begin{defi} \label{defih} Consider the two-dimensional Sierpinski gasket without two certain outmost vertices, e.g. $SG(n) \setminus \{o, a_n\}$, and
define $t_1(n)$ as the number of Hamiltonian paths on it.
\end{defi}

\begin{defi} \label{defihn} Consider the two-dimensional Sierpinski gasket without one certain outmost vertices, e.g. $SG(n) \setminus \{b_n\}$, and divide its vertices into two subsets $V_1^\prime(n)$ and $V_2^\prime(n)$. Both of these subsets contain one outmost vertex, e.g. $o \in V_1^\prime$, $a_n \in V_2^\prime(n)$.  Define $t_2(n)$ as the number of two Hamiltonian paths separately visiting all the vertices of $V_1^\prime(n)$ and $V_2^\prime(n)$, such that both of them have one end vertex located at an outmost vertex, e.g. $o$ and $a_n$.
\end{defi}

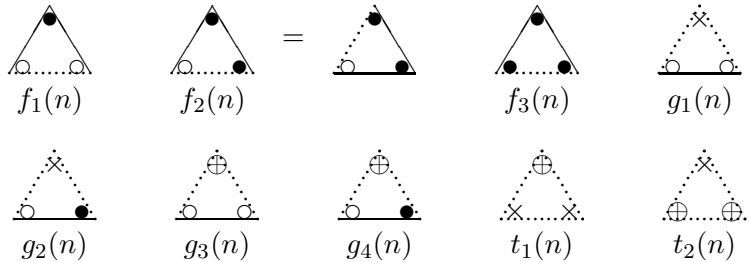
\begin{figure}[htbp]
\unitlength 1.8mm 
\begin{picture}(54,5)
\put(0,0){\line(3,5){3}}
\put(6,0){\line(-3,5){3}}
\multiput(0,0)(0.5,0){13}{\circle*{0.2}}
\multiput(1,0.5)(4,0){2}{\circle{1}}
\put(3,4){\circle*{1}}
\put(3,-2){\makebox(0,0){$f_1(n)$}}
\put(12,0){\line(3,5){3}}
\put(18,0){\line(-3,5){3}}
\multiput(12,0)(0.5,0){13}{\circle*{0.2}}
\put(13,0.5){\circle{1}}
\put(17,0.5){\circle*{1}}
\put(15,4){\circle*{1}}
\put(15,-2){\makebox(0,0){$f_2(n)$}}
\put(21,2.5){\makebox(0,0){$=$}}
\put(24,0){\line(1,0){6}}
\put(30,0){\line(-3,5){3}}
\multiput(24,0)(0.3,0.5){11}{\circle*{0.2}}
\put(25,0.5){\circle{1}}
\put(29,0.5){\circle*{1}}
\put(27,4){\circle*{1}}
\put(36,0){\line(3,5){3}}
\put(42,0){\line(-3,5){3}}
\multiput(36,0)(0.5,0){13}{\circle*{0.2}}
\multiput(37,0.5)(4,0){2}{\circle*{1}}
\put(39,4){\circle*{1}}
\put(39,-2){\makebox(0,0){$f_3(n)$}}
\put(48,0){\line(1,0){6}}
\multiput(48,0)(0.3,0.5){11}{\circle*{0.2}}
\multiput(54,0)(-0.3,0.5){11}{\circle*{0.2}}
\multiput(49,0.5)(4,0){2}{\circle{1}}
\put(51,4){\makebox(0,0){$\times$}}
\put(51,-2){\makebox(0,0){$g_1(n)$}}
\end{picture}
\vspace*{10mm}

\unitlength 1.8mm 
\begin{picture}(54,5)
\put(0,0){\line(1,0){6}}
\multiput(0,0)(0.3,0.5){11}{\circle*{0.2}}
\multiput(6,0)(-0.3,0.5){11}{\circle*{0.2}}
\put(1,0.5){\circle{1}}
\put(5,0.5){\circle*{1}}
\put(3,4){\makebox(0,0){$\times$}}
\put(3,-2){\makebox(0,0){$g_2(n)$}}
\put(12,0){\line(1,0){6}}
\multiput(12,0)(0.3,0.5){11}{\circle*{0.2}}
\multiput(18,0)(-0.3,0.5){11}{\circle*{0.2}}
\multiput(13,0.5)(4,0){2}{\circle{1}}
\put(15,4){\makebox(0,0){$\oplus$}}
\put(15,-2){\makebox(0,0){$g_3(n)$}}
\put(24,0){\line(1,0){6}}
\multiput(24,0)(0.3,0.5){11}{\circle*{0.2}}
\multiput(30,0)(-0.3,0.5){11}{\circle*{0.2}}
\put(27,4){\makebox(0,0){$\oplus$}}
\put(25,0.5){\circle{1}}
\put(29,0.5){\circle*{1}}
\put(27,-2){\makebox(0,0){$g_4(n)$}}
\multiput(36,0)(0.5,0){13}{\circle*{0.2}}
\multiput(36,0)(0.3,0.5){11}{\circle*{0.2}}
\multiput(42,0)(-0.3,0.5){11}{\circle*{0.2}}
\multiput(37,0.5)(4,0){2}{\makebox(0,0){$\times$}}
\put(39,4){\makebox(0,0){$\oplus$}}
\put(39,-2){\makebox(0,0){$t_1(n)$}}
\multiput(48,0)(0.5,0){13}{\circle*{0.2}}
\multiput(48,0)(0.3,0.5){11}{\circle*{0.2}}
\multiput(54,0)(-0.3,0.5){11}{\circle*{0.2}}
\multiput(49,0.5)(4,0){2}{\makebox(0,0){$\oplus$}}
\put(51,4){\makebox(0,0){$\times$}}
\put(51,-2){\makebox(0,0){$t_2(n)$}}
\end{picture}

\vspace*{5mm}
\caption{\footnotesize{Illustration for the quantities $f_1(n)$, $f_2(n)$, $f_3(n)$, $g_1(n)$, $g_2(n)$, $g_3(n)$, $g_4(n)$, $t_1(n)$ and $t_2(n)$. Two outmost vertices connected by a solid line belong to the same path. The outmost vertices are denoted by the following symbols: (i) an open circle corresponds to an end of a path; (ii) a solid circle corresponds to a middle point of a path; (iii) a cross is not passed by any paths; (iv) a symbol $\oplus$ corresponds to one end of a path but it is not connected with the other two outmost vertices. See text for details.}} 
\label{fghfig}
\end{figure}

The quantities defined above are illustrated in Fig. \ref{fghfig}. Note that there are two equivalent ways to draw $f_2(n)$. When $n=0$, only $f_1(0)$ and $g_1(0)$ are equal to one, and $f_2(0)=f_3(0)=g_2(0)=g_3(0)=g_4(0)=t_1(0)=t_2(0)=0$.

\section{Number of Hamiltonian cycles and Number of Hamiltonian paths with one end at origin}  
\label{sectionH0}

In this section, we first rederive the number of Hamiltonian cycles, then enumerate the number of Hamiltonian paths with one end at origin. By Eqs. (\ref{H0}) and (\ref{S}), it is sufficient to evaluate $f_1(n)$ and $f_2(n)$. We shall only consider $n \geq 1$ throughout this paper as the case $n=0$ is trivial.

\subsection{Number of Hamiltonian cycles}

\begin{figure}[htbp]
\unitlength 1.2mm 
\begin{picture}(48,10)
\put(0,0){\line(3,5){6}}
\put(12,0){\line(-3,5){6}}
\multiput(0,0)(1,0){13}{\circle*{0.2}}
\multiput(1,0.5)(10,0){2}{\circle{1}}
\put(6,9){\circle*{1}}
\put(15,5){\makebox(0,0){$=$}}
\put(18,0){\line(1,0){6}}
\put(24,0){\line(-3,5){3}}
\put(21,5){\line(3,5){3}}
\put(30,0){\line(-3,5){6}}
\multiput(18,0)(0.6,1){6}{\circle*{0.2}}
\multiput(21,5)(1,0){7}{\circle*{0.2}}
\multiput(24,0)(0.6,1){6}{\circle*{0.2}}
\multiput(24,0)(1,0){7}{\circle*{0.2}}
\multiput(19,0.5)(10,0){2}{\circle{1}}
\put(24,9){\circle*{1}}
\put(23,0.5){\circle*{1}}
\put(21,4){\circle{1}}
\multiput(22,5.5)(4,0){2}{\circle{1}}
\put(27,4){\circle{1}}
\put(25,0.5){\makebox(0,0){$\times$}}
\put(33,5){\makebox(0,0){$+$}}
\put(36,0){\line(3,5){6}}
\put(42,10){\line(3,-5){3}}
\put(45,5){\line(-3,-5){3}}
\put(42,0){\line(1,0){6}}
\multiput(36,0)(1,0){7}{\circle*{0.2}}
\multiput(42,0)(-0.6,1){6}{\circle*{0.2}}
\multiput(39,5)(1,0){7}{\circle*{0.2}}
\multiput(45,5)(0.6,-1){6}{\circle*{0.2}}
\multiput(37,0.5)(10,0){2}{\circle{1}}
\put(42,9){\circle*{1}}
\put(41,0.5){\makebox(0,0){$\times$}}
\put(39,4){\circle{1}}
\multiput(40,5.5)(4,0){2}{\circle{1}}
\put(43,0.5){\circle*{1}}
\put(45,4){\circle{1}}
\end{picture}

\caption{\footnotesize{Illustration for the expression of $f_1(n+1)$.}} 
\label{f1fig}
\end{figure}
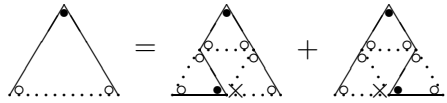

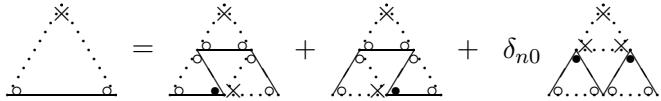
\begin{figure}[htbp]
\unitlength 1.2mm 
\begin{picture}(72,10)
\put(0,0){\line(1,0){12}}
\multiput(0,0)(0.6,1){11}{\circle*{0.2}}
\multiput(12,0)(-0.6,1){11}{\circle*{0.2}}
\multiput(1,0.5)(10,0){2}{\circle{1}}
\put(6,9){\makebox(0,0){$\times$}}
\put(15,5){\makebox(0,0){$=$}}
\put(18,0){\line(1,0){6}}
\put(24,0){\line(-3,5){3}}
\put(21,5){\line(1,0){6}}
\put(27,5){\line(3,-5){3}}
\multiput(18,0)(0.6,1){11}{\circle*{0.2}}
\multiput(24,10)(0.6,-1){6}{\circle*{0.2}}
\multiput(27,5)(-0.6,-1){6}{\circle*{0.2}}
\multiput(24,0)(1,0){7}{\circle*{0.2}}
\multiput(19,0.5)(10,0){2}{\circle{1}}
\put(24,9){\makebox(0,0){$\times$}}
\put(23,0.5){\circle*{1}}
\put(21,4){\circle{1}}
\multiput(22,5.5)(4,0){2}{\circle{1}}
\put(25,0.5){\makebox(0,0){$\times$}}
\put(27,4){\circle{1}}
\put(33,5){\makebox(0,0){$+$}}
\put(36,0){\line(3,5){3}}
\put(39,5){\line(1,0){6}}
\put(45,5){\line(-3,-5){3}}
\put(42,0){\line(1,0){6}}
\multiput(36,0)(1,0){7}{\circle*{0.2}}
\multiput(42,0)(-0.6,1){6}{\circle*{0.2}}
\multiput(39,5)(0.6,1){6}{\circle*{0.2}}
\multiput(42,10)(0.6,-1){11}{\circle*{0.2}}
\multiput(37,0.5)(10,0){2}{\circle{1}}
\put(42,9){\makebox(0,0){$\times$}}
\put(41,0.5){\makebox(0,0){$\times$}}
\put(39,4){\circle{1}}
\multiput(40,5.5)(4,0){2}{\circle{1}}
\put(43,0.5){\circle*{1}}
\put(45,4){\circle{1}}
\put(51,5){\makebox(0,0){$+$}}
\put(57,5){\makebox(0,0){$\delta_{n0}$}}
\multiput(60,0)(6,0){2}{\line(3,5){3}}
\multiput(66,0)(6,0){2}{\line(-3,5){3}}
\multiput(60,0)(1,0){13}{\circle*{0.2}}
\multiput(63,5)(1,0){7}{\circle*{0.2}}
\multiput(63,5)(0.6,1){6}{\circle*{0.2}}
\multiput(69,5)(-0.6,1){6}{\circle*{0.2}}
\multiput(61,0.5)(10,0){2}{\circle{1}}
\put(66,9){\makebox(0,0){$\times$}}
\multiput(65,0.5)(2,0){2}{\circle{1}}
\multiput(63,4)(6,0){2}{\circle*{1}}
\multiput(64,5.5)(4,0){2}{\makebox(0,0){$\times$}}
\end{picture}

\caption{\footnotesize{Illustration for the expression of $g_1(n+1)$. The last drawing only applies for $n=0$.}} 
\label{g1fig}
\end{figure}

Consider the quantities $f_1(n)$ and $g_1(n)$. The initial values are $f_1(1)=2$, $g_1(1)=3$ as illustrated in Figs. \ref{f1fig}, \ref{g1fig}. For $n \geq 1$, we have the simple relations
\beq
f_1(n+1) = 2f_1(n)^2g_1(n) \ , \qquad
g_1(n+1) = 2g_1(n)^2f_1(n) \ , 
\label{fg1}
\eeq
such that 
\beq
\frac{f_1(n+1)}{g_1(n+1)} = \frac{f_1(n)}{g_1(n)} = \frac{f_1(1)}{g_1(1)} = \frac 23 \ . 
\label{fg2}
\eeq

\vspace*{5mm}

\noindent
Hence, we have
\beq
f_1(n) = 3f_1(n-1)^3 = \cdots = 3^{\frac{3^{n-1}-1}{2}} f_1(1)^{3^{n-1}} 
= 3^{\frac{3^{n-1}-1}{2}} (2^{3^{n-1}}) = \frac{\sqrt{3}(2\sqrt{3})^{3^{n-1}}}{3}
\label{f1}
\eeq
and
\beq
g_1(n) = \frac{3f_1(n)}{2} = \frac{\sqrt{3}(2\sqrt{3})^{3^{n-1}}}{2}
\label{g1}
\eeq
for $n \ge 1$.
 
By the definition of Hamiltonian cycles, its number is given by $HC(1)=1$ and $HC(n) = f_1(n-1)^3 = f_1(n)/3$ for $n \geq 2$ as obtained by Bradley \cite{bradley86}.

\bigskip

\begin{theo} [Bradley, 1986] 
The number of Hamiltonian cycles $HC(n)$ on the Sierpinski gasket $SG(n)$ is one for $n=1$ and 
\beq
HC(n) = \frac{\sqrt{3}(2\sqrt{3})^{3^{n-1}}}{9}
\eeq
for $n \ge 2$, such that the connectivity constant $\omega_{SG}=12^{1/9}$.
\label{hc}
\end{theo}

\subsection{Number of Hamiltonian paths with one end at origin}

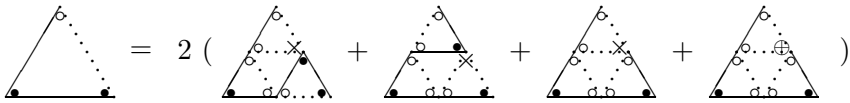
\begin{figure}[htbp]
\unitlength 1.2mm 
\begin{picture}(111,10)
\put(0,0){\line(1,0){12}}
\put(0,0){\line(3,5){6}}
\multiput(12,0)(-0.6,1){11}{\circle*{0.2}}
\multiput(1,0.5)(10,0){2}{\circle*{1}}
\put(6,9){\circle{1}}
\put(15,5){\makebox(0,0){$=$}}
\put(21,5){\makebox(0,0){$2 \ ($}}
\put(30,10){\line(-3,-5){6}}
\put(24,0){\line(1,0){6}}
\put(30,0){\line(3,5){3}}
\put(33,5){\line(3,-5){3}}
\multiput(30,10)(0.6,-1){6}{\circle*{0.2}}
\multiput(33,5)(-1,0){7}{\circle*{0.2}}
\multiput(27,5)(0.6,-1){6}{\circle*{0.2}}
\multiput(30,0)(1,0){7}{\circle*{0.2}}
\multiput(25,0.5)(10,0){2}{\circle*{1}}
\put(30,9){\circle{1}}
\put(29,0.5){\circle{1}}
\put(27,4){\circle{1}}
\put(31,0.5){\circle{1}}
\put(33,4){\circle*{1}}
\put(28,5.5){\circle{1}}
\put(32,5.5){\makebox(0,0){$\times$}}
\put(39,5){\makebox(0,0){$+$}}
\put(48,10){\line(3,-5){3}}
\put(51,5){\line(-1,0){6}}
\put(45,5){\line(-3,-5){3}}
\put(42,0){\line(1,0){12}}
\multiput(48,10)(-0.6,-1){6}{\circle*{0.2}}
\multiput(45,5)(0.6,-1){6}{\circle*{0.2}}
\multiput(48,0)(0.6,1){6}{\circle*{0.2}}
\multiput(51,5)(0.6,-1){6}{\circle*{0.2}}
\multiput(43,0.5)(10,0){2}{\circle*{1}}
\put(48,9){\circle{1}}
\put(47,0.5){\circle{1}}
\put(45,4){\circle{1}}
\put(49,0.5){\circle{1}}
\put(51,4){\makebox(0,0){$\times$}}
\put(46,5.5){\circle{1}}
\put(50,5.5){\circle*{1}}
\put(57,5){\makebox(0,0){$+$}}
\put(66,10){\line(-3,-5){6}}
\put(60,0){\line(1,0){12}}
\put(72,0){\line(-3,5){3}}
\multiput(66,10)(0.6,-1){6}{\circle*{0.2}}
\multiput(69,5)(-1,0){7}{\circle*{0.2}}
\multiput(63,5)(0.6,-1){6}{\circle*{0.2}}
\multiput(66,0)(0.6,1){6}{\circle*{0.2}}
\multiput(61,0.5)(10,0){2}{\circle*{1}}
\put(66,9){\circle{1}}
\multiput(65,0.5)(2,0){2}{\circle{1}}
\multiput(63,4)(6,0){2}{\circle{1}}
\put(64,5.5){\circle{1}}
\put(68,5.5){\makebox(0,0){$\times$}}
\put(75,5){\makebox(0,0){$+$}}
\put(84,10){\line(-3,-5){6}}
\put(78,0){\line(1,0){12}}
\put(90,0){\line(-3,5){3}}
\multiput(84,10)(0.6,-1){6}{\circle*{0.2}}
\multiput(87,5)(-1,0){7}{\circle*{0.2}}
\multiput(81,5)(0.6,-1){6}{\circle*{0.2}}
\multiput(84,0)(0.6,1){6}{\circle*{0.2}}
\multiput(79,0.5)(10,0){2}{\circle*{1}}
\put(84,9){\circle{1}}
\multiput(83,0.5)(2,0){2}{\circle{1}}
\multiput(81,4)(6,0){2}{\circle{1}}
\put(82,5.5){\circle{1}}
\put(86,5.5){\makebox(0,0){\scriptsize $\oplus$}}
\put(93,5){\makebox(0,0){$)$}}
\end{picture}

\caption{\footnotesize{Illustration for the expression of $f_2(n+1)$. The multiplication of two on the right-hand-side corresponds to the reflection symmetry with respect to the central vertical axis.}} 
\label{f2fig}
\end{figure}

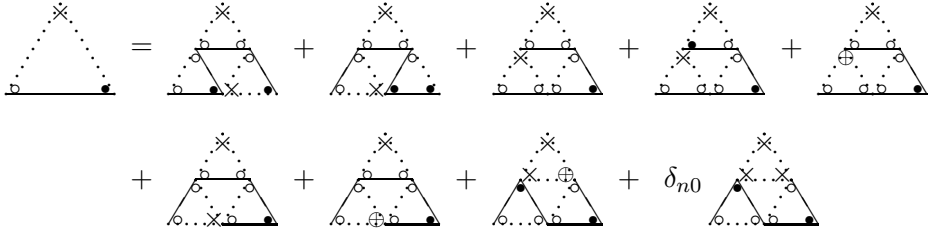
\begin{figure}[htbp]
\unitlength 1.2mm 
\begin{picture}(108,10)
\put(0,0){\line(1,0){12}}
\multiput(0,0)(0.6,1){11}{\circle*{0.2}}
\multiput(12,0)(-0.6,1){11}{\circle*{0.2}}
\put(1,0.5){\circle{1}}
\put(11,0.5){\circle*{1}}
\put(6,9){\makebox(0,0){$\times$}}
\put(15,5){\makebox(0,0){$=$}}
\put(18,0){\line(1,0){6}}
\put(24,0){\line(-3,5){3}}
\put(21,5){\line(1,0){6}}
\put(27,5){\line(3,-5){3}}
\multiput(18,0)(0.6,1){11}{\circle*{0.2}}
\multiput(24,10)(0.6,-1){6}{\circle*{0.2}}
\multiput(27,5)(-0.6,-1){6}{\circle*{0.2}}
\multiput(24,0)(1,0){7}{\circle*{0.2}}
\put(19,0.5){\circle{1}}
\put(29,0.5){\circle*{1}}
\put(24,9){\makebox(0,0){$\times$}}
\put(23,0.5){\circle*{1}}
\put(21,4){\circle{1}}
\multiput(22,5.5)(4,0){2}{\circle{1}}
\put(25,0.5){\makebox(0,0){$\times$}}
\put(27,4){\circle{1}}
\put(33,5){\makebox(0,0){$+$}}
\put(36,0){\line(3,5){3}}
\put(39,5){\line(1,0){6}}
\put(45,5){\line(-3,-5){3}}
\put(42,0){\line(1,0){6}}
\multiput(36,0)(1,0){7}{\circle*{0.2}}
\multiput(42,0)(-0.6,1){6}{\circle*{0.2}}
\multiput(39,5)(0.6,1){6}{\circle*{0.2}}
\multiput(42,10)(0.6,-1){11}{\circle*{0.2}}
\put(37,0.5){\circle{1}}
\put(47,0.5){\circle*{1}}
\put(42,9){\makebox(0,0){$\times$}}
\put(41,0.5){\makebox(0,0){$\times$}}
\put(39,4){\circle{1}}
\multiput(40,5.5)(4,0){2}{\circle{1}}
\put(43,0.5){\circle*{1}}
\put(45,4){\circle{1}}
\put(51,5){\makebox(0,0){$+$}}
\put(54,0){\line(1,0){12}}
\put(66,0){\line(-3,5){3}}
\put(63,5){\line(-1,0){6}}
\multiput(54,0)(0.6,1){11}{\circle*{0.2}}
\multiput(60,10)(0.6,-1){6}{\circle*{0.2}}
\multiput(60,0)(0.6,1){6}{\circle*{0.2}}
\multiput(60,0)(-0.6,1){6}{\circle*{0.2}}
\put(55,0.5){\circle{1}}
\put(65,0.5){\circle*{1}}
\put(60,9){\makebox(0,0){$\times$}}
\put(59,0.5){\circle{1}}
\put(57,4){\makebox(0,0){$\times$}}
\multiput(58,5.5)(4,0){2}{\circle{1}}
\put(61,0.5){\circle{1}}
\put(63,4){\circle{1}}
\put(69,5){\makebox(0,0){$+$}}
\put(72,0){\line(1,0){12}}
\put(84,0){\line(-3,5){3}}
\put(81,5){\line(-1,0){6}}
\multiput(72,0)(0.6,1){11}{\circle*{0.2}}
\multiput(78,10)(0.6,-1){6}{\circle*{0.2}}
\multiput(78,0)(0.6,1){6}{\circle*{0.2}}
\multiput(78,0)(-0.6,1){6}{\circle*{0.2}}
\put(73,0.5){\circle{1}}
\put(83,0.5){\circle*{1}}
\put(78,9){\makebox(0,0){$\times$}}
\put(77,0.5){\circle{1}}
\put(75,4){\makebox(0,0){$\times$}}
\put(76,5.5){\circle*{1}}
\put(80,5.5){\circle{1}}
\put(79,0.5){\circle{1}}
\put(81,4){\circle{1}}
\put(87,5){\makebox(0,0){$+$}}
\put(90,0){\line(1,0){12}}
\put(102,0){\line(-3,5){3}}
\put(99,5){\line(-1,0){6}}
\multiput(90,0)(0.6,1){11}{\circle*{0.2}}
\multiput(96,10)(0.6,-1){6}{\circle*{0.2}}
\multiput(96,0)(0.6,1){6}{\circle*{0.2}}
\multiput(96,0)(-0.6,1){6}{\circle*{0.2}}
\put(91,0.5){\circle{1}}
\put(101,0.5){\circle*{1}}
\put(96,9){\makebox(0,0){$\times$}}
\put(95,0.5){\circle{1}}
\put(93,4){\makebox(0,0){\scriptsize $\oplus$}}
\multiput(94,5.5)(4,0){2}{\circle{1}}
\put(97,0.5){\circle{1}}
\put(99,4){\circle{1}}
\end{picture}

\vspace*{5mm}
\begin{picture}(108,10)
\put(15,5){\makebox(0,0){$+$}}
\put(18,0){\line(3,5){3}}
\multiput(21,5)(3,-5){2}{\line(1,0){6}}
\put(27,5){\line(3,-5){3}}
\multiput(18,0)(1,0){7}{\circle*{0.2}}
\multiput(21,5)(0.6,1){6}{\circle*{0.2}}
\multiput(21,5)(0.6,-1){6}{\circle*{0.2}}
\multiput(27,5)(-0.6,1){6}{\circle*{0.2}}
\multiput(27,5)(-0.6,-1){6}{\circle*{0.2}}
\put(19,0.5){\circle{1}}
\put(29,0.5){\circle*{1}}
\put(24,9){\makebox(0,0){$\times$}}
\put(23,0.5){\makebox(0,0){$\times$}}
\multiput(21,4)(6,0){2}{\circle{1}}
\put(25,0.5){\circle{1}}
\multiput(22,5.5)(4,0){2}{\circle{1}}
\put(33,5){\makebox(0,0){$+$}}
\put(36,0){\line(3,5){3}}
\multiput(39,5)(3,-5){2}{\line(1,0){6}}
\put(45,5){\line(3,-5){3}}
\multiput(36,0)(1,0){7}{\circle*{0.2}}
\multiput(39,5)(0.6,1){6}{\circle*{0.2}}
\multiput(39,5)(0.6,-1){6}{\circle*{0.2}}
\multiput(45,5)(-0.6,1){6}{\circle*{0.2}}
\multiput(45,5)(-0.6,-1){6}{\circle*{0.2}}
\put(37,0.5){\circle{1}}
\put(47,0.5){\circle*{1}}
\put(42,9){\makebox(0,0){$\times$}}
\put(41,0.5){\makebox(0,0){\scriptsize $\oplus$}}
\multiput(39,4)(6,0){2}{\circle{1}}
\put(43,0.5){\circle{1}}
\multiput(40,5.5)(4,0){2}{\circle{1}}
\put(51,5){\makebox(0,0){$+$}}
\put(54,0){\line(3,5){3}}
\put(57,5){\line(3,-5){3}}
\put(60,0){\line(1,0){6}}
\put(66,0){\line(-3,5){3}}
\multiput(54,0)(1,0){7}{\circle*{0.2}}
\multiput(57,5)(1,0){7}{\circle*{0.2}}
\multiput(60,0)(0.6,1){6}{\circle*{0.2}}
\multiput(60,10)(0.6,-1){6}{\circle*{0.2}}
\multiput(60,10)(-0.6,-1){6}{\circle*{0.2}}
\put(55,0.5){\circle{1}}
\put(65,0.5){\circle*{1}}
\put(60,9){\makebox(0,0){$\times$}}
\multiput(59,0.5)(2,0){2}{\circle{1}}
\put(57,4){\circle*{1}}
\put(63,4){\circle{1}}
\put(58,5.5){\makebox(0,0){$\times$}}
\put(62,5.5){\makebox(0,0){\scriptsize $\oplus$}}
\put(69,5){\makebox(0,0){$+$}}
\put(75,5){\makebox(0,0){$\delta_{n0}$}}
\put(78,0){\line(3,5){3}}
\put(81,5){\line(3,-5){3}}
\put(84,0){\line(1,0){6}}
\put(90,0){\line(-3,5){3}}
\multiput(78,0)(1,0){7}{\circle*{0.2}}
\multiput(81,5)(1,0){7}{\circle*{0.2}}
\multiput(84,0)(0.6,1){6}{\circle*{0.2}}
\multiput(84,10)(0.6,-1){6}{\circle*{0.2}}
\multiput(84,10)(-0.6,-1){6}{\circle*{0.2}}
\put(79,0.5){\circle{1}}
\put(89,0.5){\circle*{1}}
\put(84,9){\makebox(0,0){$\times$}}
\multiput(83,0.5)(2,0){2}{\circle{1}}
\put(81,4){\circle*{1}}
\put(87,4){\circle{1}}
\put(82,5.5){\makebox(0,0){$\times$}}
\put(86,5.5){\makebox(0,0){$\times$}}
\end{picture}

\caption{\footnotesize{Illustration for the expression of $g_2(n+1)$. The last drawing only applies for $n=0$.}} 
\label{g2fig}
\end{figure}

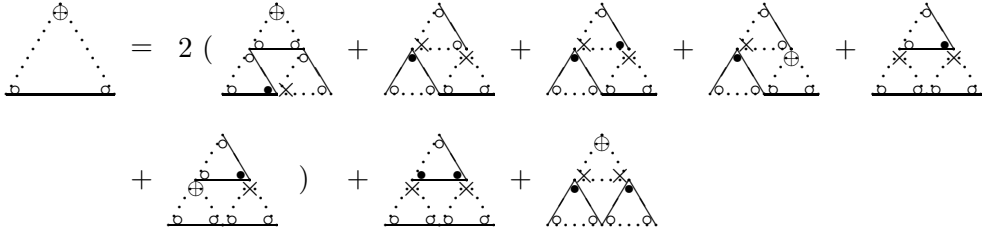
\begin{figure}[htbp]
\unitlength 1.2mm 
\begin{picture}(108,10)
\put(0,0){\line(1,0){12}}
\multiput(0,0)(0.6,1){11}{\circle*{0.2}}
\multiput(12,0)(-0.6,1){11}{\circle*{0.2}}
\multiput(1,0.5)(10,0){2}{\circle{1}}
\put(6,9){\makebox(0,0){\scriptsize $\oplus$}}
\put(15,5){\makebox(0,0){$=$}}
\put(21,5){\makebox(0,0){$2 \ ($}}
\put(24,0){\line(1,0){6}}
\put(30,0){\line(-3,5){3}}
\put(27,5){\line(1,0){6}}
\put(33,5){\line(3,-5){3}}
\multiput(24,0)(0.6,1){11}{\circle*{0.2}}
\multiput(30,10)(0.6,-1){6}{\circle*{0.2}}
\multiput(33,5)(-0.6,-1){6}{\circle*{0.2}}
\multiput(30,0)(1,0){7}{\circle*{0.2}}
\multiput(25,0.5)(10,0){2}{\circle{1}}
\put(30,9){\makebox(0,0){\scriptsize $\oplus$}}
\put(29,0.5){\circle*{1}}
\put(27,4){\circle{1}}
\multiput(28,5.5)(4,0){2}{\circle{1}}
\put(31,0.5){\makebox(0,0){$\times$}}
\put(33,4){\circle{1}}
\put(39,5){\makebox(0,0){$+$}}
\put(42,0){\line(3,5){3}}
\put(45,5){\line(3,-5){3}}
\put(48,0){\line(1,0){6}}
\put(48,10){\line(3,-5){3}}
\multiput(42,0)(1,0){7}{\circle*{0.2}}
\multiput(48,0)(0.6,1){6}{\circle*{0.2}}
\multiput(51,5)(0.6,-1){6}{\circle*{0.2}}
\multiput(45,5)(1,0){7}{\circle*{0.2}}
\multiput(45,5)(0.6,1){6}{\circle*{0.2}}
\multiput(43,0.5)(10,0){2}{\circle{1}}
\put(48,9){\circle{1}}
\multiput(47,0.5)(2,0){2}{\circle{1}}
\put(45,4){\circle*{1}}
\put(46,5.5){\makebox(0,0){$\times$}}
\put(50,5.5){\circle{1}}
\put(51,4){\makebox(0,0){$\times$}}
\put(57,5){\makebox(0,0){$+$}}
\put(60,0){\line(3,5){3}}
\put(63,5){\line(3,-5){3}}
\put(66,0){\line(1,0){6}}
\put(66,10){\line(3,-5){3}}
\multiput(60,0)(1,0){7}{\circle*{0.2}}
\multiput(66,0)(0.6,1){6}{\circle*{0.2}}
\multiput(69,5)(0.6,-1){6}{\circle*{0.2}}
\multiput(63,5)(1,0){7}{\circle*{0.2}}
\multiput(63,5)(0.6,1){6}{\circle*{0.2}}
\multiput(61,0.5)(10,0){2}{\circle{1}}
\put(66,9){\circle{1}}
\multiput(65,0.5)(2,0){2}{\circle{1}}
\put(63,4){\circle*{1}}
\put(64,5.5){\makebox(0,0){$\times$}}
\put(68,5.5){\circle*{1}}
\put(69,4){\makebox(0,0){$\times$}}
\put(75,5){\makebox(0,0){$+$}}
\put(78,0){\line(3,5){3}}
\put(81,5){\line(3,-5){3}}
\put(84,0){\line(1,0){6}}
\put(84,10){\line(3,-5){3}}
\multiput(78,0)(1,0){7}{\circle*{0.2}}
\multiput(84,0)(0.6,1){6}{\circle*{0.2}}
\multiput(87,5)(0.6,-1){6}{\circle*{0.2}}
\multiput(81,5)(1,0){7}{\circle*{0.2}}
\multiput(81,5)(0.6,1){6}{\circle*{0.2}}
\multiput(79,0.5)(10,0){2}{\circle{1}}
\put(84,9){\circle{1}}
\multiput(83,0.5)(2,0){2}{\circle{1}}
\put(81,4){\circle*{1}}
\put(82,5.5){\makebox(0,0){$\times$}}
\put(86,5.5){\circle{1}}
\put(87,4){\makebox(0,0){\scriptsize $\oplus$}}
\put(93,5){\makebox(0,0){$+$}}
\put(96,0){\line(1,0){12}}
\put(99,5){\line(1,0){6}}
\put(105,5){\line(-3,5){3}}
\multiput(96,0)(0.6,1){11}{\circle*{0.2}}
\multiput(102,0)(0.6,1){6}{\circle*{0.2}}
\multiput(102,0)(-0.6,1){6}{\circle*{0.2}}
\multiput(108,0)(-0.6,1){6}{\circle*{0.2}}
\multiput(97,0.5)(10,0){2}{\circle{1}}
\put(102,9){\circle{1}}
\multiput(101,0.5)(2,0){2}{\circle{1}}
\multiput(99,4)(6,0){2}{\makebox(0,0){$\times$}}
\put(100,5.5){\circle{1}}
\put(104,5.5){\circle*{1}}
\end{picture}

\vspace*{5mm}

\begin{picture}(108,10)
\put(15,5){\makebox(0,0){$+$}}
\put(18,0){\line(1,0){12}}
\put(21,5){\line(1,0){6}}
\put(27,5){\line(-3,5){3}}
\multiput(18,0)(0.6,1){11}{\circle*{0.2}}
\multiput(24,0)(0.6,1){6}{\circle*{0.2}}
\multiput(24,0)(-0.6,1){6}{\circle*{0.2}}
\multiput(30,0)(-0.6,1){6}{\circle*{0.2}}
\multiput(19,0.5)(10,0){2}{\circle{1}}
\put(24,9){\circle{1}}
\multiput(23,0.5)(2,0){2}{\circle{1}}
\put(21,4){\makebox(0,0){\scriptsize $\oplus$}}
\put(27,4){\makebox(0,0){$\times$}}
\put(22,5.5){\circle{1}}
\put(26,5.5){\circle*{1}}
\put(33,5){\makebox(0,0){$)$}}
\put(39,5){\makebox(0,0){$+$}}
\put(42,0){\line(1,0){12}}
\put(45,5){\line(1,0){6}}
\put(51,5){\line(-3,5){3}}
\multiput(42,0)(0.6,1){11}{\circle*{0.2}}
\multiput(48,0)(0.6,1){6}{\circle*{0.2}}
\multiput(48,0)(-0.6,1){6}{\circle*{0.2}}
\multiput(54,0)(-0.6,1){6}{\circle*{0.2}}
\multiput(43,0.5)(10,0){2}{\circle{1}}
\put(48,9){\circle{1}}
\multiput(47,0.5)(2,0){2}{\circle{1}}
\multiput(45,4)(6,0){2}{\makebox(0,0){$\times$}}
\multiput(46,5.5)(4,0){2}{\circle*{1}}
\put(57,5){\makebox(0,0){$+$}}
\put(60,0){\line(3,5){3}}
\put(63,5){\line(3,-5){3}}
\put(66,0){\line(3,5){3}}
\put(69,5){\line(3,-5){3}}
\multiput(60,0)(1,0){13}{\circle*{0.2}}
\multiput(63,5)(1,0){7}{\circle*{0.2}}
\multiput(66,10)(0.6,-1){6}{\circle*{0.2}}
\multiput(66,10)(-0.6,-1){6}{\circle*{0.2}}
\multiput(61,0.5)(10,0){2}{\circle{1}}
\put(66,9){\makebox(0,0){\scriptsize $\oplus$}}
\multiput(65,0.5)(2,0){2}{\circle{1}}
\multiput(63,4)(6,0){2}{\circle*{1}}
\multiput(64,5.5)(4,0){2}{\makebox(0,0){$\times$}}
\end{picture}

\caption{\footnotesize{Illustration for the expression of $g_3(n+1)$. The multiplication of two on the right-hand-side corresponds to the reflection symmetry with respect to the central vertical axis.}} 
\label{g3fig}
\end{figure}

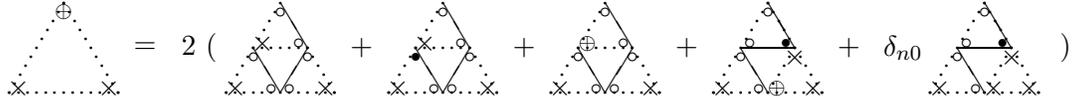
\begin{figure}[htbp]
\unitlength 1.2mm 
\begin{picture}(117,10)
\multiput(0,0)(1,0){13}{\circle*{0.2}}
\multiput(0,0)(0.6,1){11}{\circle*{0.2}}
\multiput(12,0)(-0.6,1){11}{\circle*{0.2}}
\multiput(1,0.5)(10,0){2}{\makebox(0,0){$\times$}}
\put(6,9){\makebox(0,0){\scriptsize $\oplus$}}
\put(15,5){\makebox(0,0){$=$}}
\put(21,5){\makebox(0,0){$2 \ ($}}
\put(27,5){\line(3,-5){3}}
\put(30,0){\line(3,5){3}}
\put(33,5){\line(-3,5){3}}
\multiput(24,0)(1,0){13}{\circle*{0.2}}
\multiput(24,0)(0.6,1){11}{\circle*{0.2}}
\multiput(27,5)(1,0){7}{\circle*{0.2}}
\multiput(36,0)(-0.6,1){6}{\circle*{0.2}}
\multiput(25,0.5)(10,0){2}{\makebox(0,0){$\times$}}
\put(30,9){\circle{1}}
\multiput(29,0.5)(2,0){2}{\circle{1}}
\multiput(27,4)(6,0){2}{\circle{1}}
\put(28,5.5){\makebox(0,0){$\times$}}
\put(32,5.5){\circle{1}}
\put(39,5){\makebox(0,0){$+$}}
\put(45,5){\line(3,-5){3}}
\put(48,0){\line(3,5){3}}
\put(51,5){\line(-3,5){3}}
\multiput(42,0)(1,0){13}{\circle*{0.2}}
\multiput(42,0)(0.6,1){11}{\circle*{0.2}}
\multiput(45,5)(1,0){7}{\circle*{0.2}}
\multiput(54,0)(-0.6,1){6}{\circle*{0.2}}
\multiput(43,0.5)(10,0){2}{\makebox(0,0){$\times$}}
\put(48,9){\circle{1}}
\multiput(47,0.5)(2,0){2}{\circle{1}}
\put(45,4){\circle*{1}}
\put(51,4){\circle{1}}
\put(46,5.5){\makebox(0,0){$\times$}}
\put(50,5.5){\circle{1}}
\put(57,5){\makebox(0,0){$+$}}
\put(63,5){\line(3,-5){3}}
\put(66,0){\line(3,5){3}}
\put(69,5){\line(-3,5){3}}
\multiput(60,0)(1,0){13}{\circle*{0.2}}
\multiput(60,0)(0.6,1){11}{\circle*{0.2}}
\multiput(63,5)(1,0){7}{\circle*{0.2}}
\multiput(72,0)(-0.6,1){6}{\circle*{0.2}}
\multiput(61,0.5)(10,0){2}{\makebox(0,0){$\times$}}
\put(66,9){\circle{1}}
\multiput(65,0.5)(2,0){2}{\circle{1}}
\multiput(63,4)(6,0){2}{\circle{1}}
\put(64,5.5){\makebox(0,0){\scriptsize $\oplus$}}
\put(68,5.5){\circle{1}}
\put(75,5){\makebox(0,0){$+$}}
\put(84,0){\line(-3,5){3}}
\put(81,5){\line(1,0){6}}
\put(87,5){\line(-3,5){3}}
\multiput(78,0)(1,0){13}{\circle*{0.2}}
\multiput(78,0)(0.6,1){11}{\circle*{0.2}}
\multiput(84,0)(0.6,1){6}{\circle*{0.2}}
\multiput(90,0)(-0.6,1){11}{\circle*{0.2}}
\multiput(79,0.5)(10,0){2}{\makebox(0,0){$\times$}}
\put(84,9){\circle{1}}
\put(83,0.5){\circle{1}}
\put(81,4){\circle{1}}
\put(82,5.5){\circle{1}}
\put(86,5.5){\circle*{1}}
\put(85,0.5){\makebox(0,0){\scriptsize $\oplus$}}
\put(87,4){\makebox(0,0){$\times$}}
\put(93,5){\makebox(0,0){$+$}}
\put(99,5){\makebox(0,0){$\delta_{n0}$}}
\put(108,0){\line(-3,5){3}}
\put(105,5){\line(1,0){6}}
\put(111,5){\line(-3,5){3}}
\multiput(102,0)(1,0){13}{\circle*{0.2}}
\multiput(102,0)(0.6,1){11}{\circle*{0.2}}
\multiput(108,0)(0.6,1){6}{\circle*{0.2}}
\multiput(114,0)(-0.6,1){11}{\circle*{0.2}}
\multiput(103,0.5)(10,0){2}{\makebox(0,0){$\times$}}
\put(108,9){\circle{1}}
\put(107,0.5){\circle{1}}
\put(105,4){\circle{1}}
\put(106,5.5){\circle{1}}
\put(110,5.5){\circle*{1}}
\put(109,0.5){\makebox(0,0){$\times$}}
\put(111,4){\makebox(0,0){$\times$}}
\put(117,5){\makebox(0,0){$)$}}
\end{picture}

\caption{\footnotesize{Illustration for the expression of $t_1(n+1)$. The multiplication of two on the right-hand-side corresponds to the reflection symmetry with respect to the central vertical axis. The last drawing only applies for $n=0$.}} 
\label{h1fig}
\end{figure}

Consider the quantities $f_2(n)$, $g_2(n)$, $g_3(n)$ and $t_1(n)$ for $n \geq 1$, we have the following recursion equations as illustrated in Figs. \ref{f2fig}-\ref{h1fig}. 
\beq
\left\{\begin{array}{lll}
f_2(n+1) & = & 2f_1(n)^2[g_1(n) + g_2(n) + g_3(n)] + 2f_1(n)f_2(n)g_1(n) \ , \\
g_2(n+1) & = & 2f_1(n)g_1(n)[g_1(n) + g_2(n) + g_3(n)] + f_2(n)g_1(n)^2 + f_1(n)^2t_1(n) \ , \\
g_3(n+1) & = & 2f_1(n)g_1(n)[2g_1(n) + g_2(n) + 3g_3(n)] + f_2(n)g_1(n)^2 + f_1(n)^2t_1(n) \ , \\
t_1(n+1) & = & 2g_1(n)^2[g_1(n)+g_2(n)+g_3(n)] + 2f_1(n)g_1(n)t_1(n) \ .
\end{array}
\right.
\label{f2g2g3t1}
\eeq
If we set $n=0$ in Eq. (\ref{f2g2g3t1}), only the terms contain $f_1(0)$ and $g_1(0)$ as factors on the right-hand-sides give non-zero contribution. In addition, there are special drawings shown in Figs. \ref{g2fig}, \ref{h1fig}, such that the initial values are $f_2(1)=2$, $g_2(1)=3$, $g_3(1)=4$, $t_1(1)=4$.

\vspace*{5mm}

\vspace*{5mm}

\vspace*{5mm}

Using $g_1(n)=3f_1(n)/2$ for $n \geq 1$ from Eq. (\ref{g1}) to eliminate $g_1(n)$, Eq. (\ref{f2g2g3t1}) becomes
\beq
\left\{\begin{array}{lll}
f_2(n+1) & = & 3f_1(n)^3 + 2f_1(n)^2[g_2(n) + g_3(n)] + 3f_1(n)^2f_2(n) \ , \\
g_2(n+1) & = & 3f_1(n)^2[\frac32 f_1(n) + g_2(n) + g_3(n)] + \frac94 f_1(n)^2f_2(n) + f_1(n)^2t_1(n) \ , \\
g_3(n+1) & = & 3f_1(n)^2[3f_1(n) + g_2(n) + 3g_3(n)] + \frac94 f_1(n)^2f_2(n) + f_1(n)^2t_1(n) \ , \\
t_1(n+1) & = & \frac{9f_1(n)^2}2 [\frac32 f_1(n)+g_2(n)+g_3(n)] + 3f_1(n)^2t_1(n) \ .
\end{array}
\right.
\label{fgh2}
\eeq

Define the ratios
\beq
\alpha_2(n) = \frac {f_2(n)}{f_1(n)} \ , \quad 
\beta_2(n) = \frac {g_2(n)}{f_1(n)} \ ,\quad
\beta_3(n) = \frac {g_3(n)}{f_1(n)} \ , \quad
\gamma_1(n) = \frac {t_1(n)}{f_1(n)} \ ,
\label{ab2}
\eeq
then they satisfy the recursion relations
\beq
\left\{\begin{array}{lll}
\alpha_2(n+1) & = & 1 + \alpha_2(n) + \frac23 \beta_2(n) + \frac23 \beta_3(n) \ , \\
\beta_2(n+1) & = & \frac 32 + \frac34 \alpha_2(n) + \beta_2(n) + \beta_3(n) + \frac13 \gamma_1(n) \ , \\ 
\beta_3(n+1) & = & 3 + \frac34 \alpha_2(n) + \beta_2(n) + 3\beta_3(n) + \frac13 \gamma_1(n) \ , \\
\gamma_1(n+1) & = & \frac94 + \frac32 \beta_2(n) + \frac 32\beta_3(n) + \gamma_1(n) \ , 
\end{array}
\right.
\label{Mab}
\eeq 
where the relation $f_1(n+1)=3f_1(n)^3$ in Eq. (\ref{f1}) is used.
Define the vectors $b=(1, \frac32, 3, \frac 94)^t$, $X(n)=(\alpha_2(n), \beta_2(n), \beta_3(n), \gamma_1(n))^t$, and construct the matrix
\beq
A = \left(\begin{array}{cccc}
1       & \frac23 & \frac23 & 0       \\
\frac34 & 1       & 1       & \frac13 \\
\frac34 & 1       & 3       & \frac13 \\
0       & \frac32 & \frac32 & 1  
\end{array}
\right) \ ,
\eeq
such that $X(1)=(1,3/2,2,2)^t$ and for general $n \ge 1$,
\beqs
X(n) & = & AX(n-1)+b = A^2X(n-2) + Ab + b = \cdots = A^{n-1}X(1) + \sum_{j=0}^{n-2} A^j b \cr\cr
& = & \left(\begin{array}{cccc}
(\frac{7\times 17}{2^4\times 3^3})4^n - \frac{2}{3^3} - (\frac1{2^2\times 3^2})\delta_{n1} \\
(\frac{7\times 17}{2^5\times 3^2})4^n - \frac{7}{2^2\times 3^2} + (\frac1{2^3\times 3})\delta_{n1} \\
(\frac{7\times 17}{2^4\times 3^2})4^n - \frac{47}{2^2\times 3^2}\\ 
(\frac{7\times 17}{2^6\times 3})4^n - \frac5{2^2\times 3} - \frac1{2^4}\delta_{n1} 
\end{array}
\right) \ ,
\label{Xn}
\eeqs
where $\delta_{ij}$ is the Kronecker delta function.     
Since $H_0(n) = 2f_1(n) + f_2(n) = f_1(n) \bigl( 2+\alpha_2(n) \bigr)$, we have the following theorem:

\bigskip

\begin{theo}
The number of Hamiltonian paths on the Sierpinski gasket $SG(n)$ with one end at origin for $n \geq 1$ is given by
\beq
H_0(n) = \frac{\sqrt{3}(2\sqrt{3})^{3^{n-1}}}{3} \left\{ ( \frac{7 \times 17}{2^4\times 3^3})4^n + \frac{2^2 \times 13}{3^3} - (\frac1{2^2 \times 3^2}) \delta_{n1} \right \} \ .
\label{H0s}
\eeq
\label{hpo}
\end{theo}
  
\section{Number of Hamiltonian paths}  
\label{sectionH}

\begin{figure}[htbp]
\unitlength 1.2mm 
\begin{picture}(108,10)
\put(0,0){\line(3,5){6}}
\put(12,0){\line(-3,5){6}}
\multiput(0,0)(1,0){13}{\circle*{0.2}}
\multiput(1,0.5)(10,0){2}{\circle*{1}}
\put(6,9){\circle*{1}}
\put(15,5){\makebox(0,0){$=$}}
\put(21,5){\makebox(0,0){$3 \ ($}}
\put(24,0){\line(1,0){6}}
\put(30,0){\line(-3,5){3}}
\put(27,5){\line(3,5){3}}
\put(30,10){\line(3,-5){6}}
\multiput(24,0)(0.6,1){6}{\circle*{0.2}}
\multiput(27,5)(1,0){7}{\circle*{0.2}}
\multiput(33,5)(-0.6,-1){6}{\circle*{0.2}}
\multiput(30,0)(1,0){7}{\circle*{0.2}}
\multiput(25,0.5)(10,0){2}{\circle*{1}}
\put(30,9){\circle*{1}}
\put(29,0.5){\circle*{1}}
\put(27,4){\circle{1}}
\put(31,0.5){\makebox(0,0){$\times$}}
\put(33,4){\circle{1}}
\multiput(28,5.5)(4,0){2}{\circle{1}}
\put(39,5){\makebox(0,0){$+$}}
\put(42,0){\line(3,5){6}}
\put(48,10){\line(3,-5){3}}
\put(51,5){\line(-3,-5){3}}
\put(48,0){\line(1,0){6}}
\multiput(42,0)(1,0){7}{\circle*{0.2}}
\multiput(48,0)(-0.6,1){6}{\circle*{0.2}}
\multiput(45,5)(1,0){7}{\circle*{0.2}}
\multiput(51,5)(0.6,-1){6}{\circle*{0.2}}
\multiput(43,0.5)(10,0){2}{\circle*{1}}
\put(48,9){\circle*{1}}
\put(47,0.5){\makebox(0,0){$\times$}}
\put(45,4){\circle{1}}
\put(49,0.5){\circle*{1}}
\put(51,4){\circle{1}}
\multiput(46,5.5)(4,0){2}{\circle{1}}
\put(57,5){\makebox(0,0){$+$}}
\put(60,0){\line(3,5){6}}
\put(66,10){\line(3,-5){6}}
\put(72,0){\line(-1,0){6}}
\multiput(60,0)(1,0){7}{\circle*{0.2}}
\multiput(66,0)(-0.6,1){6}{\circle*{0.2}}
\multiput(66,0)(0.6,1){6}{\circle*{0.2}}
\multiput(63,5)(1,0){7}{\circle*{0.2}}
\multiput(61,0.5)(10,0){2}{\circle*{1}}
\put(66,9){\circle*{1}}
\put(65,0.5){\makebox(0,0){$\times$}}
\put(63,4){\circle{1}}
\put(67,0.5){\circle{1}}
\put(69,4){\circle{1}}
\multiput(64,5.5)(4,0){2}{\circle{1}}
\put(75,5){\makebox(0,0){$+$}}
\put(78,0){\line(3,5){6}}
\put(84,10){\line(3,-5){6}}
\put(90,0){\line(-1,0){6}}
\multiput(78,0)(1,0){7}{\circle*{0.2}}
\multiput(84,0)(-0.6,1){6}{\circle*{0.2}}
\multiput(84,0)(0.6,1){6}{\circle*{0.2}}
\multiput(81,5)(1,0){7}{\circle*{0.2}}
\multiput(79,0.5)(10,0){2}{\circle*{1}}
\put(84,9){\circle*{1}}
\put(83,0.5){\makebox(0,0){\scriptsize $\oplus$}}
\put(81,4){\circle{1}}
\put(85,0.5){\circle{1}}
\put(87,4){\circle{1}}
\multiput(82,5.5)(4,0){2}{\circle{1}}
\put(93,5){\makebox(0,0){$+$}}
\put(96,0){\line(1,0){6}}
\put(96,0){\line(3,5){6}}
\put(102,10){\line(3,-5){6}}
\multiput(102,0)(1,0){7}{\circle*{0.2}}
\multiput(102,0)(0.6,1){6}{\circle*{0.2}}
\multiput(102,0)(-0.6,1){6}{\circle*{0.2}}
\multiput(99,5)(1,0){7}{\circle*{0.2}}
\multiput(97,0.5)(10,0){2}{\circle*{1}}
\put(102,9){\circle*{1}}
\put(101,0.5){\circle{1}}
\put(99,4){\circle{1}}
\put(103,0.5){\makebox(0,0){$\times$}}
\put(105,4){\circle{1}}
\multiput(100,5.5)(4,0){2}{\circle{1}}
\end{picture}

\vspace*{5mm}

\begin{picture}(108,10)
\put(15,5){\makebox(0,0){$+$}}
\put(18,0){\line(1,0){6}}
\put(18,0){\line(3,5){6}}
\put(24,10){\line(3,-5){6}}
\multiput(24,0)(1,0){7}{\circle*{0.2}}
\multiput(24,0)(0.6,1){6}{\circle*{0.2}}
\multiput(24,0)(-0.6,1){6}{\circle*{0.2}}
\multiput(21,5)(1,0){7}{\circle*{0.2}}
\multiput(19,0.5)(10,0){2}{\circle*{1}}
\put(24,9){\circle*{1}}
\put(23,0.5){\circle{1}}
\put(21,4){\circle{1}}
\put(25,0.5){\makebox(0,0){\scriptsize $\oplus$}}
\put(27,4){\circle{1}}
\multiput(22,5.5)(4,0){2}{\circle{1}}
\put(33,5){\makebox(0,0){$)$}}
\end{picture}

\caption{\footnotesize{Illustration for the expression of $f_3(n+1)$. The multiplication of three on the right-hand-side corresponds to the three possible orientations of $SG(n+1)$.}} 
\label{f3fig}
\end{figure}
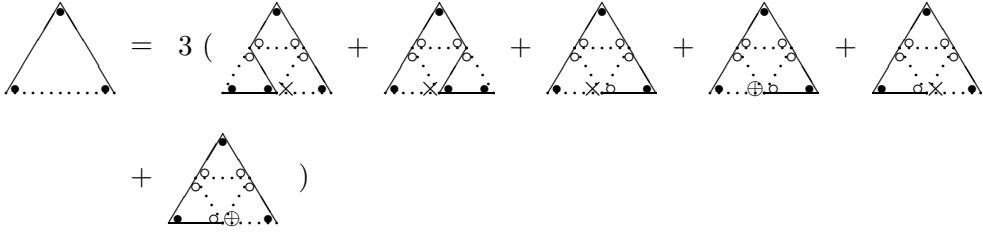

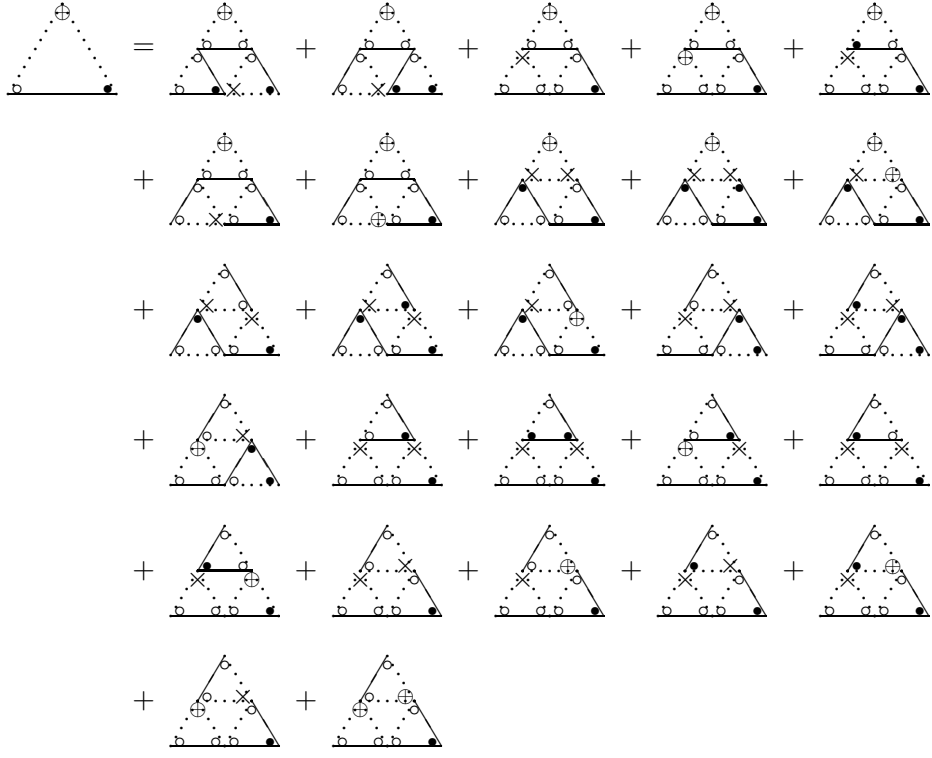
\begin{figure}[htbp]
\unitlength 1.2mm 
\begin{picture}(102,10)
\put(0,0){\line(1,0){12}}
\multiput(0,0)(0.6,1){11}{\circle*{0.2}}
\multiput(12,0)(-0.6,1){11}{\circle*{0.2}}
\put(1,0.5){\circle{1}}
\put(11,0.5){\circle*{1}}
\put(6,9){\makebox(0,0){\scriptsize $\oplus$}}
\put(15,5){\makebox(0,0){$=$}}
\put(18,0){\line(1,0){6}}
\put(24,0){\line(-3,5){3}}
\put(21,5){\line(1,0){6}}
\put(27,5){\line(3,-5){3}}
\multiput(18,0)(0.6,1){11}{\circle*{0.2}}
\multiput(24,10)(0.6,-1){6}{\circle*{0.2}}
\multiput(27,5)(-0.6,-1){6}{\circle*{0.2}}
\multiput(24,0)(1,0){7}{\circle*{0.2}}
\put(19,0.5){\circle{1}}
\put(29,0.5){\circle*{1}}
\put(24,9){\makebox(0,0){\scriptsize $\oplus$}}
\put(23,0.5){\circle*{1}}
\multiput(21,4)(6,0){2}{\circle{1}}
\multiput(22,5.5)(4,0){2}{\circle{1}}
\put(25,0.5){\makebox(0,0){$\times$}}
\put(33,5){\makebox(0,0){$+$}}
\put(36,0){\line(3,5){3}}
\put(39,5){\line(1,0){6}}
\put(45,5){\line(-3,-5){3}}
\put(42,0){\line(1,0){6}}
\multiput(36,0)(1,0){6}{\circle*{0.2}}
\multiput(42,0)(-0.6,1){6}{\circle*{0.2}}
\multiput(39,5)(0.6,1){6}{\circle*{0.2}}
\multiput(42,10)(0.6,-1){11}{\circle*{0.2}}
\put(37,0.5){\circle{1}}
\put(47,0.5){\circle*{1}}
\put(42,9){\makebox(0,0){\scriptsize $\oplus$}}
\put(41,0.5){\makebox(0,0){$\times$}}
\multiput(39,4)(6,0){2}{\circle{1}}
\multiput(40,5.5)(4,0){2}{\circle{1}}
\put(43,0.5){\circle*{1}}
\put(51,5){\makebox(0,0){$+$}}
\put(54,0){\line(1,0){12}}
\put(66,0){\line(-3,5){3}}
\put(63,5){\line(-1,0){6}}
\multiput(54,0)(0.6,1){11}{\circle*{0.2}}
\multiput(60,0)(0.6,1){6}{\circle*{0.2}}
\multiput(60,0)(-0.6,1){6}{\circle*{0.2}}
\multiput(63,5)(-0.6,1){6}{\circle*{0.2}}
\put(55,0.5){\circle{1}}
\put(65,0.5){\circle*{1}}
\put(60,9){\makebox(0,0){\scriptsize $\oplus$}}
\multiput(59,0.5)(2,0){2}{\circle{1}}
\put(57,4){\makebox(0,0){$\times$}}
\put(63,4){\circle{1}}
\multiput(58,5.5)(4,0){2}{\circle{1}}
\put(69,5){\makebox(0,0){$+$}}
\put(72,0){\line(1,0){12}}
\put(84,0){\line(-3,5){3}}
\put(81,5){\line(-1,0){6}}
\multiput(72,0)(0.6,1){11}{\circle*{0.2}}
\multiput(78,0)(0.6,1){6}{\circle*{0.2}}
\multiput(78,0)(-0.6,1){6}{\circle*{0.2}}
\multiput(81,5)(-0.6,1){6}{\circle*{0.2}}
\put(73,0.5){\circle{1}}
\put(83,0.5){\circle*{1}}
\put(78,9){\makebox(0,0){\scriptsize $\oplus$}}
\multiput(77,0.5)(2,0){2}{\circle{1}}
\put(75,4){\makebox(0,0){\scriptsize $\oplus$}}
\put(81,4){\circle{1}}
\multiput(76,5.5)(4,0){2}{\circle{1}}
\put(87,5){\makebox(0,0){$+$}}
\put(90,0){\line(1,0){12}}
\put(102,0){\line(-3,5){3}}
\put(99,5){\line(-1,0){6}}
\multiput(90,0)(0.6,1){11}{\circle*{0.2}}
\multiput(96,0)(0.6,1){6}{\circle*{0.2}}
\multiput(96,0)(-0.6,1){6}{\circle*{0.2}}
\multiput(99,5)(-0.6,1){6}{\circle*{0.2}}
\put(91,0.5){\circle{1}}
\put(101,0.5){\circle*{1}}
\put(96,9){\makebox(0,0){\scriptsize $\oplus$}}
\multiput(95,0.5)(2,0){2}{\circle{1}}
\put(93,4){\makebox(0,0){$\times$}}
\put(99,4){\circle{1}}
\put(94,5.5){\circle*{1}}
\put(98,5.5){\circle{1}}
\end{picture}

\vspace*{5mm}

\begin{picture}(102,10)
\put(15,5){\makebox(0,0){$+$}}
\put(18,0){\line(3,5){3}}
\put(21,5){\line(1,0){6}}
\put(27,5){\line(3,-5){3}}
\put(30,0){\line(-1,0){6}}
\multiput(18,0)(1,0){6}{\circle*{0.2}}
\multiput(24,0)(-0.6,1){6}{\circle*{0.2}}
\multiput(24,0)(0.6,1){6}{\circle*{0.2}}
\multiput(24,10)(0.6,-1){6}{\circle*{0.2}}
\multiput(24,10)(-0.6,-1){6}{\circle*{0.2}}
\put(19,0.5){\circle{1}}
\put(29,0.5){\circle*{1}}
\put(24,9){\makebox(0,0){\scriptsize $\oplus$}}
\put(23,0.5){\makebox(0,0){$\times$}}
\multiput(21,4)(6,0){2}{\circle{1}}
\multiput(22,5.5)(4,0){2}{\circle{1}}
\put(25,0.5){\circle{1}}
\put(33,5){\makebox(0,0){$+$}}
\put(36,0){\line(3,5){3}}
\put(39,5){\line(1,0){6}}
\put(45,5){\line(3,-5){3}}
\put(48,0){\line(-1,0){6}}
\multiput(36,0)(1,0){6}{\circle*{0.2}}
\multiput(42,0)(-0.6,1){6}{\circle*{0.2}}
\multiput(42,0)(0.6,1){6}{\circle*{0.2}}
\multiput(42,10)(0.6,-1){6}{\circle*{0.2}}
\multiput(42,10)(-0.6,-1){6}{\circle*{0.2}}
\put(37,0.5){\circle{1}}
\put(47,0.5){\circle*{1}}
\put(42,9){\makebox(0,0){\scriptsize $\oplus$}}
\put(41,0.5){\makebox(0,0){\scriptsize $\oplus$}}
\multiput(39,4)(6,0){2}{\circle{1}}
\multiput(40,5.5)(4,0){2}{\circle{1}}
\put(43,0.5){\circle{1}}
\put(51,5){\makebox(0,0){$+$}}
\put(54,0){\line(3,5){3}}
\put(57,5){\line(3,-5){3}}
\put(60,0){\line(1,0){6}}
\put(66,0){\line(-3,5){3}}
\multiput(54,0)(1,0){7}{\circle*{0.2}}
\multiput(57,5)(1,0){7}{\circle*{0.2}}
\multiput(60,0)(0.6,1){6}{\circle*{0.2}}
\multiput(60,10)(-0.6,-1){6}{\circle*{0.2}}
\multiput(60,10)(0.6,-1){6}{\circle*{0.2}}
\put(55,0.5){\circle{1}}
\put(65,0.5){\circle*{1}}
\put(60,9){\makebox(0,0){\scriptsize $\oplus$}}
\multiput(59,0.5)(2,0){2}{\circle{1}}
\put(57,4){\circle*{1}}
\put(63,4){\circle{1}}
\multiput(58,5.5)(4,0){2}{\makebox(0,0){$\times$}}
\put(69,5){\makebox(0,0){$+$}}
\put(72,0){\line(3,5){3}}
\put(75,5){\line(3,-5){3}}
\put(78,0){\line(1,0){6}}
\put(84,0){\line(-3,5){3}}
\multiput(72,0)(1,0){7}{\circle*{0.2}}
\multiput(75,5)(1,0){7}{\circle*{0.2}}
\multiput(78,0)(0.6,1){6}{\circle*{0.2}}
\multiput(78,10)(-0.6,-1){6}{\circle*{0.2}}
\multiput(78,10)(0.6,-1){6}{\circle*{0.2}}
\put(73,0.5){\circle{1}}
\put(83,0.5){\circle*{1}}
\put(78,9){\makebox(0,0){\scriptsize $\oplus$}}
\multiput(77,0.5)(2,0){2}{\circle{1}}
\put(75,4){\circle*{1}}
\put(81,4){\circle*{1}}
\multiput(76,5.5)(4,0){2}{\makebox(0,0){$\times$}}
\put(87,5){\makebox(0,0){$+$}}
\put(90,0){\line(3,5){3}}
\put(93,5){\line(3,-5){3}}
\put(96,0){\line(1,0){6}}
\put(102,0){\line(-3,5){3}}
\multiput(90,0)(1,0){7}{\circle*{0.2}}
\multiput(93,5)(1,0){7}{\circle*{0.2}}
\multiput(96,0)(0.6,1){6}{\circle*{0.2}}
\multiput(96,10)(-0.6,-1){6}{\circle*{0.2}}
\multiput(96,10)(0.6,-1){6}{\circle*{0.2}}
\put(91,0.5){\circle{1}}
\put(101,0.5){\circle*{1}}
\put(96,9){\makebox(0,0){\scriptsize $\oplus$}}
\multiput(95,0.5)(2,0){2}{\circle{1}}
\put(93,4){\circle*{1}}
\put(99,4){\circle{1}}
\put(94,5.5){\makebox(0,0){$\times$}}
\put(98,5.5){\makebox(0,0){\scriptsize $\oplus$}}
\end{picture}

\vspace*{5mm}

\begin{picture}(102,10)
\put(15,5){\makebox(0,0){$+$}}
\put(18,0){\line(3,5){3}}
\put(21,5){\line(3,-5){3}}
\put(24,0){\line(1,0){6}}
\put(24,10){\line(3,-5){3}}
\multiput(18,0)(1,0){7}{\circle*{0.2}}
\multiput(21,5)(1,0){7}{\circle*{0.2}}
\multiput(24,0)(0.6,1){6}{\circle*{0.2}}
\multiput(30,0)(-0.6,1){6}{\circle*{0.2}}
\multiput(24,10)(-0.6,-1){6}{\circle*{0.2}}
\put(19,0.5){\circle{1}}
\put(29,0.5){\circle*{1}}
\put(24,9){\circle{1}}
\multiput(23,0.5)(2,0){2}{\circle{1}}
\put(21,4){\circle*{1}}
\put(27,4){\makebox(0,0){$\times$}}
\put(22,5.5){\makebox(0,0){$\times$}}
\put(26,5.5){\circle{1}}
\put(33,5){\makebox(0,0){$+$}}
\put(36,0){\line(3,5){3}}
\put(39,5){\line(3,-5){3}}
\put(42,0){\line(1,0){6}}
\put(42,10){\line(3,-5){3}}
\multiput(36,0)(1,0){7}{\circle*{0.2}}
\multiput(39,5)(1,0){7}{\circle*{0.2}}
\multiput(42,0)(0.6,1){6}{\circle*{0.2}}
\multiput(48,0)(-0.6,1){6}{\circle*{0.2}}
\multiput(42,10)(-0.6,-1){6}{\circle*{0.2}}
\put(37,0.5){\circle{1}}
\put(47,0.5){\circle*{1}}
\put(42,9){\circle{1}}
\multiput(41,0.5)(2,0){2}{\circle{1}}
\put(39,4){\circle*{1}}
\put(45,4){\makebox(0,0){$\times$}}
\put(40,5.5){\makebox(0,0){$\times$}}
\put(44,5.5){\circle*{1}}
\put(51,5){\makebox(0,0){$+$}}
\put(54,0){\line(3,5){3}}
\put(57,5){\line(3,-5){3}}
\put(60,0){\line(1,0){6}}
\put(60,10){\line(3,-5){3}}
\multiput(54,0)(1,0){7}{\circle*{0.2}}
\multiput(57,5)(1,0){7}{\circle*{0.2}}
\multiput(60,0)(0.6,1){6}{\circle*{0.2}}
\multiput(66,0)(-0.6,1){6}{\circle*{0.2}}
\multiput(60,10)(-0.6,-1){6}{\circle*{0.2}}
\put(55,0.5){\circle{1}}
\put(65,0.5){\circle*{1}}
\put(60,9){\circle{1}}
\multiput(59,0.5)(2,0){2}{\circle{1}}
\put(57,4){\circle*{1}}
\put(63,4){\makebox(0,0){\scriptsize $\oplus$}}
\put(58,5.5){\makebox(0,0){$\times$}}
\put(62,5.5){\circle{1}}
\put(69,5){\makebox(0,0){$+$}}
\put(72,0){\line(1,0){6}}
\put(78,0){\line(3,5){3}}
\put(81,5){\line(3,-5){3}}
\put(75,5){\line(3,5){3}}
\multiput(72,0)(0.6,1){6}{\circle*{0.2}}
\multiput(75,5)(1,0){7}{\circle*{0.2}}
\multiput(75,5)(0.6,-1){6}{\circle*{0.2}}
\multiput(78,0)(1,0){7}{\circle*{0.2}}
\multiput(78,10)(0.6,-1){6}{\circle*{0.2}}
\put(73,0.5){\circle{1}}
\put(83,0.5){\circle*{1}}
\put(78,9){\circle{1}}
\multiput(77,0.5)(2,0){2}{\circle{1}}
\put(75,4){\makebox(0,0){$\times$}}
\put(81,4){\circle*{1}}
\put(76,5.5){\circle{1}}
\put(80,5.5){\makebox(0,0){$\times$}}
\put(87,5){\makebox(0,0){$+$}}
\put(90,0){\line(1,0){6}}
\put(96,0){\line(3,5){3}}
\put(99,5){\line(3,-5){3}}
\put(93,5){\line(3,5){3}}
\multiput(90,0)(0.6,1){6}{\circle*{0.2}}
\multiput(93,5)(1,0){7}{\circle*{0.2}}
\multiput(93,5)(0.6,-1){6}{\circle*{0.2}}
\multiput(96,0)(1,0){7}{\circle*{0.2}}
\multiput(96,10)(0.6,-1){6}{\circle*{0.2}}
\put(91,0.5){\circle{1}}
\put(101,0.5){\circle*{1}}
\put(96,9){\circle{1}}
\multiput(95,0.5)(2,0){2}{\circle{1}}
\put(93,4){\makebox(0,0){$\times$}}
\put(99,4){\circle*{1}}
\put(94,5.5){\circle*{1}}
\put(98,5.5){\makebox(0,0){$\times$}}
\end{picture}

\vspace*{5mm}

\begin{picture}(102,10)
\put(15,5){\makebox(0,0){$+$}}
\put(18,0){\line(1,0){6}}
\put(24,0){\line(3,5){3}}
\put(27,5){\line(3,-5){3}}
\put(21,5){\line(3,5){3}}
\multiput(18,0)(0.6,1){6}{\circle*{0.2}}
\multiput(21,5)(1,0){7}{\circle*{0.2}}
\multiput(21,5)(0.6,-1){6}{\circle*{0.2}}
\multiput(24,0)(1,0){7}{\circle*{0.2}}
\multiput(24,10)(0.6,-1){6}{\circle*{0.2}}
\put(19,0.5){\circle{1}}
\put(29,0.5){\circle*{1}}
\put(24,9){\circle{1}}
\multiput(23,0.5)(2,0){2}{\circle{1}}
\put(21,4){\makebox(0,0){\scriptsize $\oplus$}}
\put(27,4){\circle*{1}}
\put(22,5.5){\circle{1}}
\put(26,5.5){\makebox(0,0){$\times$}}
\put(33,5){\makebox(0,0){$+$}}
\put(36,0){\line(1,0){12}}
\put(39,5){\line(1,0){6}}
\put(45,5){\line(-3,5){3}}
\multiput(36,0)(0.6,1){11}{\circle*{0.2}}
\multiput(42,0)(-0.6,1){6}{\circle*{0.2}}
\multiput(42,0)(0.6,1){6}{\circle*{0.2}}
\multiput(48,0)(-0.6,1){6}{\circle*{0.2}}
\put(37,0.5){\circle{1}}
\put(47,0.5){\circle*{1}}
\put(42,9){\circle{1}}
\multiput(41,0.5)(2,0){2}{\circle{1}}
\multiput(39,4)(6,0){2}{\makebox(0,0){$\times$}}
\put(40,5.5){\circle{1}}
\put(44,5.5){\circle*{1}}
\put(51,5){\makebox(0,0){$+$}}
\put(54,0){\line(1,0){12}}
\put(57,5){\line(1,0){6}}
\put(63,5){\line(-3,5){3}}
\multiput(54,0)(0.6,1){11}{\circle*{0.2}}
\multiput(60,0)(-0.6,1){6}{\circle*{0.2}}
\multiput(60,0)(0.6,1){6}{\circle*{0.2}}
\multiput(66,0)(-0.6,1){6}{\circle*{0.2}}
\put(55,0.5){\circle{1}}
\put(65,0.5){\circle*{1}}
\put(60,9){\circle{1}}
\multiput(59,0.5)(2,0){2}{\circle{1}}
\multiput(57,4)(6,0){2}{\makebox(0,0){$\times$}}
\put(58,5.5){\circle*{1}}
\put(62,5.5){\circle*{1}}
\put(69,5){\makebox(0,0){$+$}}
\put(72,0){\line(1,0){12}}
\put(75,5){\line(1,0){6}}
\put(81,5){\line(-3,5){3}}
\multiput(72,0)(0.6,1){11}{\circle*{0.2}}
\multiput(78,0)(-0.6,1){6}{\circle*{0.2}}
\multiput(78,0)(0.6,1){6}{\circle*{0.2}}
\multiput(84,0)(-0.6,1){6}{\circle*{0.2}}
\put(73,0.5){\circle{1}}
\put(83,0.5){\circle*{1}}
\put(78,9){\circle{1}}
\multiput(77,0.5)(2,0){2}{\circle{1}}
\put(75,4){\makebox(0,0){\scriptsize $\oplus$}}
\put(81,4){\makebox(0,0){$\times$}}
\put(76,5.5){\circle{1}}
\put(80,5.5){\circle*{1}}
\put(87,5){\makebox(0,0){$+$}}
\put(90,0){\line(1,0){12}}
\put(96,10){\line(-3,-5){3}}
\put(93,5){\line(1,0){6}}
\multiput(90,0)(0.6,1){6}{\circle*{0.2}}
\multiput(96,0)(-0.6,1){6}{\circle*{0.2}}
\multiput(96,0)(0.6,1){6}{\circle*{0.2}}
\multiput(102,0)(-0.6,1){11}{\circle*{0.2}}
\put(91,0.5){\circle{1}}
\put(101,0.5){\circle*{1}}
\put(96,9){\circle{1}}
\multiput(95,0.5)(2,0){2}{\circle{1}}
\multiput(93,4)(6,0){2}{\makebox(0,0){$\times$}}
\put(94,5.5){\circle*{1}}
\put(98,5.5){\circle{1}}
\end{picture}

\vspace*{5mm}

\begin{picture}(102,10)
\put(15,5){\makebox(0,0){$+$}}
\put(18,0){\line(1,0){12}}
\put(24,10){\line(-3,-5){3}}
\put(21,5){\line(1,0){6}}
\multiput(18,0)(0.6,1){6}{\circle*{0.2}}
\multiput(24,0)(-0.6,1){6}{\circle*{0.2}}
\multiput(24,0)(0.6,1){6}{\circle*{0.2}}
\multiput(30,0)(-0.6,1){11}{\circle*{0.2}}
\put(19,0.5){\circle{1}}
\put(29,0.5){\circle*{1}}
\put(24,9){\circle{1}}
\multiput(23,0.5)(2,0){2}{\circle{1}}
\put(21,4){\makebox(0,0){$\times$}}
\put(27,4){\makebox(0,0){\scriptsize $\oplus$}}
\put(22,5.5){\circle*{1}}
\put(26,5.5){\circle{1}}
\put(33,5){\makebox(0,0){$+$}}
\put(36,0){\line(1,0){12}}
\put(48,0){\line(-3,5){3}}
\put(39,5){\line(3,5){3}}
\multiput(36,0)(0.6,1){6}{\circle*{0.2}}
\multiput(42,0)(-0.6,1){6}{\circle*{0.2}}
\multiput(42,0)(0.6,1){6}{\circle*{0.2}}
\multiput(39,5)(1,0){7}{\circle*{0.2}}
\multiput(45,5)(-0.6,1){6}{\circle*{0.2}}
\put(37,0.5){\circle{1}}
\put(47,0.5){\circle*{1}}
\put(42,9){\circle{1}}
\multiput(41,0.5)(2,0){2}{\circle{1}}
\put(39,4){\makebox(0,0){$\times$}}
\put(45,4){\circle{1}}
\put(40,5.5){\circle{1}}
\put(44,5.5){\makebox(0,0){$\times$}}
\put(51,5){\makebox(0,0){$+$}}
\put(54,0){\line(1,0){12}}
\put(66,0){\line(-3,5){3}}
\put(57,5){\line(3,5){3}}
\multiput(54,0)(0.6,1){6}{\circle*{0.2}}
\multiput(60,0)(-0.6,1){6}{\circle*{0.2}}
\multiput(60,0)(0.6,1){6}{\circle*{0.2}}
\multiput(57,5)(1,0){7}{\circle*{0.2}}
\multiput(63,5)(-0.6,1){6}{\circle*{0.2}}
\put(55,0.5){\circle{1}}
\put(65,0.5){\circle*{1}}
\put(60,9){\circle{1}}
\multiput(59,0.5)(2,0){2}{\circle{1}}
\put(57,4){\makebox(0,0){$\times$}}
\put(63,4){\circle{1}}
\put(58,5.5){\circle{1}}
\put(62,5.5){\makebox(0,0){\scriptsize $\oplus$}}
\put(69,5){\makebox(0,0){$+$}}
\put(72,0){\line(1,0){12}}
\put(84,0){\line(-3,5){3}}
\put(75,5){\line(3,5){3}}
\multiput(72,0)(0.6,1){6}{\circle*{0.2}}
\multiput(78,0)(-0.6,1){6}{\circle*{0.2}}
\multiput(78,0)(0.6,1){6}{\circle*{0.2}}
\multiput(75,5)(1,0){7}{\circle*{0.2}}
\multiput(81,5)(-0.6,1){6}{\circle*{0.2}}
\put(73,0.5){\circle{1}}
\put(83,0.5){\circle*{1}}
\put(78,9){\circle{1}}
\multiput(77,0.5)(2,0){2}{\circle{1}}
\put(75,4){\makebox(0,0){$\times$}}
\put(81,4){\circle{1}}
\put(76,5.5){\circle*{1}}
\put(80,5.5){\makebox(0,0){$\times$}}
\put(87,5){\makebox(0,0){$+$}}
\put(90,0){\line(1,0){12}}
\put(102,0){\line(-3,5){3}}
\put(93,5){\line(3,5){3}}
\multiput(90,0)(0.6,1){6}{\circle*{0.2}}
\multiput(96,0)(-0.6,1){6}{\circle*{0.2}}
\multiput(96,0)(0.6,1){6}{\circle*{0.2}}
\multiput(93,5)(1,0){7}{\circle*{0.2}}
\multiput(99,5)(-0.6,1){6}{\circle*{0.2}}
\put(91,0.5){\circle{1}}
\put(101,0.5){\circle*{1}}
\put(96,9){\circle{1}}
\multiput(95,0.5)(2,0){2}{\circle{1}}
\put(93,4){\makebox(0,0){$\times$}}
\put(99,4){\circle{1}}
\put(94,5.5){\circle*{1}}
\put(98,5.5){\makebox(0,0){\scriptsize $\oplus$}}
\end{picture}

\vspace*{5mm}

\begin{picture}(102,10)
\put(15,5){\makebox(0,0){$+$}}
\put(18,0){\line(1,0){12}}
\put(30,0){\line(-3,5){3}}
\put(21,5){\line(3,5){3}}
\multiput(18,0)(0.6,1){6}{\circle*{0.2}}
\multiput(24,0)(-0.6,1){6}{\circle*{0.2}}
\multiput(24,0)(0.6,1){6}{\circle*{0.2}}
\multiput(21,5)(1,0){7}{\circle*{0.2}}
\multiput(27,5)(-0.6,1){6}{\circle*{0.2}}
\put(19,0.5){\circle{1}}
\put(29,0.5){\circle*{1}}
\put(24,9){\circle{1}}
\multiput(23,0.5)(2,0){2}{\circle{1}}
\put(21,4){\makebox(0,0){\scriptsize $\oplus$}}
\put(27,4){\circle{1}}
\put(22,5.5){\circle{1}}
\put(26,5.5){\makebox(0,0){$\times$}}
\put(33,5){\makebox(0,0){$+$}}
\put(36,0){\line(1,0){12}}
\put(48,0){\line(-3,5){3}}
\put(39,5){\line(3,5){3}}
\multiput(36,0)(0.6,1){6}{\circle*{0.2}}
\multiput(42,0)(-0.6,1){6}{\circle*{0.2}}
\multiput(42,0)(0.6,1){6}{\circle*{0.2}}
\multiput(39,5)(1,0){7}{\circle*{0.2}}
\multiput(45,5)(-0.6,1){6}{\circle*{0.2}}
\put(37,0.5){\circle{1}}
\put(47,0.5){\circle*{1}}
\put(42,9){\circle{1}}
\multiput(41,0.5)(2,0){2}{\circle{1}}
\put(39,4){\makebox(0,0){\scriptsize $\oplus$}}
\put(45,4){\circle{1}}
\put(40,5.5){\circle{1}}
\put(44,5.5){\makebox(0,0){\scriptsize $\oplus$}}
\end{picture}

\caption{\footnotesize{Illustration for the expression of $g_4(n+1)$.}} 
\label{g4fig}
\end{figure}

\begin{figure}[htbp]
\unitlength 1.2mm 
\begin{picture}(108,10)
\multiput(0,0)(1,0){13}{\circle*{0.2}}
\multiput(0,0)(0.6,1){11}{\circle*{0.2}}
\multiput(12,0)(-0.6,1){11}{\circle*{0.2}}
\multiput(1,0.5)(10,0){2}{\makebox(0,0){\scriptsize $\oplus$}}
\put(6,9){\makebox(0,0){$\times$}}
\put(15,5){\makebox(0,0){$=$}}
\put(21,5){\makebox(0,0){$2 \ ($}}
\put(24,0){\line(1,0){6}}
\put(30,0){\line(3,5){3}}
\put(33,5){\line(-1,0){6}}
\multiput(24,0)(0.6,1){11}{\circle*{0.2}}
\multiput(30,0)(-0.6,1){6}{\circle*{0.2}}
\multiput(30,0)(1,0){7}{\circle*{0.2}}
\multiput(36,0)(-0.6,1){11}{\circle*{0.2}}
\put(25,0.5){\circle{1}}
\put(35,0.5){\makebox(0,0){\scriptsize $\oplus$}}
\put(30,9){\makebox(0,0){$\times$}}
\multiput(29,0.5)(2,0){2}{\circle{1}}
\put(27,4){\makebox(0,0){$\times$}}
\put(33,4){\circle{1}}
\multiput(28,5.5)(4,0){2}{\circle{1}}
\put(39,5){\makebox(0,0){$+$}}
\put(42,0){\line(1,0){6}}
\put(48,0){\line(3,5){3}}
\put(51,5){\line(-1,0){6}}
\multiput(42,0)(0.6,1){11}{\circle*{0.2}}
\multiput(48,0)(-0.6,1){6}{\circle*{0.2}}
\multiput(48,0)(1,0){7}{\circle*{0.2}}
\multiput(54,0)(-0.6,1){11}{\circle*{0.2}}
\put(43,0.5){\circle{1}}
\put(53,0.5){\makebox(0,0){\scriptsize $\oplus$}}
\put(48,9){\makebox(0,0){$\times$}}
\multiput(47,0.5)(2,0){2}{\circle{1}}
\put(45,4){\makebox(0,0){$\times$}}
\put(51,4){\circle{1}}
\put(46,5.5){\circle*{1}}
\put(50,5.5){\circle{1}}
\put(57,5){\makebox(0,0){$+$}}
\put(60,0){\line(1,0){6}}
\put(66,0){\line(3,5){3}}
\put(69,5){\line(-1,0){6}}
\multiput(60,0)(0.6,1){11}{\circle*{0.2}}
\multiput(66,0)(-0.6,1){6}{\circle*{0.2}}
\multiput(66,0)(1,0){7}{\circle*{0.2}}
\multiput(72,0)(-0.6,1){11}{\circle*{0.2}}
\put(61,0.5){\circle{1}}
\put(71,0.5){\makebox(0,0){\scriptsize $\oplus$}}
\put(66,9){\makebox(0,0){$\times$}}
\multiput(65,0.5)(2,0){2}{\circle{1}}
\put(63,4){\makebox(0,0){\scriptsize $\oplus$}}
\put(69,4){\circle{1}}
\put(64,5.5){\circle{1}}
\put(68,5.5){\circle{1}}
\put(75,5){\makebox(0,0){$+$}}
\put(78,0){\line(3,5){3}}
\put(81,5){\line(1,0){6}}
\put(87,5){\line(-3,-5){3}}
\multiput(78,0)(1,0){13}{\circle*{0.2}}
\multiput(84,0)(-0.6,1){6}{\circle*{0.2}}
\multiput(84,10)(-0.6,-1){6}{\circle*{0.2}}
\multiput(84,10)(0.6,-1){11}{\circle*{0.2}}
\put(79,0.5){\circle{1}}
\put(89,0.5){\makebox(0,0){\scriptsize $\oplus$}}
\put(84,9){\makebox(0,0){$\times$}}
\put(83,0.5){\makebox(0,0){$\times$}}
\put(85,0.5){\circle{1}}
\multiput(81,4)(6,0){2}{\circle{1}}
\put(82,5.5){\circle{1}}
\put(86,5.5){\circle{1}}
\put(93,5){\makebox(0,0){$+$}}
\put(96,0){\line(3,5){3}}
\put(99,5){\line(1,0){6}}
\put(105,5){\line(-3,-5){3}}
\multiput(96,0)(1,0){13}{\circle*{0.2}}
\multiput(102,0)(-0.6,1){6}{\circle*{0.2}}
\multiput(102,10)(-0.6,-1){6}{\circle*{0.2}}
\multiput(102,10)(0.6,-1){11}{\circle*{0.2}}
\put(97,0.5){\circle{1}}
\put(107,0.5){\makebox(0,0){\scriptsize $\oplus$}}
\put(102,9){\makebox(0,0){$\times$}}
\put(101,0.5){\makebox(0,0){\scriptsize $\oplus$}}
\put(103,0.5){\circle{1}}
\multiput(99,4)(6,0){2}{\circle{1}}
\put(100,5.5){\circle{1}}
\put(104,5.5){\circle{1}}
\end{picture}

\vspace*{5mm}

\begin{picture}(108,10)
\put(15,5){\makebox(0,0){$+$}}
\put(18,0){\line(3,5){3}}
\put(21,5){\line(1,0){6}}
\put(27,5){\line(-3,-5){3}}
\multiput(18,0)(1,0){13}{\circle*{0.2}}
\multiput(24,0)(-0.6,1){6}{\circle*{0.2}}
\multiput(24,10)(-0.6,-1){6}{\circle*{0.2}}
\multiput(24,10)(0.6,-1){11}{\circle*{0.2}}
\put(19,0.5){\circle{1}}
\put(29,0.5){\makebox(0,0){\scriptsize $\oplus$}}
\put(24,9){\makebox(0,0){$\times$}}
\put(23,0.5){\makebox(0,0){$\times$}}
\put(25,0.5){\circle*{1}}
\put(21,4){\circle{1}}
\put(27,4){\circle{1}}
\multiput(22,5.5)(4,0){2}{\circle{1}}
\put(33,5){\makebox(0,0){$+$}}
\put(36,0){\line(1,0){6}}
\put(42,0){\line(-3,5){3}}
\put(39,5){\line(1,0){6}}
\multiput(36,0)(0.6,1){11}{\circle*{0.2}}
\multiput(42,0)(0.6,1){6}{\circle*{0.2}}
\multiput(42,0)(1,0){7}{\circle*{0.2}}
\multiput(42,10)(0.6,-1){11}{\circle*{0.2}}
\put(37,0.5){\circle{1}}
\put(47,0.5){\makebox(0,0){\scriptsize $\oplus$}}
\put(42,9){\makebox(0,0){$\times$}}
\put(41,0.5){\circle*{1}}
\put(43,0.5){\makebox(0,0){$\times$}}
\put(39,4){\circle{1}}
\put(45,4){\makebox(0,0){$\times$}}
\multiput(40,5.5)(4,0){2}{\circle{1}}
\put(51,5){\makebox(0,0){$+$}}
\put(54,0){\line(1,0){6}}
\put(60,0){\line(-3,5){3}}
\put(57,5){\line(1,0){6}}
\multiput(54,0)(0.6,1){11}{\circle*{0.2}}
\multiput(60,0)(0.6,1){6}{\circle*{0.2}}
\multiput(60,0)(1,0){7}{\circle*{0.2}}
\multiput(60,10)(0.6,-1){11}{\circle*{0.2}}
\put(55,0.5){\circle{1}}
\put(65,0.5){\makebox(0,0){\scriptsize $\oplus$}}
\put(60,9){\makebox(0,0){$\times$}}
\put(59,0.5){\circle*{1}}
\put(61,0.5){\makebox(0,0){$\times$}}
\put(57,4){\circle{1}}
\put(63,4){\makebox(0,0){$\times$}}
\put(58,5.5){\circle{1}}
\put(62,5.5){\circle*{1}}
\put(69,5){\makebox(0,0){$+$}}
\put(72,0){\line(1,0){6}}
\put(78,0){\line(-3,5){3}}
\put(75,5){\line(1,0){6}}
\multiput(72,0)(0.6,1){11}{\circle*{0.2}}
\multiput(78,0)(0.6,1){6}{\circle*{0.2}}
\multiput(78,0)(1,0){7}{\circle*{0.2}}
\multiput(78,10)(0.6,-1){11}{\circle*{0.2}}
\put(73,0.5){\circle{1}}
\put(83,0.5){\makebox(0,0){\scriptsize $\oplus$}}
\put(78,9){\makebox(0,0){$\times$}}
\put(77,0.5){\circle*{1}}
\put(79,0.5){\makebox(0,0){$\times$}}
\put(75,4){\circle{1}}
\put(81,4){\makebox(0,0){\scriptsize $\oplus$}}
\multiput(76,5.5)(4,0){2}{\circle{1}}
\put(87,5){\makebox(0,0){$+$}}
\put(90,0){\line(3,5){3}}
\put(93,5){\line(3,-5){3}}
\put(96,0){\line(3,5){3}}
\multiput(90,0)(1,0){13}{\circle*{0.2}}
\multiput(93,5)(1,0){7}{\circle*{0.2}}
\multiput(96,10)(-0.6,-1){6}{\circle*{0.2}}
\multiput(96,10)(0.6,-1){11}{\circle*{0.2}}
\put(91,0.5){\circle{1}}
\put(101,0.5){\makebox(0,0){\scriptsize $\oplus$}}
\put(96,9){\makebox(0,0){$\times$}}
\put(95,0.5){\circle{1}}
\put(97,0.5){\circle{1}}
\put(93,4){\circle*{1}}
\put(99,4){\circle{1}}
\put(94,5.5){\makebox(0,0){$\times$}}
\put(98,5.5){\makebox(0,0){\scriptsize $\oplus$}}
\end{picture}

\vspace*{5mm}

\begin{picture}(108,10)
\put(15,5){\makebox(0,0){$+$}}
\put(18,0){\line(1,0){6}}
\put(24,0){\line(-3,5){3}}
\put(30,0){\line(-3,5){3}}
\multiput(18,0)(0.6,1){11}{\circle*{0.2}}
\multiput(21,5)(1,0){7}{\circle*{0.2}}
\multiput(24,0)(0.6,1){6}{\circle*{0.2}}
\multiput(24,0)(1,0){7}{\circle*{0.2}}
\multiput(24,10)(-0.6,-1){6}{\circle*{0.2}}
\multiput(24,10)(0.6,-1){6}{\circle*{0.2}}
\multiput(19,0.5)(10,0){2}{\circle{1}}
\put(24,9){\makebox(0,0){$\times$}}
\put(23,0.5){\circle*{1}}
\put(25,0.5){\makebox(0,0){$\times$}}
\put(21,4){\circle{1}}
\put(27,4){\circle{1}}
\put(22,5.5){\makebox(0,0){$\times$}}
\put(26,5.5){\makebox(0,0){\scriptsize $\oplus$}}
\put(33,5){\makebox(0,0){$+$}}
\put(36,0){\line(1,0){6}}
\put(42,0){\line(-3,5){3}}
\put(48,0){\line(-3,5){3}}
\multiput(36,0)(0.6,1){11}{\circle*{0.2}}
\multiput(39,5)(1,0){7}{\circle*{0.2}}
\multiput(42,0)(0.6,1){6}{\circle*{0.2}}
\multiput(42,0)(1,0){7}{\circle*{0.2}}
\multiput(42,10)(-0.6,-1){6}{\circle*{0.2}}
\multiput(42,10)(0.6,-1){6}{\circle*{0.2}}
\multiput(37,0.5)(10,0){2}{\circle{1}}
\put(42,9){\makebox(0,0){$\times$}}
\put(41,0.5){\circle*{1}}
\put(43,0.5){\makebox(0,0){$\times$}}
\put(39,4){\circle*{1}}
\put(45,4){\circle{1}}
\put(40,5.5){\makebox(0,0){$\times$}}
\put(44,5.5){\makebox(0,0){\scriptsize $\oplus$}}
\put(51,5){\makebox(0,0){$+$}}
\put(54,0){\line(1,0){6}}
\put(60,0){\line(-3,5){3}}
\put(66,0){\line(-3,5){3}}
\multiput(54,0)(0.6,1){11}{\circle*{0.2}}
\multiput(57,5)(1,0){7}{\circle*{0.2}}
\multiput(60,0)(0.6,1){6}{\circle*{0.2}}
\multiput(60,0)(1,0){7}{\circle*{0.2}}
\multiput(60,10)(-0.6,-1){6}{\circle*{0.2}}
\multiput(60,10)(0.6,-1){6}{\circle*{0.2}}
\multiput(55,0.5)(10,0){2}{\circle{1}}
\put(60,9){\makebox(0,0){$\times$}}
\put(59,0.5){\circle*{1}}
\put(61,0.5){\makebox(0,0){$\times$}}
\put(57,4){\circle{1}}
\put(63,4){\circle{1}}
\put(58,5.5){\makebox(0,0){\scriptsize $\oplus$}}
\put(62,5.5){\makebox(0,0){$\times$}}
\put(69,5){\makebox(0,0){$+$}}
\put(72,0){\line(1,0){6}}
\put(78,0){\line(-3,5){3}}
\put(84,0){\line(-3,5){3}}
\multiput(72,0)(0.6,1){11}{\circle*{0.2}}
\multiput(75,5)(1,0){7}{\circle*{0.2}}
\multiput(78,0)(0.6,1){6}{\circle*{0.2}}
\multiput(78,0)(1,0){7}{\circle*{0.2}}
\multiput(78,10)(-0.6,-1){6}{\circle*{0.2}}
\multiput(78,10)(0.6,-1){6}{\circle*{0.2}}
\multiput(73,0.5)(10,0){2}{\circle{1}}
\put(78,9){\makebox(0,0){$\times$}}
\put(77,0.5){\circle*{1}}
\put(79,0.5){\makebox(0,0){$\times$}}
\put(75,4){\circle{1}}
\put(81,4){\circle*{1}}
\put(76,5.5){\makebox(0,0){\scriptsize $\oplus$}}
\put(80,5.5){\makebox(0,0){$\times$}}
\put(87,5){\makebox(0,0){$+$}}
\put(90,0){\line(1,0){6}}
\put(96,0){\line(-3,5){3}}
\put(102,0){\line(-3,5){3}}
\multiput(90,0)(0.6,1){11}{\circle*{0.2}}
\multiput(93,5)(1,0){7}{\circle*{0.2}}
\multiput(96,0)(0.6,1){6}{\circle*{0.2}}
\multiput(96,0)(1,0){7}{\circle*{0.2}}
\multiput(96,10)(-0.6,-1){6}{\circle*{0.2}}
\multiput(96,10)(0.6,-1){6}{\circle*{0.2}}
\multiput(91,0.5)(10,0){2}{\circle{1}}
\put(96,9){\makebox(0,0){$\times$}}
\put(95,0.5){\circle*{1}}
\put(97,0.5){\makebox(0,0){$\times$}}
\put(93,4){\circle{1}}
\put(99,4){\circle{1}}
\multiput(94,5.5)(4,0){2}{\makebox(0,0){\scriptsize $\oplus$}}
\end{picture}

\vspace*{5mm}

\begin{picture}(108,10)
\put(15,5){\makebox(0,0){$+$}}
\put(18,0){\line(3,5){3}}
\put(21,5){\line(3,-5){3}}
\put(30,0){\line(-3,5){3}}
\multiput(18,0)(1,0){13}{\circle*{0.2}}
\multiput(21,5)(1,0){7}{\circle*{0.2}}
\multiput(24,0)(0.6,1){6}{\circle*{0.2}}
\multiput(24,10)(-0.6,-1){6}{\circle*{0.2}}
\multiput(24,10)(0.6,-1){6}{\circle*{0.2}}
\multiput(19,0.5)(10,0){2}{\circle{1}}
\put(24,9){\makebox(0,0){$\times$}}
\put(23,0.5){\circle{1}}
\put(25,0.5){\makebox(0,0){$\times$}}
\put(21,4){\circle*{1}}
\put(27,4){\circle{1}}
\put(22,5.5){\makebox(0,0){$\times$}}
\put(26,5.5){\makebox(0,0){\scriptsize $\oplus$}}
\put(33,5){\makebox(0,0){$+$}}
\put(36,0){\line(3,5){3}}
\put(39,5){\line(3,-5){3}}
\put(48,0){\line(-3,5){3}}
\multiput(36,0)(1,0){13}{\circle*{0.2}}
\multiput(39,5)(1,0){7}{\circle*{0.2}}
\multiput(42,0)(0.6,1){6}{\circle*{0.2}}
\multiput(42,10)(-0.6,-1){6}{\circle*{0.2}}
\multiput(42,10)(0.6,-1){6}{\circle*{0.2}}
\multiput(37,0.5)(10,0){2}{\circle{1}}
\put(42,9){\makebox(0,0){$\times$}}
\put(41,0.5){\circle{1}}
\put(43,0.5){\makebox(0,0){\scriptsize $\oplus$}}
\put(39,4){\circle*{1}}
\put(45,4){\circle{1}}
\put(40,5.5){\makebox(0,0){$\times$}}
\put(44,5.5){\makebox(0,0){\scriptsize $\oplus$}}
\put(51,5){\makebox(0,0){$+$}}
\put(54,0){\line(3,5){3}}
\put(57,5){\line(1,0){6}}
\put(60,0){\line(1,0){6}}
\multiput(54,0)(1,0){7}{\circle*{0.2}}
\multiput(60,0)(0.6,1){6}{\circle*{0.2}}
\multiput(60,0)(-0.6,1){6}{\circle*{0.2}}
\multiput(60,10)(-0.6,-1){6}{\circle*{0.2}}
\multiput(60,10)(0.6,-1){11}{\circle*{0.2}}
\multiput(55,0.5)(10,0){2}{\circle{1}}
\put(60,9){\makebox(0,0){$\times$}}
\put(59,0.5){\makebox(0,0){$\times$}}
\put(61,0.5){\circle{1}}
\put(57,4){\circle{1}}
\put(63,4){\makebox(0,0){$\times$}}
\multiput(58,5.5)(4,0){2}{\circle{1}}
\put(69,5){\makebox(0,0){$+$}}
\put(72,0){\line(3,5){3}}
\put(75,5){\line(1,0){6}}
\put(78,0){\line(1,0){6}}
\multiput(72,0)(1,0){7}{\circle*{0.2}}
\multiput(78,0)(0.6,1){6}{\circle*{0.2}}
\multiput(78,0)(-0.6,1){6}{\circle*{0.2}}
\multiput(78,10)(-0.6,-1){6}{\circle*{0.2}}
\multiput(78,10)(0.6,-1){11}{\circle*{0.2}}
\multiput(73,0.5)(10,0){2}{\circle{1}}
\put(78,9){\makebox(0,0){$\times$}}
\put(77,0.5){\makebox(0,0){$\times$}}
\put(79,0.5){\circle{1}}
\put(75,4){\circle{1}}
\put(81,4){\makebox(0,0){$\times$}}
\put(76,5.5){\circle{1}}
\put(80,5.5){\circle*{1}}
\put(87,5){\makebox(0,0){$+$}}
\put(90,0){\line(3,5){3}}
\put(93,5){\line(1,0){6}}
\put(96,0){\line(1,0){6}}
\multiput(90,0)(1,0){7}{\circle*{0.2}}
\multiput(96,0)(0.6,1){6}{\circle*{0.2}}
\multiput(96,0)(-0.6,1){6}{\circle*{0.2}}
\multiput(96,10)(-0.6,-1){6}{\circle*{0.2}}
\multiput(96,10)(0.6,-1){11}{\circle*{0.2}}
\multiput(91,0.5)(10,0){2}{\circle{1}}
\put(96,9){\makebox(0,0){$\times$}}
\put(95,0.5){\makebox(0,0){$\times$}}
\put(97,0.5){\circle{1}}
\put(93,4){\circle{1}}
\put(99,4){\makebox(0,0){\scriptsize $\oplus$}}
\multiput(94,5.5)(4,0){2}{\circle{1}}
\end{picture}

\vspace*{5mm}

\begin{picture}(108,10)
\put(15,5){\makebox(0,0){$+$}}
\put(18,0){\line(3,5){3}}
\put(21,5){\line(1,0){6}}
\put(24,0){\line(1,0){6}}
\multiput(18,0)(1,0){7}{\circle*{0.2}}
\multiput(24,0)(0.6,1){6}{\circle*{0.2}}
\multiput(24,0)(-0.6,1){6}{\circle*{0.2}}
\multiput(24,10)(-0.6,-1){6}{\circle*{0.2}}
\multiput(24,10)(0.6,-1){11}{\circle*{0.2}}
\multiput(19,0.5)(10,0){2}{\circle{1}}
\put(24,9){\makebox(0,0){$\times$}}
\put(23,0.5){\makebox(0,0){$\times$}}
\put(25,0.5){\circle*{1}}
\put(21,4){\circle{1}}
\put(27,4){\makebox(0,0){$\times$}}
\multiput(22,5.5)(4,0){2}{\circle{1}}
\put(33,5){\makebox(0,0){$+$}}
\put(36,0){\line(3,5){3}}
\put(39,5){\line(1,0){6}}
\put(42,0){\line(1,0){6}}
\multiput(36,0)(1,0){7}{\circle*{0.2}}
\multiput(42,0)(0.6,1){6}{\circle*{0.2}}
\multiput(42,0)(-0.6,1){6}{\circle*{0.2}}
\multiput(42,10)(-0.6,-1){6}{\circle*{0.2}}
\multiput(42,10)(0.6,-1){11}{\circle*{0.2}}
\multiput(37,0.5)(10,0){2}{\circle{1}}
\put(42,9){\makebox(0,0){$\times$}}
\put(41,0.5){\makebox(0,0){$\times$}}
\put(43,0.5){\circle*{1}}
\put(39,4){\circle{1}}
\put(45,4){\makebox(0,0){$\times$}}
\put(40,5.5){\circle{1}}
\put(44,5.5){\circle*{1}}
\put(51,5){\makebox(0,0){$+$}}
\put(54,0){\line(3,5){3}}
\put(57,5){\line(1,0){6}}
\put(60,0){\line(1,0){6}}
\multiput(54,0)(1,0){7}{\circle*{0.2}}
\multiput(60,0)(0.6,1){6}{\circle*{0.2}}
\multiput(60,0)(-0.6,1){6}{\circle*{0.2}}
\multiput(60,10)(-0.6,-1){6}{\circle*{0.2}}
\multiput(60,10)(0.6,-1){11}{\circle*{0.2}}
\multiput(55,0.5)(10,0){2}{\circle{1}}
\put(60,9){\makebox(0,0){$\times$}}
\put(59,0.5){\makebox(0,0){$\times$}}
\put(61,0.5){\circle*{1}}
\put(57,4){\circle{1}}
\put(63,4){\makebox(0,0){\scriptsize $\oplus$}}
\multiput(58,5.5)(4,0){2}{\circle{1}}
\put(69,5){\makebox(0,0){$+$}}
\put(72,0){\line(3,5){3}}
\put(75,5){\line(1,0){6}}
\put(78,0){\line(1,0){6}}
\multiput(72,0)(1,0){7}{\circle*{0.2}}
\multiput(78,0)(0.6,1){6}{\circle*{0.2}}
\multiput(78,0)(-0.6,1){6}{\circle*{0.2}}
\multiput(78,10)(-0.6,-1){6}{\circle*{0.2}}
\multiput(78,10)(0.6,-1){11}{\circle*{0.2}}
\multiput(73,0.5)(10,0){2}{\circle{1}}
\put(78,9){\makebox(0,0){$\times$}}
\put(77,0.5){\makebox(0,0){\scriptsize $\oplus$}}
\put(79,0.5){\circle{1}}
\put(75,4){\circle{1}}
\put(81,4){\makebox(0,0){$\times$}}
\multiput(76,5.5)(4,0){2}{\circle{1}}
\put(87,5){\makebox(0,0){$+$}}
\put(90,0){\line(3,5){3}}
\put(93,5){\line(1,0){6}}
\put(96,0){\line(1,0){6}}
\multiput(90,0)(1,0){7}{\circle*{0.2}}
\multiput(96,0)(0.6,1){6}{\circle*{0.2}}
\multiput(96,0)(-0.6,1){6}{\circle*{0.2}}
\multiput(96,10)(-0.6,-1){6}{\circle*{0.2}}
\multiput(96,10)(0.6,-1){11}{\circle*{0.2}}
\multiput(91,0.5)(10,0){2}{\circle{1}}
\put(96,9){\makebox(0,0){$\times$}}
\put(95,0.5){\makebox(0,0){\scriptsize $\oplus$}}
\put(97,0.5){\circle{1}}
\put(93,4){\circle{1}}
\put(99,4){\makebox(0,0){$\times$}}
\put(94,5.5){\circle{1}}
\put(98,5.5){\circle*{1}}
\end{picture}

\vspace*{5mm}

\begin{picture}(108,10)
\put(15,5){\makebox(0,0){$+$}}
\put(18,0){\line(3,5){3}}
\put(21,5){\line(1,0){6}}
\put(24,0){\line(1,0){6}}
\multiput(18,0)(1,0){7}{\circle*{0.2}}
\multiput(24,0)(0.6,1){6}{\circle*{0.2}}
\multiput(24,0)(-0.6,1){6}{\circle*{0.2}}
\multiput(24,10)(-0.6,-1){6}{\circle*{0.2}}
\multiput(24,10)(0.6,-1){11}{\circle*{0.2}}
\multiput(19,0.5)(10,0){2}{\circle{1}}
\put(24,9){\makebox(0,0){$\times$}}
\put(23,0.5){\makebox(0,0){\scriptsize $\oplus$}}
\put(25,0.5){\circle{1}}
\put(21,4){\circle{1}}
\put(27,4){\makebox(0,0){\scriptsize $\oplus$}}
\multiput(22,5.5)(4,0){2}{\circle{1}}
\put(33,5){\makebox(0,0){$+$}}
\put(39,5){\makebox(0,0){$\delta_{n0}$}}
\put(42,0){\line(1,0){6}}
\put(45,5){\line(3,-5){3}}
\put(51,5){\line(3,-5){3}}
\multiput(42,0)(0.6,1){11}{\circle*{0.2}}
\multiput(48,0)(1,0){7}{\circle*{0.2}}
\multiput(45,5)(1,0){7}{\circle*{0.2}}
\multiput(48,0)(0.6,1){6}{\circle*{0.2}}
\multiput(48,10)(0.6,-1){6}{\circle*{0.2}}
\multiput(43,0.5)(10,0){2}{\circle{1}}
\put(48,9){\makebox(0,0){$\times$}}
\put(47,0.5){\circle*{1}}
\put(49,0.5){\makebox(0,0){$\times$}}
\multiput(45,4)(6,0){2}{\circle{1}}
\multiput(46,5.5)(4,0){2}{\makebox(0,0){$\times$}}
\put(57,5){\makebox(0,0){$+$}}
\put(63,5){\makebox(0,0){$\delta_{n0}$}}
\put(66,0){\line(3,5){3}}
\put(69,5){\line(3,-5){3}}
\put(75,5){\line(3,-5){3}}
\multiput(66,0)(1,0){13}{\circle*{0.2}}
\multiput(69,5)(1,0){7}{\circle*{0.2}}
\multiput(72,0)(0.6,1){6}{\circle*{0.2}}
\multiput(72,0)(-0.6,1){6}{\circle*{0.2}}
\multiput(72,10)(0.6,-1){6}{\circle*{0.2}}
\multiput(72,10)(-0.6,-1){6}{\circle*{0.2}}
\multiput(67,0.5)(10,0){2}{\circle{1}}
\put(72,9){\makebox(0,0){$\times$}}
\put(71,0.5){\circle{1}}
\put(73,0.5){\makebox(0,0){$\times$}}
\put(69,4){\circle*{1}}
\put(75,4){\circle{1}}
\multiput(70,5.5)(4,0){2}{\makebox(0,0){$\times$}}
\put(81,5){\makebox(0,0){$)$}}
\end{picture}

\caption{\footnotesize{Illustration for the expression of $t_2(n+1)$. The multiplication of two on the right-hand-side corresponds to the reflection symmetry with respect to the central vertical axis. The last two drawings only apply for $n=0$.}} 
\label{h2fig}
\end{figure}
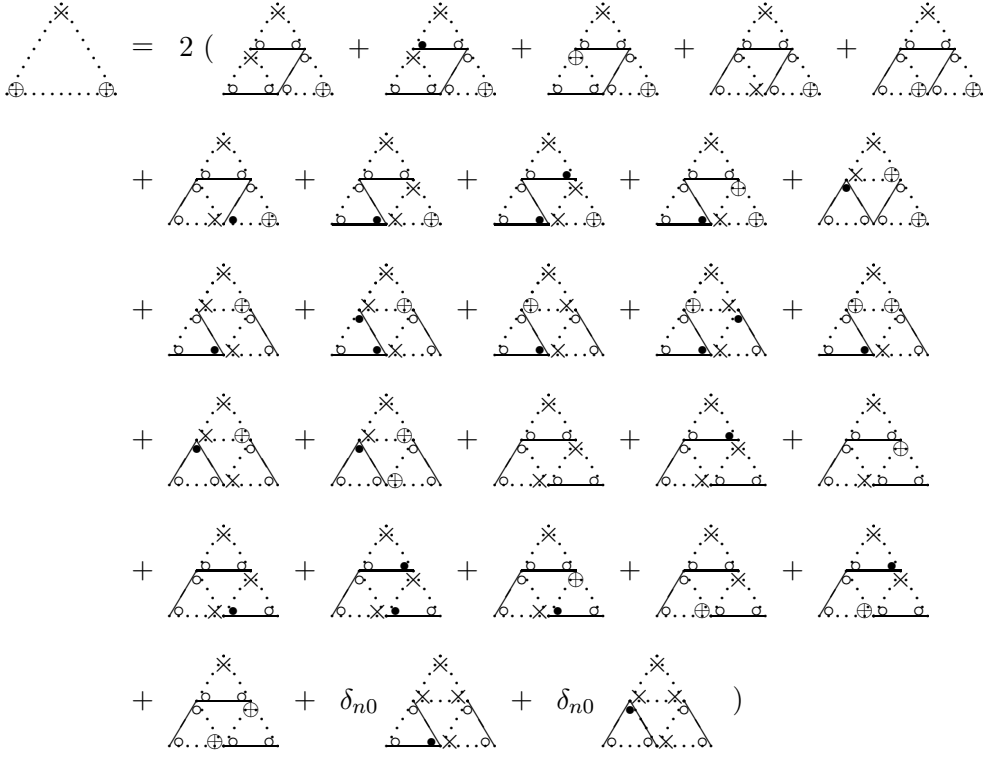

In this section, we enumerate the total number of Hamiltonian paths. As we have obtained $f_1(n)$ and $f_2(n)$ in the previous section, we need to evaluate $f_3(n)$ for $n \geq 1$ before Eq. (\ref{H}) is applied. Consider the quantities $f_3(n)$, $g_4(n)$, and $t_2(n)$ for $n \geq 1$, we have the following recursion equations as illustrated in Figs. \ref{f3fig}-\ref{h2fig}.   

\beq
\left\{\begin{array}{lll}
f_3(n+1) & = & 6f_1(n) [ f_2(n)g_2(n) + f_1(n)g_2(n) + f_1(n)g_4(n) ] \ , \\
g_4(n+1) & = & f_1(n) [ g_1(n)^2 + g_2(n)^2 ] + 4f_1(n)g_1(n) [ g_2(n) + g_3(n) + g_4(n) ] \\
& & + f_1(n)g_3(n) [ 2g_2(n) + 3g_3(n) ] + f_1(n)^2 [ t_1(n) + t_2(n) ] \\
& & + f_2(n)g_1(n) [ g_1(n) + 2g_2(n) + 2g_3(n) ] + f_1(n)f_2(n)t_1(n) \ , \\
t_2(n+1) & = & 2g_1^2(n) [ g_1(n) + 2g_2(n) + 4g_3(n) + 2g_4(n) ] \\
& & + 2g_1(n)g_3(n) [ 2g_2(n) + 3g_3(n) ] + 2g_1(n) [ g_2(n)^2 + f_2(n)t_1(n) ] \\
& & + 4f_1(n)t_1(n) [ 2g_1(n) + g_2(n) + g_3(n) ] + 4f_1(n)g_1(n)t_2(n) \ .
\end{array}
\right.
\label{f3g4t2}
\eeq
If we set $n=0$ in Eq. (\ref{f3g4t2}), only the terms contain $f_1(0)$ and $g_1(0)$ as factors on the right-hand-sides give non-zero contribution. In addition, there are special drawings shown in Fig. \ref{h2fig}, such that the initial values are $f_3(1)=0$, $g_4(1)=1$ and $t_2(1)=6$.    


\vspace*{5mm}

\vspace*{5mm}

Using $g_1(n)=3f_1(n)/2$ for $n \geq 1$ from Eq. (\ref{g1}) to eliminate $g_1(n)$, Eq. (\ref{f3g4t2}) becomes
\beq
\left\{\begin{array}{lll}
f_3(n+1) & = & 6f_1(n) [ f_2(n)g_2(n) + f_1(n)g_2(n) + f_1(n)g_4(n) ] \ , \\
g_4(n+1) & = & \frac 94 f_1(n)^3 + f_1(n)g_2(n)^2 + 6f_1(n)^2 [ g_2(n) + g_3(n) + g_4(n) ] \\
& & + f_1(n)g_3(n) [ 2g_2(n) + 3g_3(n) ] + f_1(n)^2 [ t_1(n) + t_2(n) ] \\
& & + \frac32 f_2(n)f_1(n) [ \frac 32 f_1(n) + 2g_2(n) + 2g_3(n) ] + f_1(n)f_2(n)t_1(n) \ , \\
t_2(n+1) & = & \frac 92 f_1^2(n) [ \frac 32 f_1(n) + 2g_2(n) + 4g_3(n) + 2g_4(n) ] \\
& & + 3f_1(n)g_3(n) [ 2g_2(n) + 3g_3(n) ] + 3f_1(n) [ g_2(n)^2 + f_2(n)t_1(n) ] \\
& & + 4f_1(n)t_1(n) [ 3f_1(n) + g_2(n) + g_3(n) ] + 6f_1(n)^2t_2(n) \ . \\
\end{array}
\right.
\label{fgh3}
\eeq
Define the ratios  
\beq
\alpha_3(n) = \frac{f_3(n)}{f_1(n)} \ , \quad 
\beta_4(n) = \frac{g_4(n)}{f_1(n)} \ , \quad 
\gamma_2(n) = \frac{t_2(n)}{f_1(n)} \ ,
\eeq
then they satisfy the recursion relations
\beq
\left\{\begin{array}{lll}
\alpha_3(n+1) & = & 2 [ \alpha_2(n)\beta_2(n) + \beta_2(n) + \beta_4(n) ] \ , \\
\beta_4(n+1) & = & \frac 34 + \beta_2(n) [ 2 + \frac{\beta_2(n)}3 ] + \beta_3(n) [ 2 + \frac23 \beta_2(n) + \beta_3(n) ] \\
& & + \alpha_2(n) [ \frac34 + \beta_2(n) + \beta_3(n) + \frac{\gamma_1(n)}{3} ] + 2\beta_4(n) + \frac13 [ \gamma_1(n) + \gamma_2(n) ] \ , \\
\gamma_2(n+1) & = & \frac94 + \beta_2(n) [ 3 + \beta_2(n) ] + \beta_3(n) [ 6  + 2\beta_2(n) + 3\beta_3(n) ] \\
& & + \gamma_1(n) [ 4 + \alpha_2(n) + \frac43 \beta_2(n) + \frac43 \beta_3(n) ] + 3\beta_4(n) + 2\gamma_2(n) \ .
\end{array}
\right.
\label{ab3}
\eeq
Define vectors $Y(n) = (\alpha_3(n), \beta_4(n), \gamma_2(n))^t$ and $K(n) = (K_1(n), K_2(n), K_3(n))^t$ where $K_1(n) = \alpha_3(n+1) - 2\beta_4(n)$, $K_2(n) = \beta_4(n+1) - 2\beta_4(n) - \gamma_2(n)/3$, $K_3(n) = \gamma_2(n+1) - 3\beta_4(n) - 2\gamma_2(n)$, and the matrix
\beq
B = \left(\begin{array}{ccc}
0 & 2 & 0       \\
0 & 2 & \frac13 \\
0 & 3 & 2
\end{array}
\right) \equiv P D P^{-1} \ ,
\eeq
where
\beq
D = \left(\begin{array}{ccc}
0 & 0 & 0 \\
0 & 1 & 0 \\
0 & 0 & 3
\end{array}
\right) \ , \quad 
P = \left(\begin{array}{ccc}
1 & 2  & 2 \\
0 & 1  & 3 \\
0 & -3 & 9
\end{array}
\right) \ , \quad 
P^{-1} = \left(\begin{array}{ccc}
1 & \frac{-4}3 & \frac29      \\
0 & \frac12    & \frac{-1}{6} \\
0 & \frac16    & \frac1{18}
\end{array}
\right) \ ,
\eeq
such that $Y(n+1) = K(n) + BY(n)$ for $n \geq 1$. We know $Y(1)=(0, 1/2, 3)^t$, and for $n \geq 2$,
\beqs
\lefteqn{Y(n)} \cr & = & \sum_{j=1}^{n-1} B^{n-1-j} K(j) + B^{n-1}Y(1) \cr\cr
& = & P \sum_{j=1}^{n-2} D^{n-1-j} P^{-1} K(j) + K(n-1) + P D^{n-1} P^{-1} Y(1) \cr\cr
& = & \left(\begin{array}{ccc}
1 & 2  & 2 \\
0 & 1  & 3 \\
0 & -3 & 9
\end{array}
\right) \Bigl\{ \sum_{j=1}^{n-2} \left( \begin{array}{cccc}
& 0                                                  \\
& \frac16 \bigl( 3K_2(j)-K_3(j) \bigr)               \\
& \frac{3^{n-3-j}}{2} \bigl( 3K_2(j) + K_3(j) \bigr) 
\end{array}
\right) + \left(\begin{array}{cccc}
& 0               \\
& -\frac14        \\
& \frac{3^{n-1}}4 
\end{array}
\right ) \Bigr\} + \left(\begin{array}{cccc}
& K_1(n-1) \\
& K_2(n-1) \\
& K_3(n-1)
\end{array}
\right ) \ . \cr & &  
\eeqs
From the expressions of $\alpha_2(n)$, $\beta_2(n)$, $\beta_3(n)$, $\gamma_1(n)$ given in (\ref{Xn}), $K_1(n)$, $K_2(n)$, $K_3(n)$ for $n \geq 1$ can be solved as
\beqs
K_1(n) & = & (\frac{7^2 \times 17^2}{2^8 \times 3^5}) (16)^n + (\frac{7 \times 17 \times 43}{2^5 \times 3^5}) 4^n - \frac{5^2 \times 7}{2 \times 3^5} + (\frac{37}{2^4 \times 3^3})\delta_{n1} \ , \cr\cr
K_2(n) & = & (\frac{7^2 \times 17^2}{2^7 \times 3^4}) (16)^n - (\frac{ 5 \times 7 \times 17}{2^6 \times 3^3}) 4^n - \frac{283}{2^3 \times 3^4} + (\frac{19}{2^4 \times 3^3}) \delta_{n1} \ , \cr\cr 
K_3(n) & = & (\frac{7^2 \times 17^2}{2^7 \times 3^3}) (16)^n - (\frac{7^2 \times 17}{2^6 \times 3^2}) 4^n - \frac{283}{2^3 \times 3^3} - (\frac{19}{2^4 \times 3^2}) \delta_{n1} \ .
\label{R23}
\eeqs
Hence, we have 
\beqs
\alpha_3(n) & = & \sum_{j=1}^{n-2} \bigl[ \bigl(1+3^{n-2-j} \bigr) K_2(j) + \frac13 \bigl( 3^{n-2-j}-1 \bigr) K_3(j) \bigr] + \frac{3^{n-1}-1}{2} + K_1(n-1) \cr\cr
& = & (\frac{5^2 \times 7^2\times 17^2}{2^{12} \times 3^5 \times 13}) (16)^n - (\frac{7 \times 17}{2^4 \times 3^5})4^n + (\frac{11 \times 257}{3^5 \times 2^3 \times 13}) 3^n - \frac{1009}{2^3 \times 3^5} - (\frac{1}{2^4\times 3^3}) \delta_{n2} \cr & &
\eeqs
for $n \geq 2$. Since $H(n) = 3f_1(n) + 3f_2(n) + f_3(n) = f_1(n) \bigl( 3 + 3\alpha_2(n) + \alpha_3(n) \bigr)$, we have the following theorem:

\bigskip
 
\begin{theo}
\label{hp}
The number of Hamiltonian paths on the Sierpinski gasket $SG(n)$ is twelve for $n=1$ and 
\beqs
H(n) & = & \frac{\sqrt{3}(2\sqrt{3})^{3^{n-1}}}{3} \Bigl\{ (\frac{5^2 \times 7^2 \times 17^2}{2^{12} \times 3^5 \times 13}) (16)^n + (\frac{7 \times 13 \times 17}{2^3 \times 3^5}) 4^n + (\frac{11 \times 257}{3^5 \times 2^3 \times 13}) 3^n \cr\cr
& & + \frac{4391}{2^3 \times 3^5} - (\frac{1}{2^4 \times 3^3}) \delta_{n2} \Bigr\}
\eeqs
for $n \geq 2$.
\end{theo}

To consider the asymptotic behavior when $n$ is large, let us use the symbol $A(n)\sim B(n)$ to denote $\lim_{n \rightarrow \infty} A(n)/B(n)=1$.

\begin{cor}
The asymptotic behavior for the number of Hamiltonian paths on the Sierpinski gasket $SG(n)$ when $n$ is large is given by
\beq
H(n) \sim \frac{\sqrt{3}(2\sqrt{3})^{3^{n-1}}}{3} (\frac {5^2 \times 7^2 \times 17^2}{2^{12} \times 3^5 \times 13})(16)^n \ . 
\eeq
The ratio of the number of Hamiltonian paths on the Sierpinski gasket $SG(n)$ and that with one end at origin when $n$ is large is given by
\beq
\frac {H(n)}{H_0(n)} \sim (\frac{ 5^2 \times 7 \times 17}{2^8 \times 3^2 \times 13}) 4^n \ .
\eeq
\end{cor}

Analogous to the consideration of self-avoiding paths in \cite{hattori}, consider $Z_n(\beta) = \sum_{w \in H(n)} e^{-L(w) \beta}$ with $\beta \in (0,\infty)$ and the length of $w$, denoted as $L(w)$, is equal to $\frac{3^{n+1}+1}2$ for every $w \in H(n)$. By Theorem \ref{hp}, we have the following corollary. 

\begin{cor}
\label{Zn}
There exists a critical $\beta^c = \ln \omega_{SG} = \frac{\ln (12)}{9}$ such that
\beq
\lim_{n \rightarrow \infty} Z_n(\beta) = \begin{cases} 
0      & \mbox{if} \quad \beta > \beta^c \ , \\
\infty & \mbox{if} \quad \beta \leq \beta^c \ .
\end{cases}
\eeq
Moreover, let $\beta_n = \beta^c + (\ln A_n)/3^n$ for some sequence $A_n > 0$. There is a critical $A^c_n = 2^{8n/3}$ such that
\beq
\lim_{n \rightarrow \infty} Z_n(\beta_n) = \begin{cases} 
0      & \mbox{if} \quad \lim_{n\rightarrow\infty}\frac {A_n^c}{A_n} = 0 \ , \\ 
\infty & \mbox{if} \quad \lim_{n\rightarrow\infty}\frac {A_n^c}{A_n} = \infty \ , \\
\frac{\sqrt{3}}{3} (\frac {5^2 \times 7^2 \times 17^2}{2^{12} \times 3^5 \times 13}) & \mbox{if} \quad A_n \sim A_n^c \ .
\end{cases}
\eeq
\end{cor}

\bigskip


\section{Definition for the distribution of the Hamiltonian paths on $SG(n)$ with one end at origin}
\label{distribution}

Consider the Hamiltonian paths on $SG(n)$ with one end at origin. We will study the distribution of the location of the other end. For that purpose, let us define the notation for the vertices of $SG(n)$, that is given progressively with increasing number of digits in the subscript as follows. First of all, fix the origin $o$ as the leftmost vertex, and all the Hamiltonian paths considered in the following sections have one end at $o$. Consider $SG(m)$ with $0 \le m < n$ always has $o$ as its leftmost vertex, and denote $a_m$ and $b_m$ as its rightmost and topmost vertices, respectively. $c_m$ is defined such that the vertices $a_m$, $b_m$ and $c_m$ demarcate the largest lacunary triangle of $SG(m+1)$. We then define the vertex in the middle of the line connecting $a_m$ and $a_{m+1}$ with $m \ge 1$ as $a_{m,1}$. Similarly, $a_{m,1}$ and the associated $b_{m,1}$ and $c_{m,1}$ demarcate a lacunary triangle with $b_{m,1}$ on the left and $c_{m,1}$ on the right. Next for $m \ge 2$, we append the subscript $(m,1,0)$ for the vertices of the largest lacunary inside the triangle with outmost vertices $a_m$, $a_{m,1}$, $b_{m,1}$; the subscript $(m,1,1)$ for the vertices of the largest lacunary inside the triangle with outmost vertices $a_{m,1}$, $a_{m+1}$, $c_{m,1}$; the subscript $(m,1,2)$ for the vertices of the largest lacunary inside the triangle with outmost vertices $b_{m,1}$, $c_{m,1}$, $c_m$, etc. In general for the vertices of $SG(n)$, we use the notation $x_{\vec{\gamma}}$ where $x \in \{a,b,c\}$ and the subscript $\vec{\gamma} = (\gamma_1,\gamma_2,...,\gamma_s)$ has $|\vec{\gamma}| = s$ components with $1 \leq s \leq n$, $1 \leq \gamma_1 < n$ and $\gamma_k \in \{0,1,2\}$ for $k \in \{2,3,...,s\}$.  For the vertices above the line connecting $o$ and $c_{n-1}$ on $SG(n)$, we will also use the notation $\tilde{x}_{\vec{\gamma}}$ such that it is the reflection of the vertex $x_{\vec{\gamma}}$ with respect to this line. Similarly, for the vertices above the line connecting $a_n$ and $b_{n-1}$ on $SG(n)$, we use the notation $\hat{x}_{\vec{\gamma}}$ such that it is the reflection of the vertex $x_{\vec{\gamma}}$ with respect to this line. For examples, $a_{22}=\tilde{b}_{21}$, $b_{221}=\tilde{a}_{212}$, $c_{222}=\tilde{c}_{211}$, and $b_{21}=\hat{b}_{21}$, $c_{210}=\hat{a}_{212}$, $a_{211}=\hat{c}_{211}$ on $SG(3)$. There may be more than one ways to denote a vertex. For example, $b_0$ can be written as $b_{10}$ on $SG(2)$, or $b_{200}$ on $SG(3)$, etc. The advantage of such vertex notation is that the quantities to be studied for the vertices $x_{\gamma_1,...,\gamma_s}$ with $s \ge 2$ components in the subscript on $SG(n)$ can be expressed in terms of the quantities for the vertices with $s-1$ components in the subscript on $SG(n-1)$ as discussed below.

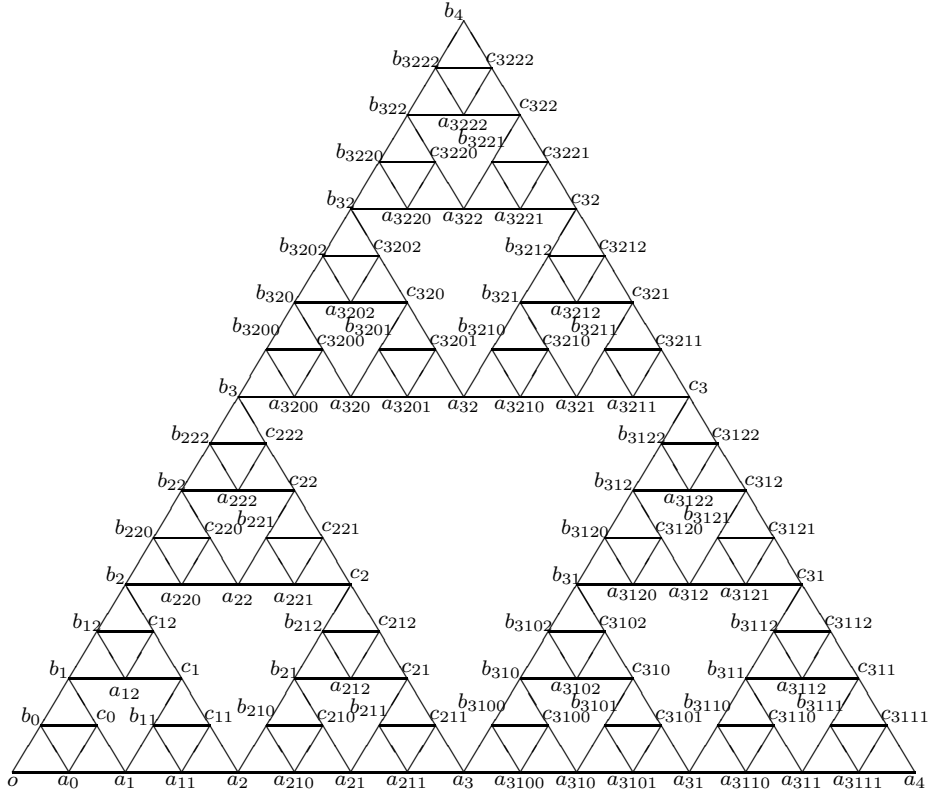
\begin{figure}[htbp]
\centering
\unitlength 1.25mm 
\begin{picture}(96,80)
\put(0,0){\line(1,0){96}}
\multiput(12,20)(48,0){2}{\line(1,0){24}}
\put(24,40){\line(1,0){48}}
\put(36,60){\line(1,0){24}}
\put(0,0){\line(3,5){48}}
\put(48,0){\line(3,5){24}}
\multiput(24,0)(48,0){2}{\line(3,5){12}}
\multiput(24,0)(48,0){2}{\line(-3,5){12}}
\put(48,0){\line(-3,5){24}}
\put(96,0){\line(-3,5){48}}
\put(48,40){\line(3,5){12}}
\put(48,40){\line(-3,5){12}}
\multiput(6,10)(24,0){4}{\line(1,0){12}}
\multiput(18,30)(48,0){2}{\line(1,0){12}}
\multiput(30,50)(24,0){2}{\line(1,0){12}}
\put(42,70){\line(1,0){12}}
\multiput(12,0)(24,0){4}{\line(3,5){6}}
\multiput(24,20)(48,0){2}{\line(3,5){6}}
\multiput(36,40)(24,0){2}{\line(3,5){6}}
\put(48,60){\line(3,5){6}}
\multiput(12,0)(24,0){4}{\line(-3,5){6}}
\multiput(24,20)(48,0){2}{\line(-3,5){6}}
\multiput(36,40)(24,0){2}{\line(-3,5){6}}
\put(48,60){\line(-3,5){6}}
\multiput(3,5)(12,0){8}{\line(1,0){6}}
\multiput(9,15)(24,0){4}{\line(1,0){6}}
\multiput(15,25)(12,0){2}{\line(1,0){6}}
\multiput(63,25)(12,0){2}{\line(1,0){6}}
\multiput(21,35)(48,0){2}{\line(1,0){6}}
\multiput(27,45)(12,0){4}{\line(1,0){6}}
\multiput(33,55)(24,0){2}{\line(1,0){6}}
\multiput(39,65)(12,0){2}{\line(1,0){6}}
\put(45,75){\line(1,0){6}}
\multiput(6,0)(12,0){8}{\line(3,5){3}}
\multiput(12,10)(24,0){4}{\line(3,5){3}}
\multiput(18,20)(12,0){2}{\line(3,5){3}}
\multiput(66,20)(12,0){2}{\line(3,5){3}}
\multiput(24,30)(48,0){2}{\line(3,5){3}}
\multiput(30,40)(12,0){4}{\line(3,5){3}}
\multiput(36,50)(24,0){2}{\line(3,5){3}}
\multiput(42,60)(12,0){2}{\line(3,5){3}}
\put(48,70){\line(3,5){3}}
\multiput(6,0)(12,0){8}{\line(-3,5){3}}
\multiput(12,10)(24,0){4}{\line(-3,5){3}}
\multiput(18,20)(12,0){2}{\line(-3,5){3}}
\multiput(66,20)(12,0){2}{\line(-3,5){3}}
\multiput(24,30)(48,0){2}{\line(-3,5){3}}
\multiput(30,40)(12,0){4}{\line(-3,5){3}}
\multiput(36,50)(24,0){2}{\line(-3,5){3}}
\multiput(42,60)(12,0){2}{\line(-3,5){3}}
\put(48,70){\line(-3,5){3}}
\put(0,-1){\makebox(0,0){$\scriptstyle o$}}
\put(6,-1){\makebox(0,0){$\scriptstyle a_0$}}
\put(12,-1){\makebox(0,0){$\scriptstyle a_1$}}
\put(18,-1){\makebox(0,0){$\scriptstyle a_{11}$}}
\put(24,-1){\makebox(0,0){$\scriptstyle a_2$}}
\put(30,-1){\makebox(0,0){$\scriptstyle a_{210}$}}
\put(36,-1){\makebox(0,0){$\scriptstyle a_{21}$}}
\put(42,-1){\makebox(0,0){$\scriptstyle a_{211}$}}
\put(48,-1){\makebox(0,0){$\scriptstyle a_3$}}
\put(54,-1){\makebox(0,0){$\scriptstyle a_{3100}$}}
\put(60,-1){\makebox(0,0){$\scriptstyle a_{310}$}}
\put(66,-1){\makebox(0,0){$\scriptstyle a_{3101}$}}
\put(72,-1){\makebox(0,0){$\scriptstyle a_{31}$}}
\put(78,-1){\makebox(0,0){$\scriptstyle a_{3110}$}}
\put(84,-1){\makebox(0,0){$\scriptstyle a_{311}$}}
\put(90,-1){\makebox(0,0){$\scriptstyle a_{3111}$}}
\put(96,-1){\makebox(0,0){$\scriptstyle a_4$}}
\put(2,6){\makebox(0,0){$\scriptstyle b_0$}}
\put(10,6){\makebox(0,0){$\scriptstyle c_0$}}
\put(14,6){\makebox(0,0){$\scriptstyle b_{11}$}}
\put(22,6){\makebox(0,0){$\scriptstyle c_{11}$}}
\put(26,7){\makebox(0,0){$\scriptstyle b_{210}$}}
\put(34.5,6){\makebox(0,0){$\scriptstyle c_{210}$}}
\put(38,7){\makebox(0,0){$\scriptstyle b_{211}$}}
\put(46.5,6){\makebox(0,0){$\scriptstyle c_{211}$}}
\put(50,7.5){\makebox(0,0){$\scriptstyle b_{3100}$}}
\put(59,6){\makebox(0,0){$\scriptstyle c_{3100}$}}
\put(62,7.5){\makebox(0,0){$\scriptstyle b_{3101}$}}
\put(71,6){\makebox(0,0){$\scriptstyle c_{3101}$}}
\put(74,7.5){\makebox(0,0){$\scriptstyle b_{3110}$}}
\put(83,6){\makebox(0,0){$\scriptstyle c_{3110}$}}
\put(86,7.5){\makebox(0,0){$\scriptstyle b_{3111}$}}
\put(95,6){\makebox(0,0){$\scriptstyle c_{3111}$}}
\put(5,11){\makebox(0,0){$\scriptstyle b_1$}}
\put(12,8.5){\makebox(0,0){$\scriptstyle a_{12}$}}
\put(19,11){\makebox(0,0){$\scriptstyle c_1$}}
\put(29,11){\makebox(0,0){$\scriptstyle b_{21}$}}
\put(36,9){\makebox(0,0){$\scriptstyle a_{212}$}}
\put(43,11){\makebox(0,0){$\scriptstyle c_{21}$}}
\put(52,11){\makebox(0,0){$\scriptstyle b_{310}$}}
\put(60,9){\makebox(0,0){$\scriptstyle a_{3102}$}}
\put(68,11){\makebox(0,0){$\scriptstyle c_{310}$}}
\put(76,11){\makebox(0,0){$\scriptstyle b_{311}$}}
\put(84,9){\makebox(0,0){$\scriptstyle a_{3112}$}}
\put(92,11){\makebox(0,0){$\scriptstyle c_{311}$}}
\put(8,16){\makebox(0,0){$\scriptstyle b_{12}$}}
\put(16,16){\makebox(0,0){$\scriptstyle c_{12}$}}
\put(31,16){\makebox(0,0){$\scriptstyle b_{212}$}}
\put(41,16){\makebox(0,0){$\scriptstyle c_{212}$}}
\put(55,16){\makebox(0,0){$\scriptstyle b_{3102}$}}
\put(65,16){\makebox(0,0){$\scriptstyle c_{3102}$}}
\put(79,16){\makebox(0,0){$\scriptstyle b_{3112}$}}
\put(89,16){\makebox(0,0){$\scriptstyle c_{3112}$}}
\put(11,21){\makebox(0,0){$\scriptstyle b_2$}}
\put(18,18.5){\makebox(0,0){$\scriptstyle a_{220}$}}
\put(24,18.5){\makebox(0,0){$\scriptstyle a_{22}$}}
\put(30,18.5){\makebox(0,0){$\scriptstyle a_{221}$}}
\put(37,21){\makebox(0,0){$\scriptstyle c_2$}}
\put(59,21){\makebox(0,0){$\scriptstyle b_{31}$}}
\put(66,19){\makebox(0,0){$\scriptstyle a_{3120}$}}
\put(72,19){\makebox(0,0){$\scriptstyle a_{312}$}}
\put(78,19){\makebox(0,0){$\scriptstyle a_{3121}$}}
\put(85,21){\makebox(0,0){$\scriptstyle c_{31}$}}
\put(13,26){\makebox(0,0){$\scriptstyle b_{220}$}}
\put(22.5,26){\makebox(0,0){$\scriptstyle c_{220}$}}
\put(26,27){\makebox(0,0){$\scriptstyle b_{221}$}}
\put(35,26){\makebox(0,0){$\scriptstyle c_{221}$}}
\put(61,26){\makebox(0,0){$\scriptstyle b_{3120}$}}
\put(71,26){\makebox(0,0){$\scriptstyle c_{3120}$}}
\put(74,27.5){\makebox(0,0){$\scriptstyle b_{3121}$}}
\put(83,26){\makebox(0,0){$\scriptstyle c_{3121}$}}
\put(17,31){\makebox(0,0){$\scriptstyle b_{22}$}}
\put(24,29){\makebox(0,0){$\scriptstyle a_{222}$}}
\put(31,31){\makebox(0,0){$\scriptstyle c_{22}$}}
\put(64,31){\makebox(0,0){$\scriptstyle b_{312}$}}
\put(72,29){\makebox(0,0){$\scriptstyle a_{3122}$}}
\put(80,31){\makebox(0,0){$\scriptstyle c_{312}$}}
\put(19,36){\makebox(0,0){$\scriptstyle b_{222}$}}
\put(29,36){\makebox(0,0){$\scriptstyle c_{222}$}}
\put(67,36){\makebox(0,0){$\scriptstyle b_{3122}$}}
\put(77,36){\makebox(0,0){$\scriptstyle c_{3122}$}}
\put(23,41){\makebox(0,0){$\scriptstyle b_3$}}
\put(30,39){\makebox(0,0){$\scriptstyle a_{3200}$}}
\put(36,39){\makebox(0,0){$\scriptstyle a_{320}$}}
\put(42,39){\makebox(0,0){$\scriptstyle a_{3201}$}}
\put(48,39){\makebox(0,0){$\scriptstyle a_{32}$}}
\put(54,39){\makebox(0,0){$\scriptstyle a_{3210}$}}
\put(60,39){\makebox(0,0){$\scriptstyle a_{321}$}}
\put(66,39){\makebox(0,0){$\scriptstyle a_{3211}$}}
\put(73,41){\makebox(0,0){$\scriptstyle c_3$}}
\put(26,47.5){\makebox(0,0){$\scriptstyle b_{3200}$}}
\put(35,46){\makebox(0,0){$\scriptstyle c_{3200}$}}
\put(38,47.5){\makebox(0,0){$\scriptstyle b_{3201}$}}
\put(47,46){\makebox(0,0){$\scriptstyle c_{3201}$}}
\put(50,47.5){\makebox(0,0){$\scriptstyle b_{3210}$}}
\put(59,46){\makebox(0,0){$\scriptstyle c_{3210}$}}
\put(62,47.5){\makebox(0,0){$\scriptstyle b_{3211}$}}
\put(71,46){\makebox(0,0){$\scriptstyle c_{3211}$}}
\put(28,51){\makebox(0,0){$\scriptstyle b_{320}$}}
\put(36,49){\makebox(0,0){$\scriptstyle a_{3202}$}}
\put(44,51){\makebox(0,0){$\scriptstyle c_{320}$}}
\put(52,51){\makebox(0,0){$\scriptstyle b_{321}$}}
\put(60,49){\makebox(0,0){$\scriptstyle a_{3212}$}}
\put(68,51){\makebox(0,0){$\scriptstyle c_{321}$}}
\put(31,56){\makebox(0,0){$\scriptstyle b_{3202}$}}
\put(41,56){\makebox(0,0){$\scriptstyle c_{3202}$}}
\put(55,56){\makebox(0,0){$\scriptstyle b_{3212}$}}
\put(65,56){\makebox(0,0){$\scriptstyle c_{3212}$}}
\put(35,61){\makebox(0,0){$\scriptstyle b_{32}$}}
\put(42,59){\makebox(0,0){$\scriptstyle a_{3220}$}}
\put(48,59){\makebox(0,0){$\scriptstyle a_{322}$}}
\put(54,59){\makebox(0,0){$\scriptstyle a_{3221}$}}
\put(61,61){\makebox(0,0){$\scriptstyle c_{32}$}}
\put(37,66){\makebox(0,0){$\scriptstyle b_{3220}$}}
\put(47,66){\makebox(0,0){$\scriptstyle c_{3220}$}}
\put(50,67.5){\makebox(0,0){$\scriptstyle b_{3221}$}}
\put(59,66){\makebox(0,0){$\scriptstyle c_{3221}$}}
\put(40,71){\makebox(0,0){$\scriptstyle b_{322}$}}
\put(48,69){\makebox(0,0){$\scriptstyle a_{3222}$}}
\put(56,71){\makebox(0,0){$\scriptstyle c_{322}$}}
\put(43,76){\makebox(0,0){$\scriptstyle b_{3222}$}}
\put(53,76){\makebox(0,0){$\scriptstyle c_{3222}$}}
\put(47,81){\makebox(0,0){$\scriptstyle b_4$}}
\end{picture}
\vspace*{3mm}
\caption{\footnotesize{The notation for the vertices of the Sierpinski gasket $SG(4)$.}} 
\label{sgvertex}
\end{figure}

Denote the number of Hamiltonian paths with one end at $o$ and the other end at vertex $x$ on $SG(n)$ as $f(n,x)$, and the distribution ratio $F(n,x) = f(n,x)/f_1(n)$. Define the probability measure 
\beq
p(n,x) = \frac{f(n,x)}{H_0(n)} \ ,
\label{goal-p}
\eeq
where $H_0(n)$ is given in Eq. (\ref{H0s}). Denote the factor 
\beq
R(n) = (\frac {7 \times 17}{2^4 \times 3^3})4^n + \frac{2^2 \times 13}{3^3} - (\frac1{2^2 \times 3^2}) \delta_{n1}
\label{Rn}
\eeq
such that $H_0(n) = f_1(n)R(n)$, then
\beq
p(n,x) = \frac{F(n,x)}{R(n)} \ .
\label{pnx}
\eeq
It is clear that the summation of $p(n,x)$ over all the vertices of $SG(n)$ should give one. The purpose of the following sections is to analyze $F(n,x)$, whose expression is simpler than $p(n,x)$ without the factor $R(n)$. We need the following definitions: (i) $g_2(n,x)$ denotes the number of $g_2(n)$ such that the end points of the Hamiltonian path are located at $o$ and $x$, and $G_2(n,x) = g_2(n,x)/g_1(n)$; (ii) $g_3(n,x)$ denotes the number of $g_3(n)$ such that one Hamiltonian path has end points located at $o$ and $a_n$, while the other Hamiltonian path has end points located at $b_n$ and $x$. Furthermore, define $u(n,x) = g_3(n,x) + g_3(n,\tilde{x})$ and $U(n,x) = u(n,x)/g_1(n)$; (iii) $t_1(n,x)$ denotes the number of  $t_1(n)$ such that the end points of the Hamiltonian path are located at $b_n$ and $x$, and $T_1(n,x) = t_1(n,x)/g_1(n)$. 

\section{Distribution $F(n,x_m)$ with $x \in \{a,b,c\}$, $0 \le m \leq n$ and mean $\ell$ displacement for these vertices} 
\label{fnxm}

\begin{figure}[htbp]
\unitlength 1.2mm 
\begin{picture}(30,10)
\put(0,0){\line(1,0){6}}
\put(6,0){\line(3,5){3}}
\put(9,5){\line(3,-5){3}}
\put(3,5){\line(3,5){3}}
\multiput(0,0)(0.6,1){6}{\circle*{0.2}}
\multiput(3,5)(1,0){7}{\circle*{0.2}}
\multiput(3,5)(0.6,-1){6}{\circle*{0.2}}
\multiput(6,0)(1,0){7}{\circle*{0.2}}
\multiput(6,10)(0.6,-1){6}{\circle*{0.2}}
\multiput(1,0.5)(10,0){2}{\circle{1}}
\put(6,9){\circle{1}}
\put(5,0.5){\circle{1}}
\put(3,4){\makebox(0,0){$\times$}}
\put(4,5.5){\circle{1}}
\put(8,5.5){\makebox(0,0){$\times$}}
\put(9,4){\circle*{1}}
\put(7,0.5){\circle{1}}
\put(15,5){\makebox(0,0){$+$}}
\put(18,0){\line(1,0){12}}
\put(21,5){\line(1,0){6}}
\put(24,10){\line(3,-5){3}}
\multiput(18,0)(0.6,1){11}{\circle*{0.2}}
\multiput(24,0)(-0.6,1){6}{\circle*{0.2}}
\multiput(24,0)(0.6,1){6}{\circle*{0.2}}
\multiput(27,5)(0.6,-1){6}{\circle*{0.2}}
\multiput(19,0.5)(10,0){2}{\circle{1}}
\put(24,9){\circle{1}}
\put(23,0.5){\circle{1}}
\put(21,4){\makebox(0,0){$\times$}}
\put(22,5.5){\circle{1}}
\put(26,5.5){\circle*{1}}
\put(25,0.5){\circle{1}}
\put(27,4){\makebox(0,0){$\times$}}
\end{picture}

\caption{\footnotesize{Illustration for the expression of $g_3(n,b_{n-1})$.}} 
\label{g3bmfig}
\end{figure}
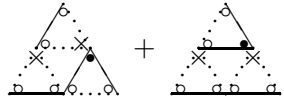

Let us consider the vertex $x_m$ on $SG(n)$ with $x \in \{a,b,c\}$ and $0 \le m \leq n$ in this section. It is easy to see that $f(n,a_m) = f(n,b_m)$ for all $m\leq n$ by symmetry, and $f(n,a_n) = f(n,b_n) = f_1(n)$ by definition. For $m=n-1 \ge 0$, we have
\beq
\left\{ 
\begin{array}{lll}
f(n,a_{n-1}) & = & f(n,b_{n-1}) = f_1(n-1)^2 g_1(n-1) = \frac 12 f_1(n) \ , \\
f(n,c_{n-1}) & = & 0 \ , \\ 
g_3(n,b_{n-1}) & = & g_3(n,c_{n-1}) = \begin{cases}
2f_1(n-1) g_1(n-1)^2 = g_1(n) & \mbox{for} \ n > 1 \ , \\
2 = \frac23 g_1(1)            & \mbox{for} \ n = 1 \ ,
\end{cases} \\
g_3(n,a_{n-1}) & = & 0 \ , 
\end{array}
\right.
\label{fg3-1}
\eeq
where the expression for $g_3(n,b_{n-1})$ is illustrated in Fig. \ref{g3bmfig}, and
\beq
\left\{ 
\begin{array}{lll}
g_2(n,a_{n-1}) & = & g_2(n,b_{n-1}) = \begin{cases}
f_1(n-1) g_1(n-1)^2 = \frac12 g_1(n) & \mbox{for} \ n > 1 \ , \\
1 = \frac{1}{3}g_1(1)                & \mbox{for} \ n = 1 \ ,
\end{cases} \\
g_2(n,c_{n-1}) & = & \begin{cases}
0                     & \mbox{for} \ n > 1 \ , \\
1 = \frac{1}{3}g_1(1) & \mbox{for} \ n = 1 \ .
\end{cases} \\ 
\end{array}
\right.
\label{g2}
\eeq

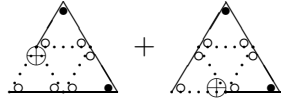
\begin{figure}[htbp]
\unitlength 1.2mm 
\begin{picture}(30,10)
\put(0,0){\line(1,0){12}}
\put(3,5){\line(3,5){3}}
\put(6,10){\line(3,-5){6}}
\multiput(0,0)(0.6,1){6}{\circle*{0.2}}
\multiput(3,5)(1,0){7}{\circle*{0.2}}
\multiput(6,0)(0.6,1){6}{\circle*{0.2}}
\multiput(6,0)(-0.6,1){6}{\circle*{0.2}}
\put(1,0.5){\circle{1}}
\put(11,0.5){\circle*{1}}
\put(6,9){\circle*{1}}
\multiput(5,0.5)(2,0){2}{\circle{1}}
\put(3,4){\makebox(0,0){$\oplus$}}
\multiput(4,5.5)(4,0){2}{\circle{1}}
\put(9,4){\circle{1}}
\put(15,5){\makebox(0,0){$+$}}
\put(18,0){\line(3,5){6}}
\put(24,0){\line(1,0){6}}
\put(24,10){\line(3,-5){6}}
\multiput(18,0)(1,0){7}{\circle*{0.2}}
\multiput(24,0)(-0.6,1){6}{\circle*{0.2}}
\multiput(24,0)(0.6,1){6}{\circle*{0.2}}
\multiput(21,5)(1,0){7}{\circle*{0.2}}
\put(19,0.5){\circle{1}}
\put(29,0.5){\circle*{1}}
\put(24,9){\circle*{1}}
\put(23,0.5){\makebox(0,0){$\oplus$}}
\put(21,4){\circle{1}}
\multiput(22,5.5)(4,0){2}{\circle{1}}
\put(25,0.5){\circle{1}}
\put(27,4){\circle{1}}
\end{picture}

\caption{\footnotesize{Illustration for the expression of $f(n,x_m)$ with $0 \le m \le n-2$.}} 
\label{fxmfig}
\end{figure}

\begin{figure}[htbp]
\unitlength 1.2mm 
\begin{picture}(30,10)
\put(0,0){\line(1,0){6}}
\put(3,5){\line(3,5){3}}
\put(6,0){\line(3,5){3}}
\put(9,5){\line(3,-5){3}}
\multiput(0,0)(0.6,1){6}{\circle*{0.2}}
\multiput(3,5)(1,0){7}{\circle*{0.2}}
\multiput(6,0)(-0.6,1){6}{\circle*{0.2}}
\multiput(9,5)(-0.6,1){6}{\circle*{0.2}}
\multiput(6,0)(1,0){7}{\circle*{0.2}}
\multiput(1,0.5)(10,0){2}{\circle{1}}
\put(6,9){\circle{1}}
\multiput(5,0.5)(2,0){2}{\circle{1}}
\put(3,4){\makebox(0,0){$\oplus$}}
\put(4,5.5){\circle{1}}
\put(8,5.5){\makebox(0,0){$\times$}}
\put(9,4){\circle*{1}}
\put(15,5){\makebox(0,0){$+$}}
\put(18,0){\line(1,0){12}}
\put(21,5){\line(1,0){6}}
\put(27,5){\line(-3,5){3}}
\multiput(18,0)(0.6,1){11}{\circle*{0.2}}
\multiput(24,0)(-0.6,1){6}{\circle*{0.2}}
\multiput(24,0)(0.6,1){6}{\circle*{0.2}}
\multiput(30,0)(-0.6,1){6}{\circle*{0.2}}
\multiput(19,0.5)(10,0){2}{\circle{1}}
\put(24,9){\circle{1}}
\multiput(23,0.5)(2,0){2}{\circle{1}}
\put(21,4){\makebox(0,0){$\oplus$}}
\put(22,5.5){\circle{1}}
\put(26,5.5){\circle*{1}}
\put(25,0.5){\circle{1}}
\put(27,4){\makebox(0,0){$\times$}}
\end{picture}

\caption{\footnotesize{Illustration for the expression of $g_3(n,x_m)$ with $0 \le m \le n-2$.}} 
\label{g3xmfig}
\end{figure}
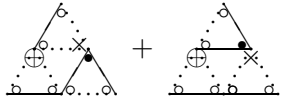

For $0 \le m \le n-2$, we have
\beq
\left\{ 
\begin{array}{lll}
f(n,x_m)
& = & f_1(n-1)^2u(n-1,x_m) \ , \\
g_3(n,x_m)
& = & 2f_1(n-1)g_{1}(n-1)g_3(n-1,x_m) 
\end{array}
\right.
\label{fg}
\eeq
for $x \in \{a,b,c\}$ as illustrated in Figs. \ref{fxmfig} and \ref{g3xmfig}, respectively.

\noindent
From Eqs. (\ref{fg1}) and (\ref{fg}), $g_3(n,x_m)$ for $0 \le m \le n-2$ can be written in general as   
\beq
g_3(n,x_m) = r(n)g_3(n-1,x_m) \ , \qquad \mbox{where} \quad 
r(n) = \frac{g_1(n)}{g_1(n-1)} \ ,
\label{rdef}
\eeq
such that $g_3(n,a_m)=0$ and
\beqs
g_3(n,b_m) = g_3(n,c_m) & = & g_3(m+1,b_m) \prod_{j=m+2}^{n} r(j) \cr\cr
& = & \begin{cases}
g_1(m+1) \prod_{j=m+2}^{n} r(j) = g_1(n)            & (0 < m < n-1) \ , \\
\frac23 g_1(1) \prod_{j=2}^{n}r(j) = \frac23 g_1(n) & (0 = m < n-1) \ ,
\end{cases}
\label{fa2}
\eeqs
where Eq. (\ref{fg3-1}) is used.
Combining Eqs. (\ref{fg3-1}) and (\ref{fa2}), we obtain
\beq
U(n,a_m) = U(n,b_m) = \begin{cases}
1       & (0 < m < n) \ , \\
\frac23 & (0 = m < n) \ ,
\end{cases}
\quad U(n,c_m) = \begin{cases}
2       & ( 0 < m < n) \ , \\
\frac43 & ( 0 = m < n) \ .
\end{cases}
\label{U1}
\eeq

Since $g_2(n,x_m) = f_1(n-1)g_1(n-1)u(n-1,x_m)$ for $0 \le m \le n-2$, we have
\beq
\label{g2'}
G_2(n,a_m) = G_2(n,b_{m}) = \begin{cases}
\frac12 & (0 < m < n) \ , \\
\frac13 & (0 = m < n) \ ,
\end{cases} \quad 
G_2(n,c_{m}) = \begin{cases}
1       & (0 < m < n-1) \ , \\
0       & (0 < m = n-1) \ , \\
\frac23 & (0 = m < n-1) \ , \\
\frac13 & (0 = m = n-1) \ ,
\end{cases}
\eeq
as Eqs. (\ref{g2}) and (\ref{fa2}) are combined. From Eq. (\ref{fg}), $f(n,x_m)$ for $m \le n-2$ is given as 
\beq
\left\{\begin{array}{lll}
f(n,a_m) & = & f(n,b_m) = f_1(n-1)^2g_3(n-1,b_m) = \begin{cases}
\frac12 f_1(n) & (0 < m < n-1) \ , \\
\frac13 f_1(n) & (0 = m < n-1) \ ,
\end{cases} \\
f(n,c_m) & = & 2f_1(n-1)^2g_3(n-1,c_m) = \begin{cases}
f_1(n)         & (0 < m < n-1) \ , \\
\frac23 f_1(n) & (0 = m < n-1) \ , \\
\end{cases}
\end{array}
\right.
\label{fa3}
\eeq
where Eqs. (\ref{fg3-1}), (\ref{fa2}) are used. Combining Eqs. (\ref{fg3-1}) and (\ref{fa3}), we have the following theorem:

\bigskip

\begin{theo}
On the Sierpinski gasket $SG(n)$, the distribution ratio $F(n,x_m)$ with $x \in \{a, b, c \}$ for integer $0 \le m \le n$ is given by
\beq
F(n,a_m) = F(n,b_m) = \begin{cases}
1       & (0 \leq m = n) \ , \\
\frac12 & (0 < m < n) \ , \\
\frac12 & (m = 0, \ n = 1) \ , \\
\frac13 & (m = 0, \ n > 1) \ ,
\end{cases} \quad
F(n,c_m) = \begin{cases}
0       & (0 \leq m = n-1) \ , \\
1       & (0 < m < n-1) \ , \\
\frac23 & (m = 0, \ n > 1) \ .
\end{cases}
\eeq
\label{Fxm}
\end{theo}

For any vertex $x \in v(SG(n))$, denote the distance between $o$ and $x$ as $|x|$, and set the the distance between the vertices $o$ and $a_0$ equal to one. For any positive $\ell > 0$, define the mean $\ell$ displacement for the vertices $x_m$ with $x \in \{a, b, c \}$ and $0 \le m \le n$ as
\beq
\xi^\prime(n,\ell) \equiv \frac {\sum_{x_j; j=0,1,2,...,n;x=a,b,c} |x_j|^\ell p(n,x_j)} {\sum_{x_j;j=0,1,2,...,n;x=a,b,c}p(n,x_j)}
= \frac {\sum_{x_j; j=0,1,2,...,n;x=a,b,c} |x_j|^\ell F(n,x_j)} {\sum_{x_j;j=0,1,2,...,n;x=a,b,c}F(n,x_j)} \ ,
\eeq
where Eq. (\ref{pnx}) is used and $p(n,c_n) = 0$ is assumed. From Theorem \ref{Fxm} for any integer $n>1$, we have
\beqs
\xi^\prime(n,\ell) & = & \frac{ 2 \times \bigl \{ \frac13 + \frac12 \times \sum_{j=1}^{n-1} 2^{j\ell} + 1 \times 2^{n\ell} \bigr \} + \frac23 \times \sqrt{3} + 1 \times \sum_{j=1}^{n-2} (2^j \times
\sqrt{3})^\ell } { 2 \times \bigl \{ \frac13 + \frac12 \times (n-1) + 1 \bigr \} + \frac23 + 1 \times (n-2)} \cr\cr
& = & \frac{ 2^{n\ell+1} + \frac23 (1 + \sqrt{3}) + \frac{2^{n\ell} \bigl( 1 + (\frac{\sqrt{3}}{2})^\ell \bigr ) - 2^\ell - (2\sqrt{3})^\ell} {2^\ell-1} }{2n + \frac13} \ .
\label{xi1}
\eeqs

\bigskip

\begin{theo}
When $n$ is large, we have the asymptotic expression
\beq
\xi^\prime(n,\ell) \sim \Bigl( 1 + \frac12 \bigl( \frac{ 1 + (\frac{\sqrt{3}}{2})^\ell} {2^\ell-1} \bigr) \Bigr) \times \frac{2^{n\ell}}{n} \ .
\eeq
\end{theo}

\begin{cor}
Consider only the vertices $x_j$ with $x \in \{a,b,c\}$ and $j = 1,2,...,n$ on $SG(n)$, and define the variable $p^\prime(n,x_j) = p(n,x_j) / \sum_{x_j; j=0,1,2,...,n;x=a,b,c} p(n,x_j)$. The mean distance and the corresponding variance of it when $n$ is large are given as

\beqs
\mu & = & \xi^\prime(n,1) \sim \frac{2^n(\frac32 + \frac{\sqrt{3}}{4})}{n} \ , \cr\cr
\sigma^2 & = & \xi^\prime(n,2) - \bigl( \xi^\prime(n,1) \bigr)^2 
\sim \frac{31}{3} \times \frac{2^{2n-3}}{n} \ .
\eeqs
\end{cor}

\section{Distribution $F(n,x_{m,2})$ with $x \in \{a,b,c\}$ and $1 \le m < n$}
\label{fnxm2}

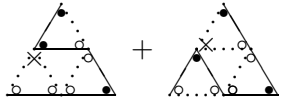
\begin{figure}[htbp]
\unitlength 1.2mm 
\begin{picture}(30,10)
\put(0,0){\line(1,0){12}}
\put(3,5){\line(1,0){6}}
\put(3,5){\line(3,5){3}}
\put(12,0){\line(-3,5){3}}
\multiput(0,0)(0.6,1){6}{\circle*{0.2}}
\multiput(6,0)(0.6,1){6}{\circle*{0.2}}
\multiput(6,0)(-0.6,1){6}{\circle*{0.2}}
\multiput(9,5)(-0.6,1){6}{\circle*{0.2}}
\put(1,0.5){\circle{1}}
\put(11,0.5){\circle*{1}}
\put(6,9){\circle*{1}}
\multiput(5,0.5)(2,0){2}{\circle{1}}
\put(3,4){\makebox(0,0){$\times$}}
\put(4,5.5){\circle*{1}}
\put(8,5.5){\circle{1}}
\put(9,4){\circle{1}}
\put(15,5){\makebox(0,0){$+$}}
\put(18,0){\line(3,5){3}}
\put(21,5){\line(3,-5){3}}
\put(24,0){\line(1,0){6}}
\put(30,0){\line(-3,5){6}}
\multiput(18,0)(1,0){7}{\circle*{0.2}}
\multiput(24,0)(0.6,1){6}{\circle*{0.2}}
\multiput(21,5)(0.6,1){6}{\circle*{0.2}}
\multiput(21,5)(1,0){7}{\circle*{0.2}}
\put(19,0.5){\circle{1}}
\put(29,0.5){\circle*{1}}
\put(24,9){\circle*{1}}
\multiput(23,0.5)(2,0){2}{\circle{1}}
\put(21,4){\circle*{1}}
\put(22,5.5){\makebox(0,0){$\times$}}
\put(26,5.5){\circle{1}}
\put(25,0.5){\circle{1}}
\put(27,4){\circle{1}}
\end{picture}

\caption{\footnotesize{Illustration for the expression of $f(n,x_{n-1,2})$.}} 
\label{fm2fig}
\end{figure}

Let us consider the vertex $x_{m,2}$ on $SG(n)$ with $x \in \{a,b,c\}$ and $1 \le m < n$ in this section. The expressions for $x_{m,1}$ can be obtained from those for $x_{m,2}$ by symmetry. For $m = n-1$, we have
\beq
\label{fxn2}
f(n,x_{n-1,2})
= f_1(n-1) g_1(n-1) f(n-1,\hat{\tilde{\hat{x}}}_{n-2}) + f_1(n-1)^2 g_2(n-1,\hat{\tilde{x}}_{n-2}) \ ,
\eeq
which is illustrated in Fig. \ref{fm2fig}, and $F(n,x_{n-1,2})$ can be obtained using Eqs. (\ref{fg3-1}) and (\ref{g2}) as given below in Theorem \ref{Fxm2}.

\begin{figure}[htbp]
\unitlength 1.2mm 
\begin{picture}(66,10)
\put(0,0){\line(3,5){3}}
\put(6,0){\line(-3,5){3}}
\put(6,0){\line(3,5){3}}
\put(12,0){\line(-3,5){3}}
\multiput(0,0)(1,0){13}{\circle*{0.2}}
\multiput(3,5)(1,0){7}{\circle*{0.2}}
\multiput(6,10)(0.6,-1){6}{\circle*{0.2}}
\multiput(6,10)(-0.6,-1){6}{\circle*{0.2}}
\multiput(1,0.5)(10,0){2}{\circle{1}}
\put(6,9){\makebox(0,0){$\oplus$}}
\multiput(5,0.5)(2,0){2}{\circle{1}}
\multiput(3,4)(6,0){2}{\circle*{1}}
\multiput(4,5.5)(4,0){2}{\makebox(0,0){$\times$}}
\put(15,5){\makebox(0,0){$+$}}
\put(18,0){\line(1,0){12}}
\put(21,5){\line(1,0){6}}
\put(27,5){\line(-3,5){3}}
\multiput(18,0)(0.6,1){11}{\circle*{0.2}}
\multiput(24,0)(0.6,1){6}{\circle*{0.2}}
\multiput(24,0)(-0.6,1){6}{\circle*{0.2}}
\multiput(30,0)(-0.6,1){6}{\circle*{0.2}}
\multiput(19,0.5)(10,0){2}{\circle{1}}
\put(24,9){\circle{1}}
\multiput(23,0.5)(2,0){2}{\circle{1}}
\multiput(21,4)(6,0){2}{\makebox(0,0){$\times$}}
\multiput(22,5.5)(4,0){2}{\circle*{1}}
\put(33,5){\makebox(0,0){$+$}}
\put(36,0){\line(1,0){6}}
\put(42,0){\line(3,5){3}}
\put(39,5){\line(3,5){3}}
\put(48,0){\line(-3,5){3}}
\multiput(36,0)(0.6,1){6}{\circle*{0.2}}
\multiput(39,5)(1,0){7}{\circle*{0.2}}
\multiput(42,0)(-0.6,1){6}{\circle*{0.2}}
\multiput(42,0)(1,0){7}{\circle*{0.2}}
\multiput(42,10)(0.6,-1){6}{\circle*{0.2}}
\multiput(37,0.5)(10,0){2}{\circle{1}}
\put(42,9){\circle{1}}
\multiput(41,0.5)(2,0){2}{\circle{1}}
\put(39,4){\makebox(0,0){$\times$}}
\put(45,4){\circle*{1}}
\put(40,5.5){\circle*{1}}
\put(44,5.5){\makebox(0,0){$\times$}}
\put(51,5){\makebox(0,0){$+$}}
\put(54,0){\line(3,5){3}}
\put(57,5){\line(3,-5){3}}
\put(60,10){\line(3,-5){3}}
\put(60,0){\line(1,0){6}}
\multiput(54,0)(1,0){7}{\circle*{0.2}}
\multiput(57,5)(1,0){7}{\circle*{0.2}}
\multiput(57,5)(0.6,1){6}{\circle*{0.2}}
\multiput(60,0)(0.6,1){6}{\circle*{0.2}}
\multiput(66,0)(-0.6,1){6}{\circle*{0.2}}
\multiput(55,0.5)(10,0){2}{\circle{1}}
\put(60,9){\circle{1}}
\multiput(59,0.5)(2,0){2}{\circle{1}}
\put(57,4){\circle*{1}}
\put(63,4){\makebox(0,0){$\times$}}
\put(58,5.5){\makebox(0,0){$\times$}}
\put(62,5.5){\circle*{1}}
\end{picture}

\vspace*{1cm}

\begin{picture}(66,10)
\put(0,0){\line(3,5){3}}
\put(3,5){\line(1,0){6}}
\put(6,0){\line(3,5){3}}
\put(6,0){\line(1,0){6}}
\multiput(0,0)(1,0){7}{\circle*{0.2}}
\multiput(3,5)(0.6,1){6}{\circle*{0.2}}
\multiput(6,10)(0.6,-1){11}{\circle*{0.2}}
\multiput(6,0)(-0.6,1){6}{\circle*{0.2}}
\multiput(1,0.5)(10,0){2}{\circle{1}}
\put(6,9){\makebox(0,0){$\oplus$}}
\put(5,0.5){\makebox(0,0){$\times$}}
\put(7,0.5){\circle*{1}}
\multiput(3,4)(6,0){2}{\circle{1}}
\multiput(4,5.5)(4,0){2}{\circle{1}}
\put(15,5){\makebox(0,0){$+$}}
\put(18,0){\line(1,0){6}}
\put(24,0){\line(-3,5){3}}
\put(21,5){\line(1,0){6}}
\put(27,5){\line(3,-5){3}}
\multiput(18,0)(0.6,1){11}{\circle*{0.2}}
\multiput(24,0)(0.6,1){6}{\circle*{0.2}}
\multiput(24,0)(1,0){7}{\circle*{0.2}}
\multiput(24,10)(0.6,-1){6}{\circle*{0.2}}
\multiput(19,0.5)(10,0){2}{\circle{1}}
\put(24,9){\makebox(0,0){$\oplus$}}
\put(23,0.5){\circle*{1}}
\put(25,0.5){\makebox(0,0){$\times$}}
\multiput(21,4)(6,0){2}{\circle{1}}
\multiput(22,5.5)(4,0){2}{\circle{1}}
\put(33,5){\makebox(0,0){$+$}}
\put(36,0){\line(3,5){3}}
\put(39,5){\line(3,-5){3}}
\put(42,10){\line(3,-5){3}}
\put(42,0){\line(1,0){6}}
\multiput(36,0)(1,0){7}{\circle*{0.2}}
\multiput(39,5)(1,0){7}{\circle*{0.2}}
\multiput(39,5)(0.6,1){6}{\circle*{0.2}}
\multiput(42,0)(0.6,1){6}{\circle*{0.2}}
\multiput(45,5)(0.6,-1){6}{\circle*{0.2}}
\multiput(37,0.5)(10,0){2}{\circle{1}}
\put(42,9){\circle{1}}
\multiput(41,0.5)(2,0){2}{\circle{1}}
\put(39,4){\circle*{1}}
\put(45,4){\makebox(0,0){$\oplus$}}
\put(40,5.5){\makebox(0,0){$\times$}}
\put(44,5.5){\circle{1}}
\put(51,5){\makebox(0,0){$+$}}
\put(54,0){\line(1,0){12}}
\put(57,5){\line(3,5){3}}
\put(57,5){\line(1,0){6}}
\multiput(54,0)(0.6,1){6}{\circle*{0.2}}
\multiput(57,5)(1,0){7}{\circle*{0.2}}
\multiput(60,0)(-0.6,1){6}{\circle*{0.2}}
\multiput(60,0)(0.6,1){6}{\circle*{0.2}}
\multiput(60,10)(0.6,-1){11}{\circle*{0.2}}
\multiput(55,0.5)(10,0){2}{\circle{1}}
\put(60,9){\circle{1}}
\multiput(59,0.5)(2,0){2}{\circle{1}}
\put(57,4){\makebox(0,0){$\times$}}
\put(63,4){\makebox(0,0){$\oplus$}}
\put(58,5.5){\circle*{1}}
\put(62,5.5){\circle{1}}
\end{picture}

\caption{\footnotesize{Illustration for the expression of $u(m+1,x_{m,2})$.}} 
\label{uxm2fig}
\end{figure}
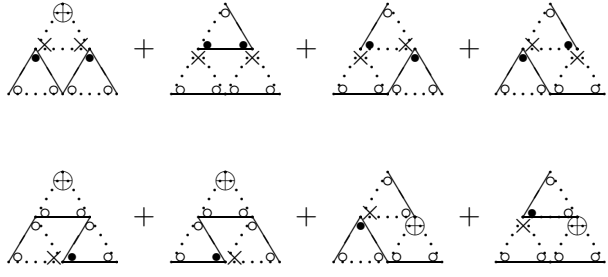

Next consider $m \le n-2$, we have
\beq
\label{f2}
f(n,x_{m,2}) = f_1(n-1)^2 u(n-1,x_{m,2}) \ ,
\eeq
similar to the first line of Eq. (\ref{fg}). It is then necessary to derive $u(n-1,x_{m,2})$ using Eq. (\ref{rdef}) so that
\beq
\label{xx}
u(n-1,x_{m,2}) = u(m+1,x_{m,2}) \prod_{j=m+2}^{n-1} r(j) \ ,
\eeq
where the product is set to one when $n=m+2$ and
%
\beqs
u(m+1,x_{m,2})
& = & f_1(m)^2 t_1(m,x_{m-1}) + g_1(m)^2 f(m,\hat{x}_{m-1}) \cr
& & + f_1(m) g_1(m) \bigl( g_2(m,\tilde{\hat{x}}_{m-1}) + g_2(m,\hat{x}_{m-1}) + 2u(m,x_{m-1}) \bigr) 
\label{u}
\eeqs
as illustrated in Fig. \ref{uxm2fig}.

\noindent
The term $t_1(m,x_{m-1})$ is given by 
\beqs
\label{hn}
t_1(m,b_{m-1}) & = & t_1(m,c_{m-1}) = \begin{cases}
g_1(m-1)^3 = \frac32 f_1(m-1)g_1(m-1)^2 = \frac34 g_1(m)     & (m > 1) \ , \\
1 = \frac13 g_1(1) & (m = 1) \ ,
\end{cases} \cr\cr 
t_1(m,a_{m-1}) & = & \begin{cases}
0                  & (m > 1) \ , \\
2 = \frac23 g_1(1) & (m = 1) \ .
\end{cases}
\eeqs
Substituting the results in Eqs. (\ref{fg3-1}), (\ref{g2}), (\ref{U1}) and (\ref{hn}) into Eq. (\ref{u}), we have
\beqs
u(m+1,a_{m,2}) & = & f_1(m)^2 t_1(m,a_{m-1}) + g_1(m)^2 f(m,c_{m-1}) \cr
& & + 2f_1(m) g_1(m) \bigl( g_2(m,c_{m-1}) + u(m,a_{m-1}) \bigr) \cr
& = & \begin{cases}
g_1(m+1)               & (m > 1) \ , \\
44 = \frac{11}9 g_1(2) & (m = 1) \ ,
\end{cases}
\eeqs
where $g_1(2) = 36$, and
\beqs
u(m+1,b_{m,2}) & = & f_1(m)^2 t_1(m,b_{m-1}) + g_1(m)^2 f(m,b_{m-1}) \cr
& & + f_1(m) g_1(m) \bigl( g_2(m,a_{m-1}) + g_2(m,b_{m-1}) + 2u(m,b_{m-1}) \bigr) \cr
& = & \begin{cases}
2g_1(m+1)                 & (m > 1) \ , \\
49 = \frac{49}{36} g_1(2) & (m = 1) \ ,
\end{cases}
\eeqs
\beqs
u(m+1,c_{m,2}) & = & f_1(m)^2 t_1(m,c_{m-1}) + g_1(m)^2 f(m,a_{m-1}) \cr
& & + f_1(m) g_1(m) \bigl( g_2(m,b_{m-1}) + g_2(m,a_{m-1}) + 2u(m,c_{m-1}) \bigr) \cr
& = & \begin{cases}
3 g_1(m+1)                & (m>1) \ , \\
73 = \frac{73}{36} g_1(2) & (m=1) \ . 
\end{cases}
\eeqs
It follows that Eq. (\ref{xx}) becomes
\beqs
u(n-1,a_{m,2}) & = & \begin{cases}
g_1(n-1)            & (m > 1) \ , \\
\frac{11}9 g_1(n-1) & (m = 1) \ ,
\end{cases} \quad
u(n-1,b_{m,2}) = \begin{cases}
2g_1(n-1)              & (m > 1) \ , \\
\frac{49}{36} g_1(n-1) & (m = 1) \ ,
\end{cases} \cr
u(n-1,c_{m,2}) & = & \begin{cases}
3 g_1(n-1)            & (m > 1) \ , \\
\frac{73}{36}g_1(n-1) & (m = 1) \ ,
\end{cases}
\eeqs
for $m \le n-2$.
Combining above results, we obtain the following theorem:

\bigskip

\begin{theo}
\label{Fxm2}
On the Sierpinski gasket $SG(n)$ with $n \ge 2$, the distribution ratio $F(n,x_{n-1,\gamma_2})$ with $x \in \{a, b, c \}$ and $\gamma_2 \in \{1,2\}$ is given by
\beqs
F(n,a_{n-1,2}) & = & F(n,c_{n-1,2}) = F(n,b_{n-1,1}) = F(n,c_{n-1,1}) = \begin{cases}
\frac12      & (n > 2) \ , \\
\frac{5}{12} & (n = 2) \ ,
\end{cases} \cr\cr
F(n,b_{n-1,2}) & = & F(n,a_{n-1,1}) = \begin{cases}
0       & (n > 2) \ , \\
\frac16 & (n = 2) \ ,
\end{cases}
\eeqs
and the distribution ratio $F(n,x_{m,\gamma_2})$ with integer $m \le n-2$, $x \in \{a, b, c \}$ and $\gamma_2 \in \{1,2\}$ is given by
\beqs
F(n,a_{m,2}) & = & F(n,b_{m,1}) = \begin{cases}
\frac12       & (m > 1) \ , \\
\frac{11}{18} & (m = 1) \ ,
\end{cases} \quad
F(n,b_{m,2}) = F(n,a_{m,1}) = \begin{cases}
1             & (m > 1) \ , \\
\frac{49}{72} & (m = 1) \ ,
\end{cases} \cr\cr
F(n,c_{m,2}) & = & F(n,c_{m,1}) = \begin{cases}
\frac32       & (m > 1) \ , \\
\frac{73}{72} & (m = 1) \ .
\end{cases}
\eeqs
\end{theo}

For $m = n-1$, we also have
\beq
\label{g2xn2}
g_2(n,x_{n-1,2})
= f_1(n-1) g_1(n-1) g_2(n-1,\hat{\tilde{\hat{x}}}_{n-2}) + f_1(n-1)^2 t_1(n-1,\tilde{\hat{x}}_{n-2}) \ ,
\eeq
which is illustrated in Fig. \ref{g2m2fig}, and $G_2(n,x_{n-1,2})$ can be obtained using Eqs. (\ref{g2}) and (\ref{hn}) as given below.

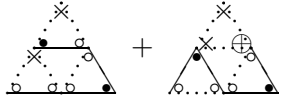
\begin{figure}[htbp]
\unitlength 1.2mm 
\begin{picture}(30,10)
\put(0,0){\line(1,0){12}}
\put(3,5){\line(1,0){6}}
\put(12,0){\line(-3,5){3}}
\multiput(0,0)(0.6,1){11}{\circle*{0.2}}
\multiput(6,0)(0.6,1){6}{\circle*{0.2}}
\multiput(6,0)(-0.6,1){6}{\circle*{0.2}}
\multiput(9,5)(-0.6,1){6}{\circle*{0.2}}
\put(1,0.5){\circle{1}}
\put(11,0.5){\circle*{1}}
\put(6,9){\makebox(0,0){$\times$}}
\multiput(5,0.5)(2,0){2}{\circle{1}}
\put(3,4){\makebox(0,0){$\times$}}
\put(9,4){\circle{1}}
\put(4,5.5){\circle*{1}}
\put(8,5.5){\circle{1}}
\put(15,5){\makebox(0,0){$+$}}
\put(18,0){\line(3,5){3}}
\put(21,5){\line(3,-5){3}}
\put(24,0){\line(1,0){6}}
\put(30,0){\line(-3,5){3}}
\multiput(18,0)(1,0){7}{\circle*{0.2}}
\multiput(24,0)(0.6,1){6}{\circle*{0.2}}
\multiput(21,5)(0.6,1){6}{\circle*{0.2}}
\multiput(21,5)(1,0){7}{\circle*{0.2}}
\multiput(27,5)(-0.6,1){6}{\circle*{0.2}}
\put(19,0.5){\circle{1}}
\put(29,0.5){\circle*{1}}
\put(24,9){\makebox(0,0){$\times$}}
\multiput(23,0.5)(2,0){2}{\circle{1}}
\put(21,4){\circle*{1}}
\put(27,4){\circle{1}}
\put(22,5.5){\makebox(0,0){$\times$}}
\put(26,5.5){\makebox(0,0){$\oplus$}}
\end{picture}

\caption{\footnotesize{Illustration for the expression of $g_2(n,x_{n-1,2})$.}} 
\label{g2m2fig}
\end{figure}

\begin{figure}[htbp]
\unitlength 1.5mm 
\begin{picture}(48,10)
\put(0,0){\line(1,0){12}}
\put(0,0){\line(3,5){6}}
\put(12,0){\line(-3,5){6}}
\put(3,5){\line(1,0){6}}
\put(6,0){\line(3,5){3}}
\put(6,0){\line(-3,5){3}}
\scriptsize
\put(6,-1){\makebox(0,0){1/2}}
\put(12,-1){\makebox(0,0){1}}
\put(1,6){\makebox(0,0){1/2}}
\put(10,6){\makebox(0,0){0}}
\put(5,11){\makebox(0,0){1}}
\normalsize
\put(6,-5){\makebox(0,0){$F(1,x)$}}

\put(18,0){\line(1,0){12}}
\put(18,0){\line(3,5){6}}
\put(30,0){\line(-3,5){6}}
\put(21,5){\line(1,0){6}}
\put(24,0){\line(3,5){3}}
\put(24,0){\line(-3,5){3}}
\scriptsize
\put(24,-1){\makebox(0,0){1/3}}
\put(19,6){\makebox(0,0){1/3}}
\put(29,6){\makebox(0,0){1/3}}
\normalsize
\put(24,-5){\makebox(0,0){$G_2(1,x)$}}

\put(36,0){\line(1,0){12}}
\put(36,0){\line(3,5){6}}
\put(48,0){\line(-3,5){6}}
\put(39,5){\line(1,0){6}}
\put(42,0){\line(3,5){3}}
\put(42,0){\line(-3,5){3}}
\scriptsize
\put(42,-1){\makebox(0,0){2/3}}
\put(37,6){\makebox(0,0){2/3}}
\put(47,6){\makebox(0,0){4/3}}
\normalsize
\put(42,-5){\makebox(0,0){$U(1,x)$}}
\end{picture}

\vspace*{5mm}
\caption{\footnotesize{The distribution ratios $F(1,x)$, $G_2(1,x)$ and $U(1,x)$ for $SG(1)$.}} 
\label{FG2U1fig}
\end{figure}
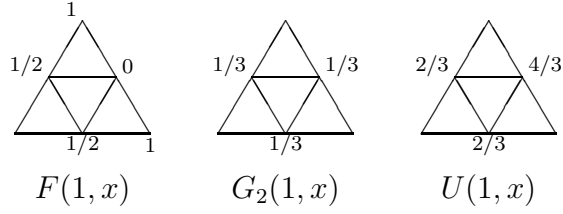

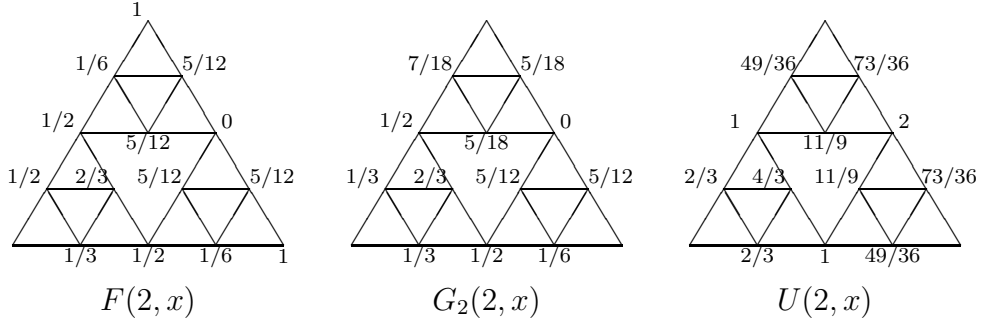
\begin{figure}[htbp]
\unitlength 1.5mm
\begin{picture}(84,20)
\put(0,0){\line(1,0){24}}
\put(0,0){\line(3,5){12}}
\put(24,0){\line(-3,5){12}}
\put(6,10){\line(1,0){12}}
\put(12,0){\line(3,5){6}}
\put(12,0){\line(-3,5){6}}
\multiput(3,5)(12,0){2}{\line(1,0){6}}
\multiput(6,0)(12,0){2}{\line(3,5){3}}
\multiput(6,0)(12,0){2}{\line(-3,5){3}}
\put(9,15){\line(1,0){6}}
\put(12,10){\line(3,5){3}}
\put(12,10){\line(-3,5){3}}
\scriptsize
\put(6,-1){\makebox(0,0){1/3}}
\put(12,-1){\makebox(0,0){1/2}}
\put(18,-1){\makebox(0,0){1/6}}
\put(24,-1){\makebox(0,0){1}}
\put(1,6){\makebox(0,0){1/2}}
\put(7,6){\makebox(0,0){2/3}}
\put(13,6){\makebox(0,0){5/12}}
\put(23,6){\makebox(0,0){5/12}}
\put(4,11){\makebox(0,0){1/2}}
\put(12,9){\makebox(0,0){5/12}}
\put(19,11){\makebox(0,0){0}}
\put(7,16){\makebox(0,0){1/6}}
\put(17,16){\makebox(0,0){5/12}}
\put(11,21){\makebox(0,0){1}}
\normalsize
\put(12,-5){\makebox(0,0){$F(2,x)$}}

\put(30,0){\line(1,0){24}}
\put(30,0){\line(3,5){12}}
\put(54,0){\line(-3,5){12}}
\put(36,10){\line(1,0){12}}
\put(42,0){\line(3,5){6}}
\put(42,0){\line(-3,5){6}}
\multiput(33,5)(12,0){2}{\line(1,0){6}}
\multiput(36,0)(12,0){2}{\line(3,5){3}}
\multiput(36,0)(12,0){2}{\line(-3,5){3}}
\put(39,15){\line(1,0){6}}
\put(42,10){\line(3,5){3}}
\put(42,10){\line(-3,5){3}}
\scriptsize
\put(36,-1){\makebox(0,0){1/3}}
\put(42,-1){\makebox(0,0){1/2}}
\put(48,-1){\makebox(0,0){1/6}}
\put(31,6){\makebox(0,0){1/3}}
\put(37,6){\makebox(0,0){2/3}}
\put(43,6){\makebox(0,0){5/12}}
\put(53,6){\makebox(0,0){5/12}}
\put(34,11){\makebox(0,0){1/2}}
\put(42,9){\makebox(0,0){5/18}}
\put(49,11){\makebox(0,0){0}}
\put(37,16){\makebox(0,0){7/18}}
\put(47,16){\makebox(0,0){5/18}}
\normalsize
\put(42,-5){\makebox(0,0){$G_2(2,x)$}}

\put(60,0){\line(1,0){24}}
\put(60,0){\line(3,5){12}}
\put(84,0){\line(-3,5){12}}
\put(66,10){\line(1,0){12}}
\put(72,0){\line(3,5){6}}
\put(72,0){\line(-3,5){6}}
\multiput(63,5)(12,0){2}{\line(1,0){6}}
\multiput(66,0)(12,0){2}{\line(3,5){3}}
\multiput(66,0)(12,0){2}{\line(-3,5){3}}
\put(69,15){\line(1,0){6}}
\put(72,10){\line(3,5){3}}
\put(72,10){\line(-3,5){3}}
\scriptsize
\put(66,-1){\makebox(0,0){2/3}}
\put(72,-1){\makebox(0,0){1}}
\put(78,-1){\makebox(0,0){49/36}}
\put(61,6){\makebox(0,0){2/3}}
\put(67,6){\makebox(0,0){4/3}}
\put(73,6){\makebox(0,0){11/9}}
\put(83,6){\makebox(0,0){73/36}}
\put(64,11){\makebox(0,0){1}}
\put(72,9){\makebox(0,0){11/9}}
\put(79,11){\makebox(0,0){2}}
\put(67,16){\makebox(0,0){49/36}}
\put(77,16){\makebox(0,0){73/36}}
\normalsize
\put(72,-5){\makebox(0,0){$U(2,x)$}}
\end{picture}

\vspace*{5mm}
\caption{\footnotesize{The distribution ratios $F(2,x)$, $G_2(2,x)$ and $U(2,x)$ for $SG(2)$.}} 
\label{FG2U2fig}
\end{figure}

\begin{lemma}
On the Sierpinski gasket $SG(n)$ with $n \ge 2$, the distribution ratio $G_2(n,x_{n-1,2})$ with $x \in \{a, b, c \}$ is given by
\beq
G_2(n,a_{n-1,2}) = G_2(n,c_{n-1,2}) = \begin{cases}
\frac12      & (n > 2) \ , \\
\frac{5}{18} & (n = 2) \ ,
\end{cases} \quad  
G_2(n,b_{n-1,2}) = \begin{cases}
0            & (n > 2) \ , \\
\frac{7}{18} & (n = 2) \ .
\end{cases} 
\eeq
\end{lemma}
The values for $G_2(n,x_{n-1,1})$ and $G_2(n,x_{m,\gamma_2})$ with $m \le n-2$, $x \in \{a,b,c\}$ and $\gamma_2 \in \{1,2\}$ and equal to $F(n,x_{n-1,1})$ and $F(n,x_{m,\gamma_2})$, respectively, as discussed in the following section. The distribution ratios $F(n,x)$, $G_2(n,x)$ and $U(n,x)$ for $n=1, 2$ are shown in Figs. \ref{FG2U1fig} and \ref{FG2U2fig}, respectively.

\section{Distribution $F(n,x_{\vec{\gamma}})$ with $|\vec{\gamma}| \ge 3$ and mean $\ell$ displacement for all vertices on $SG(n)$} 
\label{fnxg}

Using the vertex notation given in Section \ref{distribution}, the subscript of a vertex $x_{\vec{\gamma}}$ on $SG(n)$ is given as $\vec{\gamma} = (\gamma_1, \gamma_2, \gamma_3, \cdots, \gamma_s)$ where $1 \le \gamma_1 \le n-1$ and $\gamma_k \in \{0,1,2\}$ for $k=2,3,...,s$. As the vertices $x_m$ and $x_{m,1}$, $x_{m,2}$, i.e. $s = |\vec{\gamma}| = 1,2$, on $SG(n)$ with $x \in \{a,b,c\}$ and $m \le n$ have been given in the previous two sections, the purpose of this section is to consider the other vertex $x_{\vec{\gamma}}$ with $3 \leq |\vec{\gamma}| \leq n$ and $x \in \{a,b,c\}$. We have to derive $F(n,x_{\vec{\gamma}})$ for the vertices with $\gamma_1=n-1$, $\gamma_2=2$ on $SG(n)$ and the vertices with $\gamma_1 < n-1$. That is, we should consider the vertices within the upper $SG(n-1)$ and the lower-left $SG(n-1)$ that constitute $SG(n)$. For the vertices with $\gamma_1=n-1$, $\gamma_2=1$, i.e. the vertices within the lower-right $SG(n-1)$, we use $F(n,x_{\vec{\gamma}})= F(n,\tilde{x}_{\vec{\gamma}})$ by symmetry. A quantity like $F(n,x_{\vec{\gamma}})$ on $SG(n)$ will be expressed in terms of the quantities on $SG(n-1)$ as the recursion relation. Let us denote the subscript $\vec{\gamma}^\prime$ for the corresponding vertex $x_{\vec{\gamma}^\prime}$ on $SG(n-1)$. When the first component of $\vec{\gamma}$, i.e. $\gamma_1$, is equal to $n-1$ and $\gamma_2 \in \{1,2\}$, we have $\vec{\gamma}^\prime = (n-2,\gamma_3,\cdots,\gamma_s)$ with $|\vec{\gamma}^\prime| = |\vec{\gamma}|-1$; while the first component of $\vec{\gamma}$ is smaller than $n-1$, we have $\vec{\gamma}^\prime = \vec{\gamma}$.

By the argument of Eqs. (\ref{fxn2}) and (\ref{f2}), we have
\beq
\begin{cases}  f(n,x_{\vec{\gamma}}) = f_1(n-1)g_1(n-1)f(n-1,\hat{\tilde{\hat{x}}}_{\vec{\gamma}^\prime}) + f_1(n-1)^2g_2(n-1,\hat{\tilde{x}}_{\vec{\gamma}^\prime}) \\
\hspace*{9.4cm} \mbox{if} \quad \gamma_1 = n-1, \ \gamma_2 = 2 \ , \\
f(n,x_{\vec{\gamma}}) = f_1(n-1)^2u(n-1,x_{\vec{\gamma}^\prime}) \qquad \qquad \qquad \qquad \mbox{if} \quad \gamma_1 < n-1 \ ,
\end{cases}
\label{fx}
\eeq  
to obtain the following lemma:
\begin{lemma}
\label{lemmaF}
On the Sierpinski gasket $SG(n)$, the distribution ratio $F(n,x_{\vec{\gamma}})$ with $3 \le |\vec{\gamma}| \le n$ is given by
\beq
\begin{cases} 
F(n,x_{\vec{\gamma}}) = \frac12 F(n-1,\hat{\tilde{\hat{x}}}_{\vec{\gamma}^\prime}) + \frac12 G_2(n-1,\hat{\tilde{x}}_{\vec{\gamma}^\prime}) & \mbox{if} \quad \gamma_1=n-1, \ \gamma_2 = 2 \ , \\
F(n,x_{\vec{\gamma}}) = \frac12 U(n-1,x_{\vec{\gamma}^\prime}) & \mbox{if} \quad \gamma_1<n-1 \ .
\end{cases}
\label{goal}
\eeq
\end{lemma}
To evaluate (\ref{goal}), we need the following lemmas. 

\begin{lemma}
\label{lemmaG2} 
On the Sierpinski gasket $SG(n)$, the distribution ratio $G_2(n,x_{\vec{\gamma}})$ is given as follows:
(i) When $\gamma_1=n-1$ and $\gamma_2=1$ with $|\vec{\gamma}| \geq 2$, we have
\beq
\label{g2-01}
G_2(n,x_{\vec{\gamma}}) = F(n,x_{\vec{\gamma}}) \ .
\eeq
(ii) When $\gamma_1<n-1$, we have
\beq
\label{gu}
G_2(n,x_{\vec{\gamma}}) = F(n,x_{\vec{\gamma}}) = \frac12 U(n-1,x_{\vec{\gamma}^\prime}) \ , \quad  
U(n,x_{\vec{\gamma}}) = U(n-1,x_{\vec{\gamma}^\prime}) \ .
\eeq
(iii) When $\gamma_1=n-1$ and $\gamma_2=2$ with $|\vec{\gamma}| \geq 3$, we have
\beq
\label{g2-s3}
G_2(n,x_{\vec{\gamma}}) = \begin{cases}
G_2(n-1,\hat{\tilde{\hat{x}}}_{\vec{\gamma}^\prime}) & \mbox{if} \quad \gamma_3=1, 2 \ , \\
\frac12 \bigl( G_2(n-1,\hat{\tilde{x}}_{\vec{\gamma}^\prime}) + G_2(n-1,\hat{\tilde{\hat{x}}}_{\vec{\gamma}^\prime}) \bigr ) & \mbox{if} \quad \gamma_3=0 \ .
\end{cases}
\eeq  
\end{lemma}
 
\begin{lemma}
\label{lemmaU}
On the Sierpinski gasket $SG(n)$, the distribution ratio $U(n,x_{\vec{\gamma}})$ for $\gamma_1=n-1$ and $\gamma_2=2$ with $|\vec{\gamma}| \geq 3$ is given by
\beq
\label{u-1}
U(n,x_{\vec{\gamma}}) = \begin{cases}
U(n-1,\hat{x}_{\vec{\gamma}^\prime}) + U(n-1,x_{\vec{\gamma}^\prime})  & \mbox{if} \quad \gamma_3=2 \ , \\
G_2(n-1,\hat{x}_{\vec{\gamma}^\prime}) + G_2(n-1,\tilde{\hat{x}}_{\vec{\gamma}^\prime}) + U(n-1,x_{\vec{\gamma}^\prime}) & \mbox{if} \quad \gamma_3 = 0, 1 \ .
\end{cases} 
\eeq 
\end{lemma}
 
\bigskip   
    
\noindent {\it Proof of Lemma~\ref{lemmaG2}:}  
Consider the vertex with $\gamma_1=n-1$ and $\gamma_2=1$ first. We have
%
\beqs
\begin{cases} 
f(n,x_{\vec{\gamma}}) = f_1(n-1)g_1(n-1)f(n-1,\hat{x}_{\vec{\gamma}^\prime}) + f_1(n-1)^2 g_2(n-1,\hat{x}_{\vec{\gamma}^\prime}) \ , \\
g_2(n,x_{\vec{\gamma}}) = g_1(n-1)^2f(n-1,\hat{x}_{\vec{\gamma}^\prime}) + f_1(n-1)g_1(n-1)g_2(n-1,\hat{x}_{\vec{\gamma}^\prime}) \ ,
\end{cases}
\label{fg2x}
\eeqs
%
which leads to (\ref{g2-01}). 

For the vertex with $\gamma_1<n-1$, we have
\begin{align}
\label{u-4}
\begin{cases} 
f(n,x_{\vec{\gamma}}) = f_1(n-1)^2 u(n-1,x_{\vec{\gamma}^\prime}) \ , \\
g_2(n,x_{\vec{\gamma}}) = f_1(n-1)g_1(n-1)u(n-1,x_{\vec{\gamma}^\prime}) \ , \\
g_3(n,x_{\vec{\gamma}}) = 2f_1(n-1)g_1(n-1)g_3(n-1,x_{\vec{\gamma}^\prime}) \ ,
\end{cases}
\end{align} 
similar to Eq. (\ref{fg}), such that Eq. (\ref{gu}) is proved.

For Eq. (\ref{g2-s3}) with $\gamma_1=n-1$ and $\gamma_2=2$, we have
\beq
g_2(n,x_{\vec{\gamma}}) = f_1(n-1)g_1(n-1)g_2(n-1,\hat{\tilde{\hat{x}}}_{\vec{\gamma}^\prime}) + f_1(n-1)^2t_1(n-1,\tilde{\hat{x}}_{\vec{\gamma}^\prime}) \ ,
\label{g2-0}
\eeq
similar to Eq. (\ref{g2xn2}), so that
\beq
\label{g2-1}
G_2(n,x_{\vec{\gamma}}) = \frac12 G_2(n-1,\hat{\tilde{\hat{x}}}_{\vec{\gamma}^\prime}) + \frac13 T_1(n-1,\tilde{\hat{x}}_{\vec{\gamma}^\prime}) \ .
\eeq
The recursion relations for $t_1(n,x_{\vec{\gamma}})$ can be derived as 
\beq
\begin{cases}
t_1(n,x_{\vec{\gamma}}) = g_1(n-1)^2g_2(n-1,\hat{\tilde{x}}_{\vec{\gamma}^\prime}) + f_1(n-1)g_1(n-1)t_1(n-1,\tilde{x}_{\vec{\gamma}^\prime}) \quad \mbox{if} \quad \gamma_1<n-1 \ , \\
t_1(n,x_{\vec{\gamma}}) = g_1(n-1)^2g_2(n-1,\tilde{x}_{\vec{\gamma}^\prime}) + f_1(n-1)g_1(n-1)t_1(n-1,\hat{x}_{\vec{\gamma}^\prime}) \\
\hspace*{11.3cm} \mbox{if} \quad \gamma_1=n-1, \ \gamma_2=1 \ , \\
t_1(n,x_{\vec{\gamma}}) = g_1(n-1)^2 u(n-1,\hat{x}_{\vec{\gamma}^\prime}) \qquad \qquad \qquad \qquad \qquad \qquad \ \ \mbox{if} \quad \gamma_1=n-1, \ \gamma_2=2 \ ,
\end{cases}
\label{t1}
\eeq
or equivalently,
\beq
\label{h}
\begin{cases}
T_1(n,x_{\vec{\gamma}}) = \frac34 G_2(n-1,\hat{\tilde{x}}_{\vec{\gamma}^\prime}) + \frac12 T_1(n-1,\tilde{x}_{\vec{\gamma}^\prime}) & \mbox{if} \quad \gamma_1<n-1 \ , \\
T_1(n,x_{\vec{\gamma}}) = \frac34 G_2(n-1,\tilde{x}_{\vec{\gamma}^\prime}) + \frac12 T_1(n-1,\hat{x}_{\vec{\gamma}^\prime}) & \mbox{if} \quad \gamma_1=n-1, \ \gamma_2=1 \ , \\
T_1(n,x_{\vec{\gamma}}) = \frac34 U(n-1,\hat{x}_{\vec{\gamma}^\prime}) & \mbox{if} \quad \gamma_1=n-1, \ \gamma_2=2 \ .
\end{cases}
\eeq
For the vertex $x_{\vec{\gamma}}$ with $\gamma_1<n-1$, the corresponding $\hat{x}_{\vec{\gamma}}$ is located in the upper $SG(n-1)$, so that Eq. (\ref{g2-1}) can be applied to give
\beq
\frac32 G_2(n,\hat{x}_{\vec{\gamma}}) = \frac34 G_2(n-1,\hat{\tilde{x}}_{\vec{\gamma}^\prime}) + \frac12 T_1(n-1,\tilde{x}_{\vec{\gamma}^\prime}) = T_1(n,x_{\vec{\gamma}}) \ ,
\label{g2a}
\eeq
where the last equality follows from the first line of (\ref{h}). For the vertex $x_{\vec{\gamma}}$ with $\gamma_1=n-1$ and $\gamma_2=1$, the corresponding $\tilde{x}_{\vec{\gamma}}$ is located in the upper $SG(n-1)$, so that Eq. (\ref{g2-1}) can be applied to give
\beqs
\frac32 G_2(n,\tilde{x}_{\vec{\gamma}}) & = & \frac34 G_2(n-1,\hat{\tilde{\hat{\tilde{x}}}}_{\vec{\gamma}^\prime}) + \frac12 T_1(n-1,\tilde{\hat{\tilde{x}}}_{\vec{\gamma}^\prime}) \cr\cr
& = & \frac34 G_2(n-1,\tilde{\hat{x}}_{\vec{\gamma}^\prime}) + \frac 12 T_1(n-1,x_{\vec{\gamma}^\prime}) = T_1(n,\hat{x}_{\vec{\gamma}}) \ ,
\label{g2b}
\eeqs
where the last equality follows from the second line of (\ref{h}), and the relations $\hat{\tilde{\hat{\tilde{x}}}}_{\vec{\gamma}} = \tilde{\hat{x}}_{\vec{\gamma}}$ and  $T_1(n-1,x_{\vec{\gamma}^\prime}) = T_1(n-1,\tilde{\hat{\tilde{x}}}_{\vec{\gamma}^\prime})$ are used. For the vertex $x_{\vec{\gamma}}$ with $\gamma_1=n-1$ and $\gamma_2=2$, the corresponding $\hat{x}_{\vec{\gamma}}$ is located in the lower-left $SG(n-1)$, so that Eq. (\ref{u-4}) can be applied to give
\beq
\frac32 G_2(n,\hat{x}_{\vec{\gamma}}) = \frac34 U(n,\hat{x}_{\vec{\gamma}^\prime}) = T_1(n,x_{\vec{\gamma}}) \ ,
\label{g2c}
\eeq
where the last equality follows from the third line of (\ref{h}).
Combining Eqs. (\ref{g2a})-(\ref{g2c}) for the vertex $x_{\vec{\gamma}}$ with $|\vec{\gamma}| \geq 2$, we obtain the following relation:
\beq
\label{u-7}
T_1(n,x_{\vec{\gamma}}) = 
\begin{cases}
\frac32 G_2(n,\hat{x}_{\vec{\gamma}}) & \mbox{if} \quad \gamma_1<n-1 \quad \mbox{or} \quad \gamma_1=n-1, \ \gamma_2=2 \ , \\
\frac32 G_2(n,\tilde{\hat{x}}_{\vec{\gamma}}) & \mbox{if}\quad \gamma_1=n-1, \ \gamma_2=1 \ .
\end{cases}
\eeq
For the vertex $x_{\vec{\gamma}}$ with $\gamma_1<n-1$, or $\gamma_1=n-1$, $\gamma_2=1$, or $\gamma_1=n-1$, $\gamma_2=2$, the corresponding $\tilde{\hat{x}}_{\vec{\gamma}}$ is located in the lower-right $SG(n-1)$, or upper $SG(n-1)$, or lower-left $SG(n-1)$, respectively. Therefore, the last term in Eq. (\ref{g2-1}) becomes
\beq
\frac13 T_1(n-1,\tilde{\hat{x}}_{\vec{\gamma}^\prime}) = 
\begin{cases}
\frac12 G_2(n-1,\hat{\tilde{\hat{x}}}_{\vec{\gamma}^\prime}) & \mbox{if} \quad \gamma_3=1,2 \ , \\
\frac12 G_2(n-1, \tilde{\hat{\tilde{\hat{x}}}}_{\vec{\gamma}^\prime}) = \frac12 G_2(n-1,\hat{\tilde{x}}_{\vec{\gamma}^\prime}) & \mbox{if} \quad \gamma_3=0 \ ,
\end{cases}
\eeq
such that Eq. (\ref{g2-s3}) is proved.
\EndProof

\bigskip  
  
\noindent {\it Proof of Lemma~\ref{lemmaU}:}  
As $U(n,x_{\vec{\gamma}})$ for $\gamma_1 < n-1$ has been given in Eq. (\ref{gu}) and $U(n,x_{\vec{\gamma}}) = U(n,\tilde x_{\vec{\gamma}})$ by definition, we only consider the vertices with $\gamma_1=n-1$ and $\gamma_2=2$ in this lemma.
By the same argument of (\ref{u}), we have 
\beqs
u(n,x_{\vec{\gamma}}) & = & f_1(n-1)^2t_1(n-1,x_{\vec{\gamma}^\prime}) + g_1(n-1)^2f(n-1,\hat{x}_{\vec{\gamma}^\prime}) \cr
& & + f_1(n-1)g_1(n-1) \bigl( g_2(n-1,\hat{x}_{\vec{\gamma}^\prime}) + g_2(n-1,\tilde{\hat{x}}_{\vec{\gamma}^\prime}) + 2u(n-1,x_{\vec{\gamma}^\prime}) \bigr) \ ,
\eeqs
or equivalently,
\beq
\label{ug3}
U(n,x_{\vec{\gamma}}) = \frac13 T_1(n-1,x_{\vec{\gamma}^\prime}) + \frac12 F(n-1,\hat{x}_{\vec{\gamma}^\prime}) + \frac12 G_2(n-1,\hat{x}_{\vec{\gamma}^\prime}) + \frac12 G_2(n-1,\tilde{\hat{x}}_{\vec{\gamma}^\prime}) + U(n-1,x_{\vec{\gamma}^\prime}) \ .
\eeq 
 
Similar to the definition of $\vec{\gamma}^\prime$, let us denote the subscript $\vec{\gamma}^{\prime\prime}$ for the vertex on $SG(n-2)$. When the first component of $\vec{\gamma}^\prime$, i.e. $\gamma_1$, is equal to $n-2$ and $\gamma_3 \in \{1,2\}$, we have $\vec{\gamma}^{\prime\prime} = (n-3,\gamma_4,\cdots,\gamma_s)$ with $|\vec{\gamma}^{\prime\prime}| = |\vec{\gamma}^\prime|-1$; while $\gamma_3 = 0$, we have $\vec{\gamma}^{\prime\prime} = \vec{\gamma}^\prime$. Consider the vertex $x_{\vec{\gamma}^\prime}$ with $\gamma_3=2$ first. The corresponding $\hat{x}_{\vec{\gamma}^\prime}$ and $\tilde{\hat{x}}_{\vec{\gamma}^\prime}$ are both located in the lower-left $SG(n-2)$. Using the third line of Eq. (\ref{t1}) and the first two lines of Eq. (\ref{u-4}), we have
%
\beqs
\label{xr1}
\left\{ \begin{array}{lll}
t_1(n-1,x_{\vec{\gamma}^\prime}) & = & g_1(n-2)^2u(n-2,\hat{x}_{\vec{\gamma}^{\prime\prime}}) = \frac34 g_1(n-1)U(n-2,\hat{x}_{\vec{\gamma}^{\prime\prime}}) \ , \\
f(n-1,\hat{x}_{\vec{\gamma}^\prime}) & = & f_1(n-2)^2 u(n-2,\hat{x}_{\vec{\gamma}^{\prime\prime}}) = \frac12 f_1(n-1)U(n-2,\hat{x}_{\vec{\gamma}^{\prime\prime}}) \ , \\
g_2(n-1,\hat{x}_{\vec{\gamma}^\prime}) & = & g_2(n-1,\tilde{\hat{x}}_{\vec{\gamma}^\prime}) = f_1(n-2)g_1(n-2)u(n-2,\hat{x}_{\vec{\gamma}^{\prime\prime}}) \cr
& = & \frac12 g_1(n-1)U(n-2,\hat{x}_{\vec{\gamma}^{\prime\prime}}) \ ,
\end{array} \right .
\eeqs
%
where $U(n-2,\hat{x}_{\vec{\gamma}^{\prime\prime}}) = U(n-1,\hat{x}_{\vec{\gamma}^{\prime}})$ as given in Eq. (\ref{gu}). Substituting Eq. (\ref{xr1}) into Eq. (\ref{ug3}), the first line of (\ref{u-1}) for $\gamma_3=2$ is proved.
 
Next consider the vertex $x_{\vec{\gamma}^\prime}$ with $\gamma_3=0$. The corresponding $\hat{x}_{\vec{\gamma}^\prime}$ and $\tilde{\hat{x}}_{\vec{\gamma}^\prime}$ are located in the upper $SG(n-2)$ and lower-right $SG(n-2)$, respectively. Using the first line of Eq. (\ref{t1}), the first line of Eq. (\ref{fx}), Eq. (\ref{g2-0}), and the second line of Eq. (\ref{fg2x}), we have
\beqs
\begin{cases} 
t_1(n-1,x_{\vec{\gamma}^\prime}) = g_1(n-2)^2g_2(n-2,\hat{\tilde{x}}_{\vec{\gamma}^{\prime\prime}}) + f_1(n-2)g_1(n-2)t_1(n-2,\tilde{x}_{\vec{\gamma}^{\prime\prime}}) \ , \\
f(n-1,\hat{x}_{\vec{\gamma}^\prime}) = f_1(n-2)g_1(n-2)f(n-2,\hat{\tilde{\hat{x}}}_{\vec{\gamma}^{\prime\prime}}) + f_1(n-2)^2 g_2(n-2,\hat{\tilde{\hat{x}}}_{\vec{\gamma}^{\prime\prime}}) \ , \\
g_2(n-1,\hat{x}_{\vec{\gamma}^\prime}) = f_1(n-2)g_1(n-2)g_2(n-2,\hat{\tilde{x}}_{\vec{\gamma}^{\prime\prime}}) + f_1(n-2)^2t_1(n-2,\tilde{x}_{\vec{\gamma}^{\prime\prime}}) \ , \\
g_2(n-1,\tilde{\hat{x}}_{\vec{\gamma}^\prime}) = g_1(n-2)^2f(n-2,\hat{\tilde{\hat{x}}}_{\vec{\gamma}^{\prime\prime}}) + f_1(n-2)g_1(n-2)g_2(n-2,\hat{\tilde{\hat{x}}}_{\vec{\gamma}^{\prime\prime}}) \ ,
\end{cases}
\eeqs
or equivalently,
\beqs
\label{xx3}
\begin{cases} 
T_1(n-1,x_{\vec{\gamma}^\prime}) = \frac34 G_2(n-2,\hat{\tilde{x}}_{\vec{\gamma}^{\prime\prime}}) + \frac12 T_1(n-2,\tilde{x}_{\vec{\gamma}^{\prime\prime}}) \ , \\
F(n-1,\hat{x}_{\vec{\gamma}_\prime}) = \frac12 F(n-2,\hat{\tilde{\hat{x}}}_{\vec{\gamma}^{\prime\prime}}) + \frac12 G_2(n-2,\hat{\tilde{\hat{x}}}_{\vec{\gamma}^{\prime\prime}}) \ , \\ G_2(n-1,\hat{x}_{\vec{\gamma}^\prime}) = \frac12 G_2(n-2,\hat{\tilde{x}}_{\vec{\gamma}^{\prime\prime}}) + \frac13 T_1(n-2,\tilde{x}_{\vec{\gamma}^{\prime\prime}}) \ , \\
G_2(n-1,\tilde{\hat{x}}_{\vec{\gamma}^\prime}) = \frac12 F(n-2,\hat{\tilde{\hat{x}}}_{\vec{\gamma}^{\prime\prime}}) + \frac12 G_2(n-2,\hat{\tilde{\hat{x}}}_{\vec{\gamma}^{\prime\prime}}) \ .
\end{cases}
\eeqs
It follows that $T_1(n-1,x_{\vec{\gamma}^\prime}) = \frac32 G_2(n-1,\hat{x}_{\vec{\gamma}^\prime})$, $F(n-1,\hat{x}_{\vec{\gamma}^\prime}) = G_2(n-1,\tilde{\hat{x}}_{\vec{\gamma}^\prime})$, and Eq. (\ref{ug3}) becomes the second line of Eq. (\ref{u-1}) for $\gamma_3=0$.  

Finally, consider the vertex $x_{\vec{\gamma}^\prime}$ with $\gamma_3=1$. The corresponding $\hat{x}_{\vec{\gamma}^\prime}$ and $\tilde{\hat{x}}_{\vec{\gamma}^\prime}$ are located in the lower-right $SG(n-2)$ and upper $SG(n-2)$, respectively. Using the second line of Eq. (\ref{t1}) and Eqs. (\ref{fg2x}), (\ref{g2-0}), we have
\beqs
\begin{cases} 
t_1(n-1,x_{\vec{\gamma}^\prime}) = g_1(n-2)^2g_2(n-2,\tilde{x}_{\vec{\gamma}_{\prime\prime}}) + f_1(n-2)g_1(n-2)t_1(n-2,\hat{x}_{\vec{\gamma}^{\prime\prime}}) \ , \\
f(n-1,\hat{x}_{\vec{\gamma}^\prime}) = f_1(n-2)g_1(n-2)f(n-2,x_{\vec{\gamma}^{\prime\prime}}) + f_1(n-2)^2g_2(n-2,x_{\vec{\gamma}^{\prime\prime}}) \ , \\
g_2(n-1,\hat{x}_{\vec{\gamma}^\prime}) = g_1(n-2)^2f(n-2,x_{\vec{\gamma}_{\prime\prime}}) + f_1(n-2)g_1(n-2)g_2(n-2,x_{\vec{\gamma}^{\prime\prime}}) \ , \\
g_2(n-1,\tilde{\hat{x}}_{\vec{\gamma}^\prime}) = f_1(n-2)g_1(n-2)g_2(n-2,\tilde{x}_{\vec{\gamma}^{\prime\prime}}) + f_1(n-2)^2t_1(n-2,\hat{\tilde{x}}_{\vec{\gamma}^{\prime\prime}}) \ .
\end{cases}
\eeqs
Because $t_1(n-2,\hat{\tilde{x}}_{\vec{\gamma}^{\prime\prime}}) = t_1(n-2,\hat{x}_{\vec{\gamma}^{\prime\prime}})$, it follows that $T_1(n-1,x_{\vec{\gamma}^\prime}) = \frac32 G_2(n-1,\tilde{\hat{x}}_{\vec{\gamma}^\prime})$, $F(n-1,\hat{x}_{\vec{\gamma}^\prime}) = G_2(n-1,\hat{x}_{\vec{\gamma}^\prime})$, and Eq. (\ref{ug3}) becomes the second line of Eq. (\ref{u-1}) for $\gamma_3=1$. %
\EndProof

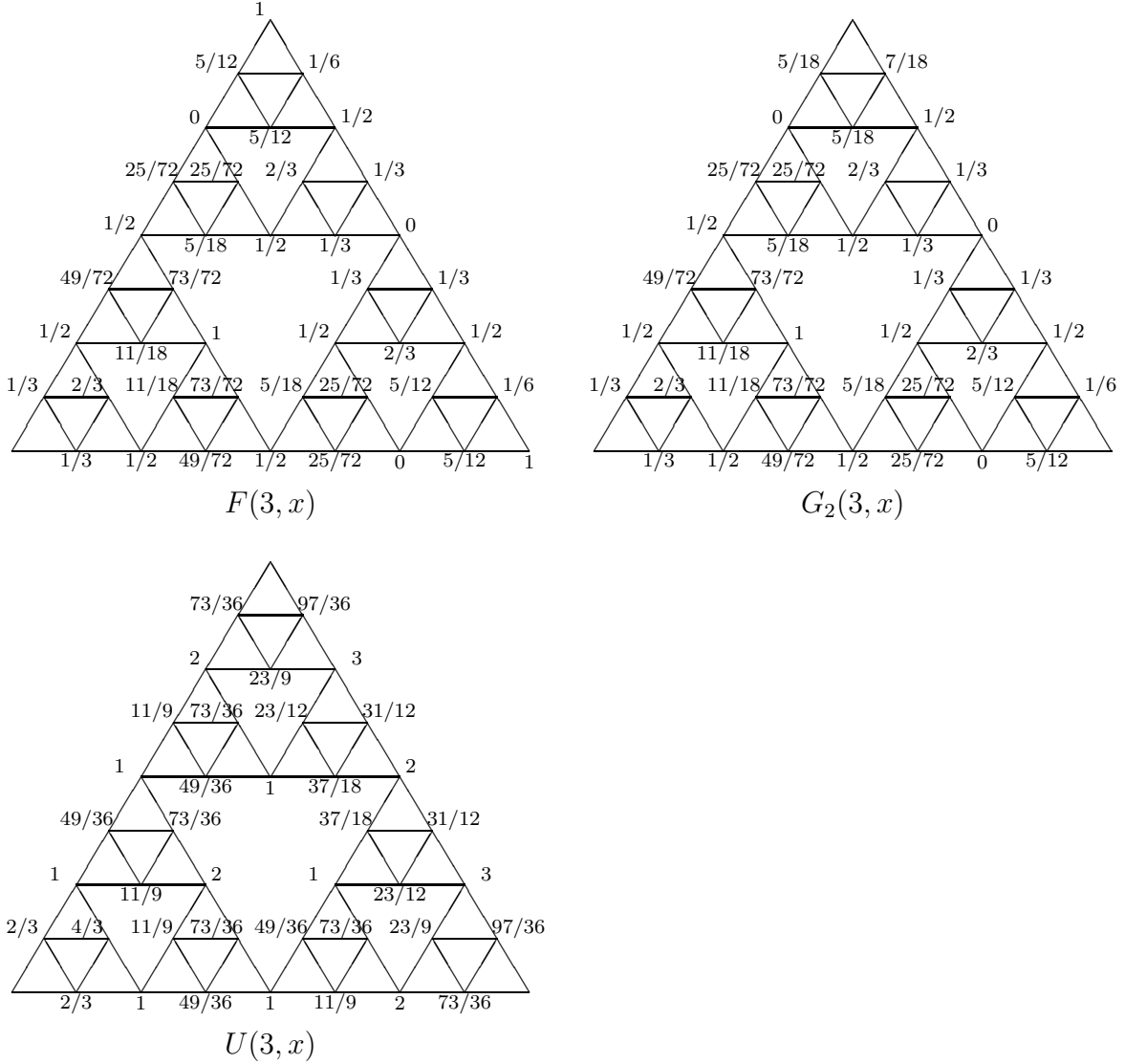
\begin{figure}[htbp]
\unitlength 1.5mm
\begin{picture}(102,40)
\put(0,0){\line(1,0){48}}
\put(12,20){\line(1,0){24}}
\put(0,0){\line(3,5){24}}
\put(24,0){\line(3,5){12}}
\put(24,0){\line(-3,5){12}}
\put(48,0){\line(-3,5){24}}
\put(6,10){\line(1,0){12}}
\put(30,10){\line(1,0){12}}
\put(18,30){\line(1,0){12}}
\put(12,0){\line(3,5){6}}
\put(36,0){\line(3,5){6}}
\put(24,20){\line(3,5){6}}
\put(12,0){\line(-3,5){6}}
\put(36,0){\line(-3,5){6}}
\put(24,20){\line(-3,5){6}}
\multiput(3,5)(12,0){4}{\line(1,0){6}}
\multiput(6,0)(12,0){4}{\line(3,5){3}}
\multiput(6,0)(12,0){4}{\line(-3,5){3}}
\multiput(9,15)(24,0){2}{\line(1,0){6}}
\multiput(12,10)(24,0){2}{\line(3,5){3}}
\multiput(12,10)(24,0){2}{\line(-3,5){3}}
\multiput(15,25)(12,0){2}{\line(1,0){6}}
\multiput(18,20)(12,0){2}{\line(3,5){3}}
\multiput(18,20)(12,0){2}{\line(-3,5){3}}
\put(21,35){\line(1,0){6}}
\put(24,30){\line(3,5){3}}
\put(24,30){\line(-3,5){3}}
\scriptsize
\put(6,-1){\makebox(0,0){1/3}}
\put(12,-1){\makebox(0,0){1/2}}
\put(18,-1){\makebox(0,0){49/72}}
\put(24,-1){\makebox(0,0){1/2}}
\put(30,-1){\makebox(0,0){25/72}}
\put(36,-1){\makebox(0,0){0}}
\put(42,-1){\makebox(0,0){5/12}}
\put(48,-1){\makebox(0,0){1}}
\put(1,6){\makebox(0,0){1/3}}
\put(7,6){\makebox(0,0){2/3}}
\put(13,6){\makebox(0,0){11/18}}
\put(19,6){\makebox(0,0){73/72}}
\put(25,6){\makebox(0,0){5/18}}
\put(31,6){\makebox(0,0){25/72}}
\put(37,6){\makebox(0,0){5/12}}
\put(47,6){\makebox(0,0){1/6}}
\put(4,11){\makebox(0,0){1/2}}
\put(12,9){\makebox(0,0){11/18}}
\put(19,11){\makebox(0,0){1}}
\put(28,11){\makebox(0,0){1/2}}
\put(36,9){\makebox(0,0){2/3}}
\put(44,11){\makebox(0,0){1/2}}
\put(7,16){\makebox(0,0){49/72}}
\put(17,16){\makebox(0,0){73/72}}
\put(31,16){\makebox(0,0){1/3}}
\put(41,16){\makebox(0,0){1/3}}
\put(10,21){\makebox(0,0){1/2}}
\put(18,19){\makebox(0,0){5/18}}
\put(24,19){\makebox(0,0){1/2}}
\put(30,19){\makebox(0,0){1/3}}
\put(37,21){\makebox(0,0){0}}
\put(13,26){\makebox(0,0){25/72}}
\put(19,26){\makebox(0,0){25/72}}
\put(25,26){\makebox(0,0){2/3}}
\put(35,26){\makebox(0,0){1/3}}
\put(17,31){\makebox(0,0){0}}
\put(24,29){\makebox(0,0){5/12}}
\put(32,31){\makebox(0,0){1/2}}
\put(19,36){\makebox(0,0){5/12}}
\put(29,36){\makebox(0,0){1/6}}
\put(23,41){\makebox(0,0){1}}
\normalsize
\put(24,-5){\makebox(0,0){$F(3,x)$}}

\put(54,0){\line(1,0){48}}
\put(66,20){\line(1,0){24}}
\put(54,0){\line(3,5){24}}
\put(78,0){\line(3,5){12}}
\put(78,0){\line(-3,5){12}}
\put(102,0){\line(-3,5){24}}
\put(60,10){\line(1,0){12}}
\put(84,10){\line(1,0){12}}
\put(72,30){\line(1,0){12}}
\put(66,0){\line(3,5){6}}
\put(90,0){\line(3,5){6}}
\put(78,20){\line(3,5){6}}
\put(66,0){\line(-3,5){6}}
\put(90,0){\line(-3,5){6}}
\put(78,20){\line(-3,5){6}}
\multiput(57,5)(12,0){4}{\line(1,0){6}}
\multiput(60,0)(12,0){4}{\line(3,5){3}}
\multiput(60,0)(12,0){4}{\line(-3,5){3}}
\multiput(63,15)(24,0){2}{\line(1,0){6}}
\multiput(66,10)(24,0){2}{\line(3,5){3}}
\multiput(66,10)(24,0){2}{\line(-3,5){3}}
\multiput(69,25)(12,0){2}{\line(1,0){6}}
\multiput(72,20)(12,0){2}{\line(3,5){3}}
\multiput(72,20)(12,0){2}{\line(-3,5){3}}
\put(75,35){\line(1,0){6}}
\put(78,30){\line(3,5){3}}
\put(78,30){\line(-3,5){3}}
\scriptsize
\put(60,-1){\makebox(0,0){1/3}}
\put(66,-1){\makebox(0,0){1/2}}
\put(72,-1){\makebox(0,0){49/72}}
\put(78,-1){\makebox(0,0){1/2}}
\put(84,-1){\makebox(0,0){25/72}}
\put(90,-1){\makebox(0,0){0}}
\put(96,-1){\makebox(0,0){5/12}}
\put(55,6){\makebox(0,0){1/3}}
\put(61,6){\makebox(0,0){2/3}}
\put(67,6){\makebox(0,0){11/18}}
\put(73,6){\makebox(0,0){73/72}}
\put(79,6){\makebox(0,0){5/18}}
\put(85,6){\makebox(0,0){25/72}}
\put(91,6){\makebox(0,0){5/12}}
\put(101,6){\makebox(0,0){1/6}}
\put(58,11){\makebox(0,0){1/2}}
\put(66,9){\makebox(0,0){11/18}}
\put(73,11){\makebox(0,0){1}}
\put(82,11){\makebox(0,0){1/2}}
\put(90,9){\makebox(0,0){2/3}}
\put(98,11){\makebox(0,0){1/2}}
\put(61,16){\makebox(0,0){49/72}}
\put(71,16){\makebox(0,0){73/72}}
\put(85,16){\makebox(0,0){1/3}}
\put(95,16){\makebox(0,0){1/3}}
\put(64,21){\makebox(0,0){1/2}}
\put(72,19){\makebox(0,0){5/18}}
\put(78,19){\makebox(0,0){1/2}}
\put(84,19){\makebox(0,0){1/3}}
\put(91,21){\makebox(0,0){0}}
\put(67,26){\makebox(0,0){25/72}}
\put(73,26){\makebox(0,0){25/72}}
\put(79,26){\makebox(0,0){2/3}}
\put(89,26){\makebox(0,0){1/3}}
\put(71,31){\makebox(0,0){0}}
\put(78,29){\makebox(0,0){5/18}}
\put(86,31){\makebox(0,0){1/2}}
\put(73,36){\makebox(0,0){5/18}}
\put(83,36){\makebox(0,0){7/18}}
\normalsize
\put(78,-5){\makebox(0,0){$G_2(3,x)$}}
\end{picture}

\vspace*{1.5cm}

\begin{picture}(102,40)
\put(0,0){\line(1,0){48}}
\put(12,20){\line(1,0){24}}
\put(0,0){\line(3,5){24}}
\put(24,0){\line(3,5){12}}
\put(24,0){\line(-3,5){12}}
\put(48,0){\line(-3,5){24}}
\put(6,10){\line(1,0){12}}
\put(30,10){\line(1,0){12}}
\put(18,30){\line(1,0){12}}
\put(12,0){\line(3,5){6}}
\put(36,0){\line(3,5){6}}
\put(24,20){\line(3,5){6}}
\put(12,0){\line(-3,5){6}}
\put(36,0){\line(-3,5){6}}
\put(24,20){\line(-3,5){6}}
\multiput(3,5)(12,0){4}{\line(1,0){6}}
\multiput(6,0)(12,0){4}{\line(3,5){3}}
\multiput(6,0)(12,0){4}{\line(-3,5){3}}
\multiput(9,15)(24,0){2}{\line(1,0){6}}
\multiput(12,10)(24,0){2}{\line(3,5){3}}
\multiput(12,10)(24,0){2}{\line(-3,5){3}}
\multiput(15,25)(12,0){2}{\line(1,0){6}}
\multiput(18,20)(12,0){2}{\line(3,5){3}}
\multiput(18,20)(12,0){2}{\line(-3,5){3}}
\put(21,35){\line(1,0){6}}
\put(24,30){\line(3,5){3}}
\put(24,30){\line(-3,5){3}}
\scriptsize
\put(6,-1){\makebox(0,0){2/3}}
\put(12,-1){\makebox(0,0){1}}
\put(18,-1){\makebox(0,0){49/36}}
\put(24,-1){\makebox(0,0){1}}
\put(30,-1){\makebox(0,0){11/9}}
\put(36,-1){\makebox(0,0){2}}
\put(42,-1){\makebox(0,0){73/36}}
\put(1,6){\makebox(0,0){2/3}}
\put(7,6){\makebox(0,0){4/3}}
\put(13,6){\makebox(0,0){11/9}}
\put(19,6){\makebox(0,0){73/36}}
\put(25,6){\makebox(0,0){49/36}}
\put(31,6){\makebox(0,0){73/36}}
\put(37,6){\makebox(0,0){23/9}}
\put(47,6){\makebox(0,0){97/36}}
\put(4,11){\makebox(0,0){1}}
\put(12,9){\makebox(0,0){11/9}}
\put(19,11){\makebox(0,0){2}}
\put(28,11){\makebox(0,0){1}}
\put(36,9){\makebox(0,0){23/12}}
\put(44,11){\makebox(0,0){3}}
\put(7,16){\makebox(0,0){49/36}}
\put(17,16){\makebox(0,0){73/36}}
\put(31,16){\makebox(0,0){37/18}}
\put(41,16){\makebox(0,0){31/12}}
\put(10,21){\makebox(0,0){1}}
\put(18,19){\makebox(0,0){49/36}}
\put(24,19){\makebox(0,0){1}}
\put(30,19){\makebox(0,0){37/18}}
\put(37,21){\makebox(0,0){2}}
\put(13,26){\makebox(0,0){11/9}}
\put(19,26){\makebox(0,0){73/36}}
\put(25,26){\makebox(0,0){23/12}}
\put(35,26){\makebox(0,0){31/12}}
\put(17,31){\makebox(0,0){2}}
\put(24,29){\makebox(0,0){23/9}}
\put(32,31){\makebox(0,0){3}}
\put(19,36){\makebox(0,0){73/36}}
\put(29,36){\makebox(0,0){97/36}}
\normalsize
\put(24,-5){\makebox(0,0){$U(3,x)$}}
\end{picture}

\vspace*{5mm}
\caption{\footnotesize{The distributions $F(3,x)$, $G_2(3,x)$ and $U(3,x)$ for $SG(3)$.}} 
\label{FG2U3fig}
\end{figure}

The distributions $F(n,x)$, $G_2(n,x)$ and $U(n,x)$ for $n=3$ are shown in Fig. \ref{FG2U3fig}.

From Lemmas~\ref{lemmaF} to \ref{lemmaU}, we have the
following proposition:

\begin{propo}
\label{keyprop}
For almost all the vertex $x_{\vec{\gamma}}$ on the Sierpinski gasket $SG(n)$ with $n \geq 2$, we have $F(n,x_{\vec{\gamma}}) = G_2(n,x_{\vec{\gamma}})$, except the outmost vertices $a_n$, $b_n$, and the three vertices $x_{\vec{\gamma}^0}$ with $\vec{\gamma}^0=(n-1,2,2,...,2)$ and $x \in \{a,b,c\}$. The values for the vertices $x_{\vec{\gamma}^0}$ are given by 
\beqs
F(n,a_{\vec{\gamma}^0}) & = & \frac 5{12} \ , \cr\cr 
F(n,b_{\vec{\gamma}^0}) & = & \begin{cases} \frac 5{12} & \mbox{for odd} \ n \ , \\
\frac 16 & \mbox{for even} \ n \ ,
\end{cases} \quad 
F(n,c_{\vec{\gamma}^0}) = \begin{cases} \frac 16 & \mbox{for odd} \ n \ , \\
\frac 5{12} & \mbox{for even} \ n \ ,
\end{cases}
\label{Fs}
\eeqs
and
\beqs
G_2(n,a_{\vec{\gamma}^0}) & = & \frac 5{18} \ , \cr\cr 
G_2(n,b_{\vec{\gamma}^0}) & = & \begin{cases} \frac 5{18} & \mbox{for odd} \ n \ , \\
\frac 7{18} & \mbox{for even} \ n \ ,
\end{cases} \quad 
G_2(n,c_{\vec{\gamma}^0}) = \begin{cases} \frac 7{18} & \mbox{for odd} \ n \ , \\
\frac 5{18} & \mbox{for even} \ n \ .
\end{cases}
\label{G2s}
\eeqs
\end{propo}

\noindent {\it Proof:} 
For the outmost vertices $a_n$, $b_n$, we know $F(n,a_n) = F(n,b_n) = 1$, while there are no values assigned for $G_2(n,a_n)$ and $G_2(n,b_n)$ by definition.
From Lemma~\ref{lemmaG2}, we already know that $G_2(n,x_{\vec{\gamma}}) = F(n,x_{\vec{\gamma}})$ for the vertices with $\gamma_1 < n-1$ and the vertices with $\gamma_1=n-1$ and $\gamma_2=1$. We should only consider the vertices with $\gamma_1=n-1$ and $\gamma_2=2$ in this proof. 

Let us verify the values for the three special vertices $x_{\vec{\gamma}^0}$ with $\gamma_1=n-1$ and $\gamma_2=\gamma_3=\cdots=2$ first. Eqs. (\ref{Fs}) and (\ref{G2s}) are correct for $n=2, 3$ as shown in Figs. \ref{FG2U2fig} and \ref{FG2U3fig}. It is easy to obtain Eq. (\ref{G2s}) using the upper line of Eq. (\ref{g2-s3}). Eq. (\ref{Fs}) can also be verified by the upper line of Eq. (\ref{goal}) because $\hat{\tilde{x}}_{\vec{\gamma}^\prime}$ corresponds to the vertices with subscript $(n-1,1,1,\cdots)$ such that $G_2(n,\hat{\tilde{x}}_{\vec{\gamma}^\prime}) = F(n,\hat{\tilde{x}}_{\vec{\gamma}^\prime}) = F(n,\hat{\tilde{\hat{x}}}_{\vec{\gamma}^\prime})$.

Now consider the vertices with $\gamma_1=n-1$, $\gamma_2=2$ other than the three special vertices discussed above. For the vertices with $\gamma_1=n-1$, $\gamma_2=2$, $\gamma_3=0$, $F(n,x_{\vec{\gamma}}) = G_2(n,x_{\vec{\gamma}})$ can be verified by comparing the lower line of Eq. (\ref{g2-s3}) with the upper line of Eq. (\ref{goal}). Both of them contain the term $\frac12 G_2(n-1,\hat{\tilde{x}}_{\vec{\gamma}^\prime})$, while $F(n-1,\hat{\tilde{\hat{x}}}_{\vec{\gamma}^\prime}) = G_2(n-1,\hat{\tilde{\hat{x}}}_{\vec{\gamma}^\prime})$ for such vertices according to Eq. (\ref{g2-01}). For the vertices with $\gamma_1=n-1$, $\gamma_2=2$, $\gamma_3=1$, compare the upper lines of Eqs. (\ref{g2-s3}) and (\ref{goal}). We know $F(n-1,\hat{\tilde{\hat{x}}}_{\vec{\gamma}^\prime}) = G_2(n-1,\hat{\tilde{\hat{x}}}_{\vec{\gamma}^\prime})$ for such vertices according to Eq. (\ref{gu}), while $G_2(n-1,\hat{\tilde{x}}_{\vec{\gamma}^\prime}) = G_2(n-1,\tilde{\hat{\tilde{x}}}_{\vec{\gamma}^\prime}) = G_2(n-1,\hat{\tilde{\hat{x}}}_{\vec{\gamma}^\prime})$ for these vertices. Finally for the vertices with $\gamma_1=n-1$, $\gamma_2=2$ and $\gamma_3=2$ other than the three special vertices discussed above, again compare the upper lines of Eqs. (\ref{g2-s3}) and (\ref{goal}). We have $F(n-1,\hat{\tilde{\hat{x}}}_{\vec{\gamma}^\prime}) = G_2(n-1,\hat{\tilde{\hat{x}}}_{\vec{\gamma}^\prime})$ for such vertices by induction, while $G_2(n-1,\hat{\tilde{x}}_{\vec{\gamma}^\prime}) = F(n-1,\hat{\tilde{x}}_{\vec{\gamma}^\prime}) = F(n-1,\tilde{\hat{\tilde{x}}}_{\vec{\gamma}^\prime}) =
F(n-1,\hat{\tilde{\hat{x}}}_{\vec{\gamma}^\prime})$ for these vertices.
\EndProof

For any positive $\ell > 0$, define the mean $\ell$ displacement $\xi(n,\ell)$ as
\beq
\label{xi}
\xi(n,\ell) = \sum_{x\in v(SG(n))} |x|^\ell p(n,x) \ , \ \mbox{and} \qquad 
\psi(n,\ell) = \sum_{x\in v(SG(n))} |x|^\ell F(n,x) \ ,
\eeq
where $\xi(n,\ell) = \psi(n,\ell) / R(n)$ according to Eq. (\ref{pnx}).
We would like to study the behavior of $\xi(n,\ell)$ when $n$ is large. For that purpose, let us also define
\beq
\phi(n,\ell) = \sum_{x\in V(SG(n))}|x|^\ell U(n,x) \ , \qquad 
{\cal U}(n) = \sum_{x_{\vec{\gamma}}\in v(SG(n)),|\vec{\gamma}|>2} U(n,x) \ .
\eeq
We shall first derive upper and lower bounds for $\psi(n,\ell)$ and $\phi(n,\ell)$ as follows.
By the first line of Eq. (\ref{goal}) and Proposition~\ref{keyprop}, for every $n \geq 3$, the summation of the value of $F(n,x_{\vec{\gamma}})$ over most of the vertices in the upper $SG(n-1)$ is given as
\beq
\sum_{x_{\vec{\gamma}}\in v(SG(n)),
|\vec{\gamma}|>2,\gamma_1=n-1,\gamma_2=2} F(n,x_{\vec{\gamma}}) = \sum_{x_{\vec{\gamma}}\in v(SG(n-1))} F(n-1,x_{\vec{\gamma}}) + O(1) = R(n-1) + O(1) \ ,
\eeq
where from Eq. (\ref{Rn}) $R(n)=(\frac {7 \times 17}{2^4 \times 3^3})4^n + O(1)$ when $n$ is large, and the number of neglected vertices with $|\vec{\gamma}| = 1,2$ in the upper $SG(n-1)$ has order one. By Eq. (\ref{goal}), we have
\beq
\sum_{x_{\vec{\gamma}}\in v(SG(n))} F(n,x_{\vec{\gamma}}) = R(n) = \frac12 {\cal U}(n-1) + 2R(n-1) + O(n) \ ,
\eeq
because the number of neglected vertices with $|\vec{\gamma}| = 1,2$ has order $n$, so that when $n$ is large
\beq
{\cal U}(n-1) = 2 \bigl( R(n) - 2R(n-1) \bigr) + O(n) = R(n) + O(n) \ .
\eeq
For all the vertices $x_{\vec{\gamma}} \in v(SG(n))$  with $|\vec{\gamma}|>2,\gamma_1=n-1,\gamma_2=2$, the distance between $x_{\vec{\gamma}}$ and $o$, i.e. $|x_{\vec{\gamma}}|$, is larger than $2^{n-1}$ and less than $2^n$. Using the inequality: $2^{(n-1)\ell} \leq |x_{\vec{\gamma}}|^\ell \leq 2^{n \ell}$ for $\ell>0$, we have the upper and lower bounds for $\psi(n,\ell)$ as
\beq
\frac12 \phi(n-1,\ell) + 2 \Bigl \{ 2^{(n-1)\ell} \bigl[ R(n-1) + O(n) \bigr] \Bigr\} \le \psi(n,\ell) \leq \frac12 \phi(n-1,\ell) + 2 \Bigl\{ 2^{n \ell} \bigl[ R(n-1) + O(n) \bigr] \Bigr \} \ ,
\eeq
i.e.
\beqs
& & \frac12 \phi(n-1,\ell) + 2^{n\ell-\ell+1} \bigl[ (\frac {7 \times 17}{2^4 \times 3^3})4^{n-1} + O(n) \bigr ] \cr
& & \le \psi(n,\ell) \le \frac12 \phi(n-1,\ell) + 2^{n\ell+1} \bigl[ (\frac {7 \times 17}{2^4 \times 3^3})4^{n-1} + O(n) \bigr ] \ .
\label{psi}
\eeqs
Similarly, using Eqs. (\ref{gu}), (\ref{u-1}) and Proposition \ref{keyprop}, we have the upper and lower bounds for $\phi(n,\ell)$ as
\beqs
& & \phi(n-1,\ell) + 2^{(n-1)\ell} \times  2\Bigl\{ 2R(n-1) + {\cal U}(n-1) + O(n) \Bigr\} \cr
& & \le \phi(n,\ell) \le \phi(n-1,\ell) + 2^{n\ell} \times 2\Bigl\{ 2R(n-1) + {\cal U}(n-1) + O(n) \Bigr\} \ ,
\eeqs
i.e.
\beqs
& & \phi(n-1,\ell) + 2^{(n-1)\ell+1} \bigl[ 6 \times (\frac {7 \times 17}{2^4 \times 3^3}) \times 4^{n-1} + O(n) \bigr] \cr
& & \le \phi(n,\ell) \le \phi(n-1,\ell) + 2^{n\ell+1} \bigl[ 6 \times (\frac {7 \times 17}{2^4 \times 3^3}) \times 4^{n-1} + O(n) \bigr] \ .
\label{phi}
\eeqs
By induction, Eq. (\ref{phi}) becomes
\beqs
\label{phin}
(\frac {7 \times 17}{2^2 \times 3^2}) \Bigl\{ \frac {(2^{2+\ell})^n}{2^{2+\ell}-1} + 2^{(n-1)\ell}O(n)  \Bigr\} \leq \phi(n,\ell) 
\leq (\frac {7 \times 17}{2^2 \times 3^2}) \Bigl\{ \frac {2^\ell (2^{2+\ell})^n}{2^{2+\ell}-1} + 2^{(n-1)\ell}O(n)  \Bigr\} \ .
\eeqs
Substituting Eq. (\ref{phin}) into Eq. (\ref{psi}), we have
\beqs
& & (\frac {7 \times 17}{2^4 \times 3^3}) \times 2^{(\ell + 2)n - \ell} \Bigl\{ \frac {2^{1+\ell}+1}{2^{2+\ell}-1} + 4^{-(n-1)} O(n) \Bigr\} \cr
& \leq & \psi(n,\ell) \le (\frac {7 \times 17}{2^4 \times 3^3}) \times 2^{(\ell+2)n} \Bigl\{ \frac
{2^{1+\ell}+1}{2^{2+\ell}-1} + 4^{-(n-1)} O(n)  \Bigr\} 
\eeqs
for large $n$, which leads to the following theorem:

\bigskip

\begin{theo}
When $n$ is large, the mean $\ell$ displacement is bounded as
\beqs
2^{(n-1)\ell} \Bigl\{ \frac
{2^{1+\ell}+1}{2^{2+\ell}-1} + 4^{-n} O(n) \Bigr\} \leq
\xi(n,\ell) \leq 2^{n\ell} \Bigl\{ \frac
{2^{1+\ell}+1}{2^{2+\ell}-1} + 4^{-n} O(n) \Bigr\} \ .
\eeqs
\end{theo}

Similar to the consideration in \cite{hattori2}, we can calculate the value of $\ln \xi(n,\ell) / \ln L(w)$ in the large $n$ limit, where $L(w)$ is defined above Corollary \ref{Zn}.

\begin{cor} 
In the limit of infinite $n$, the exponent for the mean $\ell$ displacement is given as
\beq
\lim_{n\rightarrow\infty} \frac {\ln \xi(n,\ell)}{\ln [(3^{n+1}+1)/2]} = \ell \ \frac{\ln 2}{\ln 3} \ .
\eeq
\end{cor}

\acknowledgments

The authors would like to thank Weigen Yan for helpful discussion and Chuan-Hung Chen for use of computer facility.  The research of S.C.C. was partially supported by the NSC grant NSC-97-2112-M-006-007-MY3 and NSC-98-2119-M-002-001. The research of L.C.C was partially supported by TJ \& MY Foundation and NSC grant.
L.C.C. would like to thank PIMS, university of British Columbia for the hospitality.

\end{document}